\documentclass[galaxies,article,accept,pdftex,moreauthors]{Definitions/mdpi} 
\firstpage{1} 
\pubvolume{12}
\issuenum{3}
\articlenumber{25}
\pubyear{2024}
\copyrightyear{2024}
\externaleditor{\textls[-25]{Academic Editor: Luigina Feretti}}
\datereceived{2 April 2024} 
\daterevised{9 May 2024} 
\dateaccepted{10 May 2024} 
\datepublished{15 May 2024}
\hreflink{https://doi.org/10.3390/\linebreak galaxies12030025} 
\pdfoutput=1 

\usepackage{tablefootnote}
\usepackage{mathabx}
\usepackage{multicol}   
\usepackage{bm}		
\usepackage[normalem]{ulem} 
\usepackage{longtable}
\usepackage{pdfpages}
\usepackage{subcaption}
\usepackage{pdflscape}
\newcommand\farcm{\hbox{$.\mkern-4mu^\prime$}}

\Title{{High--redshift quasars at $z \geq 3$: radio variability and MPS/GPS candidates}}

\TitleCitation{High--redshift quasars at $z \geq 3$}

\Author{Yulia Sotnikova $^{1,2,3}$*\orcidA{}, Alexander Mikhailov $^{1}$\orcidE{}, Timur Mufakharov $^{1,2,3}$\orcidB{}, Tao An $^{4,5}$\orcidC{}, Dmitry Kudryavtsev $^{1}$\orcidD{}, Marat Mingaliev $^{1,3,6}$\orcidK{}, Roman Udovitskiy $^{1}$, Anastasia Kudryashova $^{1}$\orcidG{}, Vlad Stolyarov $^{1,7}$\orcidF{} and Tamara Semenova$^{1}$\orcidH{}}

\AuthorNames{Yu. Sotnikova, A. Mikhailov and T. Mufakharov}

\AuthorCitation{Sotnikova, Yu.; Mikhailov, A.; Mufakharov, T.}

\address{%
$^{1}$ \quad Special Astrophysical Observatory of the Russian Academy of Sciences, Nizhny Arkhyz, 369167, Russia\\
$^{2}$ \quad Institute for Nuclear Research of the Russian Academy of Sciences, Moscow 117312, Russia\\
$^{3}$ \quad Kazan Federal University, 18 Kremlyovskaya St, Kazan 420008, Russia\\
$^{4}$ \quad Shanghai Astronomical Observatory, Chinese Academy of Sciences, 80 Nandan Road, Shanghai 200030, P. R. China\\
$^{5}$ \quad Key Laboratory of Radio Astronomy and Technology, Chinese Academy of Sciences, A20 Datun Road, Beijing, 100101, P. R. China\\
$^{6}$ \quad Institute of Applied Astronomy, Russian Academy of Sciences, St. Petersburg 191187, Russia\\
$^{7}$ \quad Astrophysics Group, Cavendish Laboratory, University of Cambridge, Cambridge, CB3 0HE, UK
}

\corres{Correspondence: lacerta999@gmail.com}

\abstract{We present a study of the radio variability of bright, $S_{1.4}\geq100$ mJy, high-redshift  quasars at $z\geq3$ on timescales up to 30--40 years. The study involved simultaneous RATAN-600 measurements at frequencies of 2.3, 4.7, 8.2, 11.2, and 22.3 GHz in  2017--2020. In addition, data from the literature were used. We have found that the variability index, $V_S$, which quantifies the normalized difference between the maximum and minimum flux density while accounting for measurement uncertainties, ranges from 0.02 to 0.96 for the quasars. Approximately half of the objects in the sample exhibit a variability index within the range of 0.25 to 0.50, comparable to that observed in blazars at lower redshifts. The distribution of $V_S$ at 22.3 GHz is significantly different from that at 2.3--11.2 GHz, which may be attributed to the fact that a compact AGN core dominates at the source's rest frame frequencies greater than 45 GHz, leading to higher variability indices obtained at 22.3 GHz (the $V_S$ distribution peaks around 0.4) compared to the lower frequencies (the $V_S$ distribution at 2.3 and 4.7 GHz peaks around 0.1--0.2). Several source groups with distinctive variability characteristics were found using cluster analysis of quasars. We propose 7 new candidates for gigahertz peaked-spectrum (GPS) sources and 5 new megahertz peaked-spectrum (MPS) sources based on their spectrum shape and variability features. Only 6 out of 23 sources previously reported as GPS demonstrate a low variability level typical of classical GPS sources ($V_{S} < 0.25$) at 4.7--22.3~GHz. When excluding the highly variable peaked-spectrum blazars, we expect no more than 20\% of the sources in the sample to be GPS candidates and no more than 10\% to be MPS candidates.}


\begin{document}

\section{Introduction}

Quasars \citep{1963Natur.197.1040S} are highly variable celestial objects \citep{1963ApJ...138...30M} that emit radiation across the electromagnetic spectrum. In the radio band, their variability \citep{1965Sci...148.1458D,1968ARA&A...6..417K} is often attributed to the presence of outflows or jets that emanate from the central region of a galaxy \citep{1970Natur.227.1303R}. These jets can exhibit a range of behaviors, including propagating shocks (e.g., \citep{1985ApJ...298..114M,2021IAUS..359...27N}), precession (e.g., \citep{2009ApJ...697.1621G,2020MNRAS.499.5765H}), bending (e.g., \citep{1999A&A...347...30V}), and interaction with gas clouds (e.g., \citep{2019MNRAS.485.2710J}). In addition, several other phenomena may contribute to the observed flux density fluctuations in quasars, such as accretion disk instabilities (e.g., \citep{2013MNRAS.434.3487A,2014MNRAS.443...58W}), interstellar scintillation (e.g., \citep{1998MNRAS.294..307W,2019MNRAS.489.5365K}), and lensing by the ionized structures (e.g., \citep{1987Natur.326..675F,2013A&A...555A..80P}).

Quasars at high redshifts play a crucial role in understanding the evolution of galaxies and the influence of active galactic nuclei (AGNs) in shaping the Universe. However, they are rarely detected in the radio band due to their weak flux densities. These quasars typically exhibit either a steep, flat, or peaked radio spectrum \citep{2000A&AS..143..303D,2006MNRAS.366...58D,2016MNRAS.463.3260C}. The peaked-spectrum sources are classified as high-frequency peakers (HFPs) as well as megahertz and gigahertz peaked-spectrum (MPS and GPS) sources \citep{1991ApJ...380...66O,2016MNRAS.459.2455C,2021A&ARv..29....3O} based on the location of the peak frequency in their radio spectra. The peaked-spectrum (PS) sources represent an early stage in the radio source evolution, where HFPs may evolve to GPSs, and some GPSs may further develop into extended FRI/II radio galaxies  \citep{1990A&A...231..333F,2000MNRAS.319..445S,2012ApJ...760...77A}. Some other GPS sources exhibit intermittent and short-lived radio emission, offering insights into the dynamic nature of quasars and AGN evolution \citep{1996ApJ...460..612R}. \citet{2015MNRAS.450.1477C} proposed to search for high-redshift ($z>2$) sources by identifying MPS sources that are compact on scales of tens of milliarcseconds; though later \citet{2017MNRAS.467.2039C} found a roughly equal proportion of flat, steep, and peaked radio spectra among 30 quasars at $z>4.5$. 

In a study of the radio spectra of over a hundred radio-loud $z\geq3$ quasars, we found that almost half of the sample can be classified as PS sources, either GPS or MPS \citep{2021MNRAS.508.2798S}. A significant fraction of PS sources in this sample are identified as blazars which are more easily detectable due to the beamed emission from their jets compared to unbeamed radio galaxies and AGNs. For blazar-type sources, beaming not only amplifies the flux density but can also shift the peak frequencies to higher values, contributing in certain cases to their classification as HFPs. Variability serves as an important indicator to distinguish between blazars and unbeamed PS sources. Genuine GPS sources exhibit low levels of variability (e.g., \citep{1998A&AS..131..303S,1998PASP..110..493O,1999MmSAI..70..117S,2009AN....330..128T,2012A&A...544A..25M,2013AstBu..68..262M,2019AstBu..74..348S,2021AN....342.1195S}), which may be attributed to the expansion of the compact lobes. This expansion usually does not change the shape of the spectrum, although the peak frequency might shift \citep{2016AN....337..120D}. In contrast, blazars exhibit strong and rapid flux variations in a wide range of timescales and frequencies, often accompanied by changes in the spectrum shape, such as a peaked spectrum during flaring activity (e.g., \citep{2005BaltA..14..413K,2010MNRAS.408.1075O,2021AN....342.1195S}). By classifying GPS/MPS sources and blazars among high-redshift radio sources, we can gain insights into the underlying physical mechanisms that govern their formation and evolution, which may have profound implications for our understanding of the early Universe. In this study we determine the incidence of GPS and MPS sources in the early Universe and assess the degree to which they are contaminated by blazars.

VLBI studies of high-redshift quasars reveal two major morphological patterns: a core--jet structure, indicative of relativistic beaming of the jet, and a compact double-lobed morphology, the so-called compact symmetric objects (CSOs) with their jets lying on the plane of the sky and relativistic beaming is not significant. Typically, quasars with the core--jet morphology have flat radio spectra at gigahertz frequencies, while CSOs have peaked radio spectra, small sizes ($<1$ kpc) and are classified as MPS/GPS sources \citep{1991ApJ...380...66O,2016MNRAS.459.2455C,2021A&ARv..29....3O}). 

To identify blazars, GPS, and MPS sources, multi-frequency radio measurements provide fundamental insights into the amplitude, time scale, and frequency evolution of the flux density variability. The features of variability can be used to identify blazar candidates as well as GPS and MPS sources. The aim of this paper is to study the radio variability properties of bright quasars at $z\geq3$ on historical timescales of 30--40 yrs. We used the multi-frequency RATAN-600 measurements obtained from 2017 to 2020 quasi-simultaneously at frequencies of 2.3--22.3 GHz, complemented by available literature data. Utilising the dataset compiled by \citet{2021MNRAS.508.2798S}, we present the average variability parameters for the quasars, discuss flux density variations for individual objects, and classify GPS/MPS sources according to their long-term spectral behaviour.


\section{Sample}

The sample contains 101 quasars selected from the NRAO VLA Sky Survey \citep[NVSS,][]{1998AJ....115.1693C} in the declination range from $-34^{\degree}$ to $+49^{\degree}$ with the flux density $S_{1.4}\geq100$ mJy and with redshifts $z\geq 3$. The spectroscopic redshifts were obtained by cross-identification in the National Aeronautics and Space Administration (NASA)/Infrared Processing and Analysis Center (IPAC) Extragalactic Database (NED).\!\footnote{NED: \url{https://ned.ipac.caltech.edu}} 
Observations of the sample were conducted with the RATAN-600 radio telescope during 2017--2020 at frequencies of 1.2, 2.3, 4.7, 8.2, 11.2, and 22.3 GHz quasi-simultaneously, within a 5--6 minute interval. The flux densities were measured with a standard error of about 5--20\% at 4.7, 8.2, 11.2 GHz and 10--35\% at 1.2, 2.3, and 22.3 GHz \citep{2021MNRAS.508.2798S}. The parameters of the receivers are shown in Table~\ref{tab:par}. The detection limit for a RATAN-600 single sector is approximately 5 mJy at 4.7 GHz (integration time is about 3s) under good conditions and at the average antenna elevation angle (${\rm Dec}\sim0^{\circ}$). 

\begin{table}[H]
\caption{RATAN-600 continuum radiometer parameters: the central frequency $f_0$, the bandwidth $\Delta f_0$, the detection limit for point sources per transit $\Delta F$, and the angular resolution, measured by the full width at half maximum (FWHM) along RA and Dec, calculated for the average angles.} 
\label{tab:par}
\newcolumntype{C}{>{\centering\arraybackslash}X}
\begin{tabularx}{\textwidth}{CCCCC}
\toprule
\textbf{$f_{0}$} & \textbf{$\Delta f_{0}$} & \textbf{$\Delta F$} &  \textbf{FWHM$_{\rm {RA}}$}  & \textbf{FWHM$_{\rm{Dec}}$} \\
\textbf{(GHz)} & \textbf{(GHz)} & \textbf{(mJy/beam)} & \textbf{(arcmin)}  & \textbf{(arcmin)} \\
\midrule
 $22.3$ & $2.5$  &  $50$ & $0\farcm17$ & $1\farcm6$  \\ 
 $11.2$ & $1.4$  &  $15$ & $0\farcm34$ & $3\farcm2$ \\ 
 $8.2$  & $1.0$  &  $10$ & $0\farcm47$ & $4\farcm4$   \\ 
 $4.7$  & $0.6$  &  $8$  & $0\farcm81$ & $7\farcm6$   \\ 
 $2.3$  & $0.08$  &  $40$ & $1\farcm71$ & $15\farcm8$  \\ 
 $1.2$  & $0.08$ &  $200$ & $3\farcm07$ & $27\farcm2$ \\ 
\bottomrule
\end{tabularx}
\end{table} 

Our analysis of the radio spectra, which incorporate both the RATAN-600 measurements and the literature data, revealed that 47 out of 101 quasars (46\%) exhibit peaked radio spectra, while 24\% have flat spectra,\!\footnote{Spectral index estimated from the power law $S \sim \nu^\alpha$} and 15\% have steep ones. A small number of the sources have complex, upturn, or inverted spectra. In this paper the spectral indices are used in terms of $\alpha_{\rm low}$ and $\alpha_{\rm high}$ as they were defined in \cite{2021MNRAS.508.2798S}. For peaked and upturn spectra, $\alpha_{\rm low}$ and $\alpha_{\rm high}$ are determined lower and higher the frequency where the spectral slope changes its sign from positive to negative or vice versa. For the rest of spectral types, the low-frequency spectral index is calculated using decimetre-wavelength measurements, such as Giant Metrewave Radio Telescope Sky Survey (TGSS, \cite{2017A&A...598A..78I}), GaLactic and Extragalactic All-sky Murchison Widefield Array (GLEAM, \cite{2017MNRAS.464.1146H}), or  Westerbork Northern Sky Survey (WENSS, \cite{1997A&AS..124..259R}). The high-frequency spectral index is estimated using centimetre catalogue data points. A spectrum is considered flat if the spectral index is $-0.5 \leq \alpha \leq 0$, and inverted if $\alpha > 0$. The spectrum is assumed to be complex for indeterminate shapes with two or more maxima or minima. The spectrum is called steep or ultra-steep for $-1.1 < \alpha < -0.5$ and $\alpha \leq -1.1$ respectively.

The list of the sources and their characteristics are presented in Table~\ref{TableA1}: Col.~1 is the source name; Col.~2 is the redshift; Col.~3 is the average RATAN-600 flux density at 4.7~GHz; Cols.~4--5 are the radio morphology and its reference; Cols.~6--7 are the spectral indices, calculated in \cite{2021MNRAS.508.2798S}; Col.~8 is the radio spectrum type; Col.~9 is the blazar type; Col.~10 is the number of cluster. Forty-eight quasars in the source list are classified as blazars according to the ``Open Universe for Blazars---Reference list V2.0''.\footnote{\url{https://openuniverse.asi.it/OU4Blazars/MasterListV2/}}, which includes 6077 blazars identified through radio, X-ray, and $\gamma$-ray surveys as well as objects discovered based on their multi-frequency SED properties. This database serves as the largest compilation of blazars to date \citep{2019A&A...631A.116G}. It contains the blazars from the Roma-BZCAT catalogue \citep{2015Ap&SS.357...75M}, the classification of which we used in our research. Most of the quasars of the sample (88 of 101) have been observed using VLBI at 2, 5, 8, or 15 GHz, with the majority of them (66 of 88) exhibiting a core--jet morphology.

Compared to the sample in \cite{2021MNRAS.508.2798S} we have excluded the source J1600$+$0412 (87GB 155733.8$+$042120). J1600$+$0412 has $z=0.79$ in the latest data release of the Sloan Digital Sky Survey \citep{2015MNRAS.452.4153A,2018A&A...613A..51P}, but previously its redshift was miscalculated as $z=3.11$ \citep{2003ARep...47..458A}. 

There are six gravitationally lensed quasars in the sample.\!\footnote{marked by the letter ``a'' in Table~\ref{TableA1}} Their structure is unresolved by RATAN-600 because the maximum separations between them are in the range of $0.54^{\second}-3.4^{\second}$ which is less than the angular resolution at any RATAN-600 frequency. The notes for them are given in Appendix~\ref{sec:gravlensed}. 

\subsection{The literature data}
\label{sec:literature}
The literature data and their use in our study have been previously described in detail in \cite{2021MNRAS.508.2798S}. The vast majority of the external radio continuum measurements were obtained from the astrophysical CATalogs support System (CATS database,\!\footnote{\url{https://www.sao.ru/cats}} \cite{2005BSAO...58..118V,2009DatSJ...8...34V}). These data allow us to estimate radio flux variability over a time period of up to 30-40 years at the most representative observation frequency of 4.7 GHz. The external data were adopted within 10\% ranges of the RATAN-600 five frequencies. For example, at $\nu = 4.7$ GHz we used all data within $4.23$--$5.17$ GHz. The majority of the quasars (94/101) has more than 30~years of observing coverage, and 78 out of 101 have more than 10~radio measurements at 4.7 GHz (Table~\ref{TableA2}). The median values of the number of measurements (N$_{\rm obs}$) and the years of observations at 4.7 and 22.3 GHz ($t$) for the sample are 16/35 and 9/14 respectively (Table~\ref{table:obs}). In our study we define the timescale of radio monitoring in the rest frame as $t_{\rm rest} = t_{\rm obs} / (1+z)$; if not specified, the timescale is used in the observer's frame of reference.

\begin{table}[H]
\caption{\label{table:obs} The number of quasars (N) observed at 4.7 and 22.3~GHz with the median values of observed epochs N$_{\rm obs}$ and monitoring timescales in the observer's ($t_{\rm obs}$) and source's ($t_{\rm rest}$) frames of reference in years.}
\newcolumntype{C}{>{\centering\arraybackslash}X}
\begin{tabularx}{\textwidth}{CCCCC}
\toprule
 \textbf{Frequency} & \textbf{N} & \textbf{N$_{\rm obs}$} & \textbf{$t_{\rm obs}$} & \textbf{$t_{\rm rest}$} \\
 \textbf{(GHz)} & & & \textbf{(yrs)} & \textbf{(yrs)} \\
\midrule
4.7  & 101 & 16 & 35 & 8  \\
22.3 & 52  & 9  & 14 & 3  \\
\bottomrule
\end{tabularx}
\end{table} 

\section{Multi--band radio variability}

\subsection{Variability estimates}
In order to characterise the variability properties of the quasars at different frequencies, we have computed the variability and modulation indices and the fractional variability. The first and third characteristics take into account measurement uncertainties, while the modulation index and the fractional variability are less sensitive to outliers. The variability index was calculated using the formula from \cite{1992ApJ...399...16A}:
\begin{equation}
\label{eq:Var}
V_{S}=\frac{(S_{\rm max}-\sigma_{S_{\rm max}})-(S_{\rm min}+\sigma_{S_{\rm min}})}
{(S_{\rm max}-\sigma_{S_{\rm max}})+(S_{\rm min}+\sigma_{S_{\rm min}})}
\end{equation}
where $S_{\rm max}$ and $S_{\rm min}$ are the maximum and minimum flux densities over all epochs of observations; $\sigma_{S_{\rm max}}$ and $\sigma_{S_{\rm min}}$ are their errors. This formula prevents one from overestimating the variability when there are observations with large uncertainties in the data. We obtain a negative value of $V_{S}$ in the case where the flux error is greater than the observed scatter in the data. 

The modulation index, defined as the standard deviation of flux density $\sigma_{S}$ divided by the mean flux density $\bar S$, was calculated as in \cite{2003A&A...401..161K}:

\begin{equation}
    M = \frac{\sigma_{S}}{\bar S}
\end{equation}

The fractional variability $F_{S}$ is defined as in \cite{2003MNRAS.345.1271V}:
\begin{equation}
\label{eq:frac}
F_{S}=\sqrt{\frac{V^2-\bar\sigma^{2}_{err}}{\bar S^2}}
\end{equation}
where $V^2$ is the variance, $\bar S$ is the mean flux density, and $\sigma_{err}$ 
is the root mean square error. The uncertainty of $F_{S}$ is determined as:
\begin{equation}
\label{eq:frerr}
\bigtriangleup F_{S}=\sqrt{\left(\sqrt{\frac{1}{2N}}\frac{\bar\sigma^{2}_{err}}{F_{S}*\bar S^2}\right)^2 + \left(\sqrt{\frac{\bar \sigma^{2}_{err}}{N}}\frac{1}{\bar S}\right)^2}
\end{equation}
where $N$ is the number of observations.

\begin{figure*}
\centerline{\includegraphics[width=\textwidth]{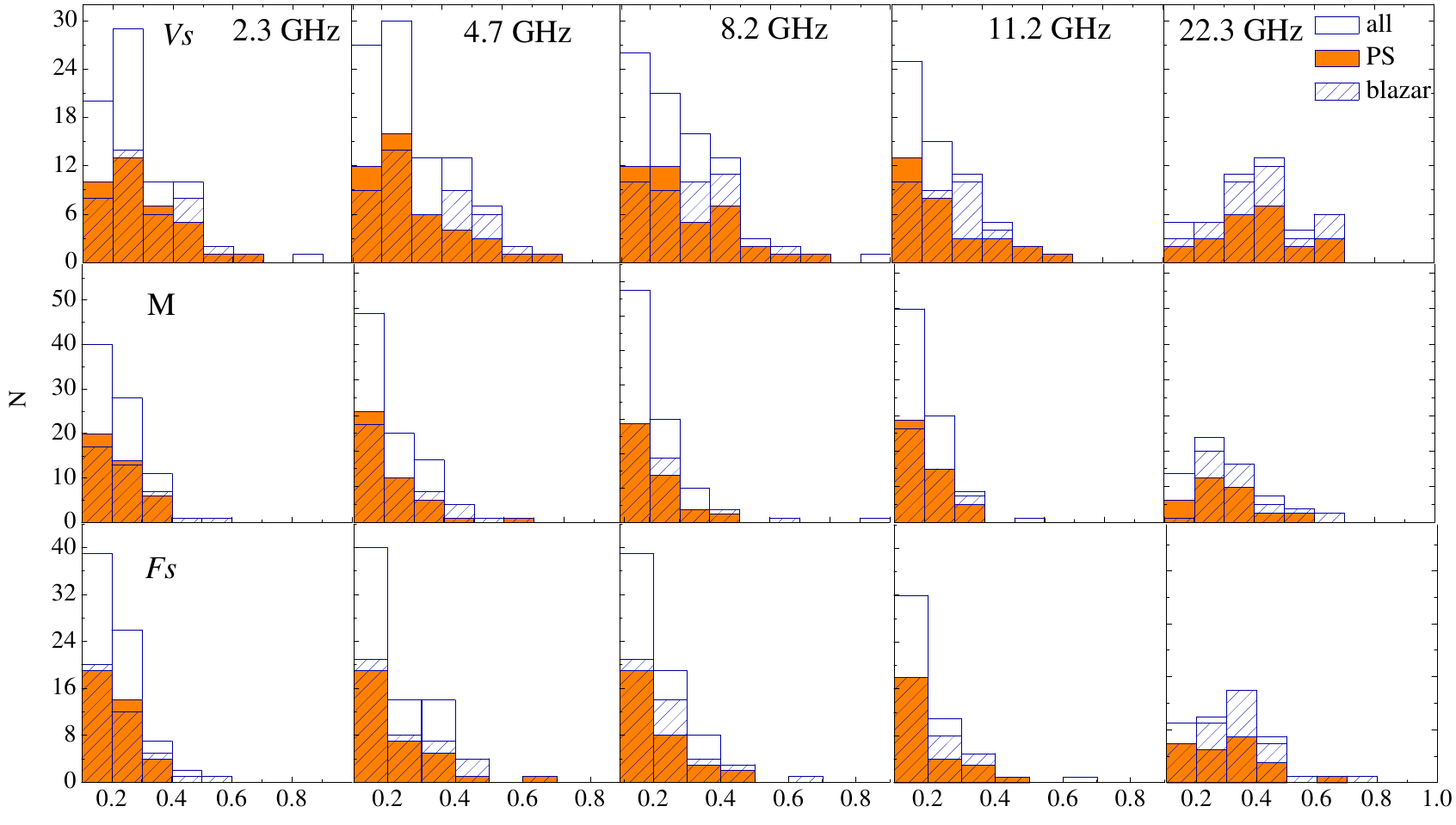}}
\caption{The variability (top row), modulation (middle row), and fractional variability (bottom row) index distributions for the whole sample, peaked-spectrum (PS) sources, and blazars.}
\label{fig:var_distr}
\end{figure*}

The distributions of the variability index, modulation index, and fractional variability at 2.3--22.3 GHz are shown in Fig.~\ref{fig:var_distr}. At 1.2 GHz we did not obtain sufficient number of measurements to estimate variability. Table~\ref{TableA2} lists these parameters for each quasar, while their statistics are given in Table~\ref{table1}. The negative variability indices have been excluded from the calculation and distribution diagrams.

Light curves of the quasars obtained with RATAN-600 are presented in Appendix D. For PS candidates we included additional RATAN-600 measurements of 2006--2017 from \cite{2012A&A...544A..25M,2019AstBu..74..348S}. Their radio spectra, compiled using both the RATAN-600 and CATS radio data, are also presented in Appendix D. 

\begin{figure}
\centerline{\includegraphics[width=0.85\textwidth]{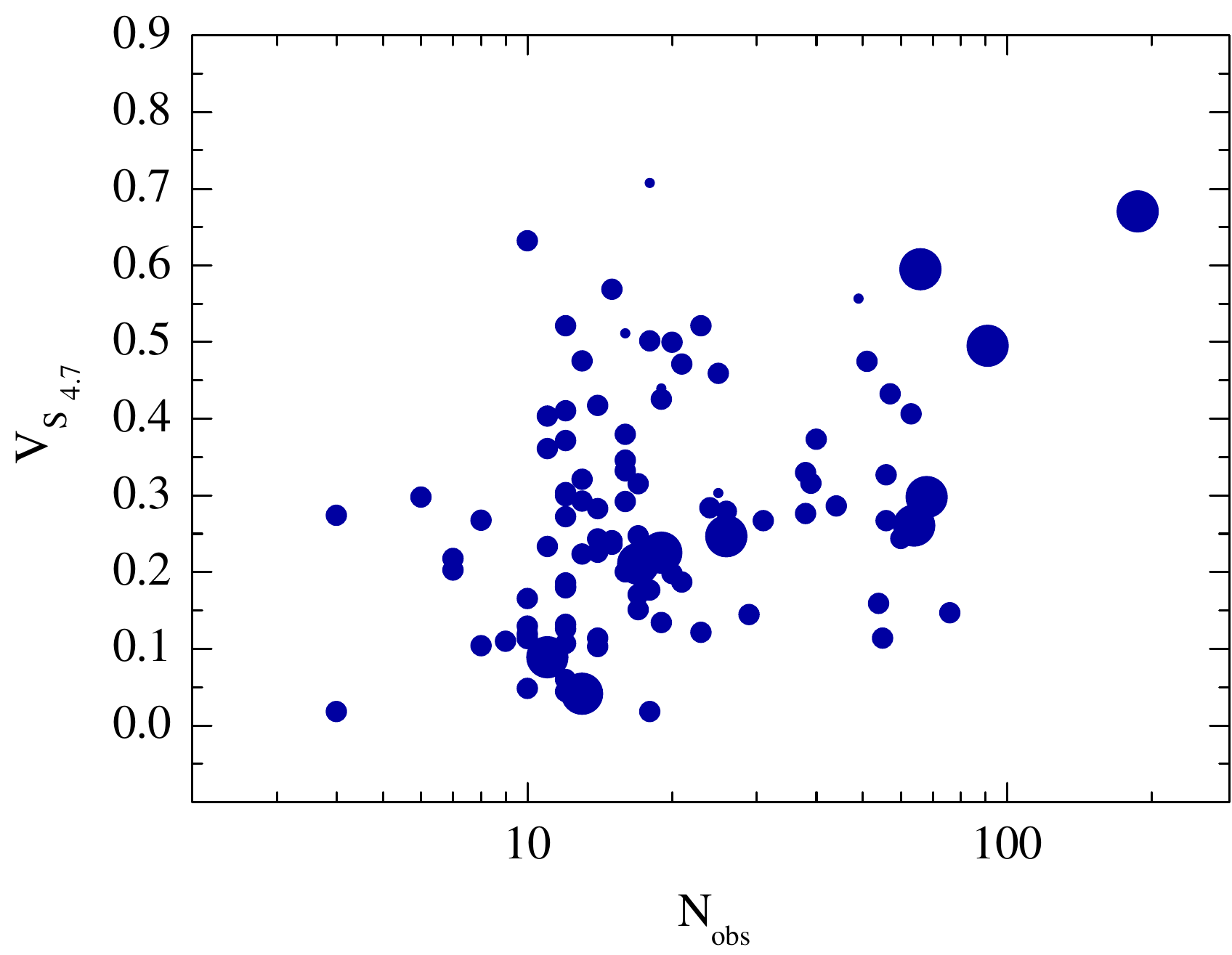}}
\caption{The variability index $V_{S_{4.7}}$ versus the number of observations N$_{\rm obs}$ for the quasars. The size of the symbols is proportional to years of monitoring in the rest frame, we define three scales: $t_{\rm rest} \leq~5$, $5 < t_{\rm rest} \leq 10$, and $t_{\rm rest} >10$ years. Bright quasars have been observed more frequently and for a longer period.}
\label{fig:N}
\end{figure}

The number of observing epochs N$_{\rm obs}$ and the timescale $t$ play a key role in exploring AGN variability. Previous studies have known that variability tends to increase with a greater number of observations, as sparser sampling may lead to missed peaks of variability \citep{2000AJ....120.2278T, 2007AJ....133.1947N, 2024arXiv240202283K}. This finding is verified by our observations, i.e., we found a correlation between N$_{\rm obs}$ and the variability index at 4.7 GHz with Pearson's $r=0.35$, p-val.~$<10^{-3}$ (Fig.~\ref{fig:N}). The observations of quasars in our sample are not homogeneous and differ in timescales (see the parameters $t_{4.7}$ and N$_{4.7}$ in Table~\ref{TableA2}), affecting the revealed variability level. The largest number of observations at 4.7 GHz is N$_{\rm obs}=187$ ($V_{S_{4.7}}=0.67$) for the blazar J0646$+$4451.

We note that the distributions of $M$ and $F_{S}$ are similar (see Fig.~\ref{fig:var_distr}). Both these parameters are suitable for a large number of observations N$_{\rm obs}$; however, unlike the modulation index, the fractional variability takes measurement errors into account. Therefore, the third parameter, $V_{S}$, can be used for any number of observations. For these reasons, we will use the variability index $V_{S}$ and the fractional variability $F_{S}$ in our analysis, as these parameters can be effectively utilized regardless of the number of observations. We note that we see in our sample a correlation between variability index and fractional variability.

\begin{table}[H]
\caption{\label{table1} The median, mean, maximum, and minimum values of the modulation index $M$, variability index $V_{S}$, and fractional variability $F_{S}$ calculated for different frequency bands (see text).}
\newcolumntype{C}{>{\centering\arraybackslash}X}
\begin{tabularx}{\textwidth}{CCCCCC}
\toprule
     & \textbf{N} & \textbf{median} & \textbf{mean} &  \textbf{min} & \textbf{max} \\
\midrule
$M_{2.3}$     & 94 & 0.18 & 0.20$\pm$0.09 & 0.01 & 0.53 \\
$V_{S_{2.3}}$ & 87 & 0.23 & 0.25$\pm$0.15 & 0.01 & 0.86 \\
$F_{S_{2.3}}$ & 80 & 0.20 & 0.21$\pm$0.09 & 0.03 & 0.59 \\
\midrule
$M_{4.7}$     & 101 & 0.17 & 0.20$\pm$0.11 & 0.05 & 0.62 \\
$V_{S_{4.7}}$ & 101 & 0.27 & 0.28$\pm$0.15 & 0.02 & 0.71  \\
$F_{S_{4.7}}$ & 90  & 0.17 & 0.20$\pm$0.12 & 0.01 & 0.66 \\
\midrule
$M_{8.2}$     & 100 & 0.17 & 0.21$\pm$0.12 & 0.03 & 1.00 \\
$V_{S_{8.2}}$ & 95  & 0.23 & 0.27$\pm$0.17 & 0.02 & 0.96 \\
$F_{S_{8.2}}$ & 84 & 0.17 & 0.21$\pm$0.14 & 0.04 & 0.97 \\
\midrule
$M_{11.2}$     & 92 & 0.17 & 0.18$\pm$0.09 & 0.01 & 0.57 \\
$V_{S_{11.2}}$ & 77 & 0.18 & 0.21$\pm$0.14 & 0.02 & 0.66 \\
$F_{S_{11.2}}$ & 66 & 0.14 & 0.17$\pm$0.11 & 0.05 & 0.63 \\
\midrule
$M_{22.3}$     & 55 & 0.28 & 0.30$\pm$0.13 & 0.09 & 0.68 \\
$V_{S_{22.3}}$ & 51 & 0.36 & 0.35$\pm$0.18 & 0.04 & 0.68 \\
$F_{S_{22.3}}$ & 45 & 0.31 & 0.31$\pm$0.15 & 0.05 & 0.72 \\
\bottomrule
\end{tabularx}
\end{table}

\subsection{Refractive interstellar scintillation}
\label{sec:RISS}
The variability observed in this study can be caused by either intrinsic (related to the source properties) or extrinsic factors arising from the interaction of radio emission from a source with an inhomogeneous propagation medium.

The flickers caused by scattering on the inhomogeneities of the interplanetary medium have a characteristic timescale of the order of a second or less \citep{2018MNRAS.473.2965M}, so they are smoothed out during RATAN-600 observations due to the relatively long scan time of 3--5 minutes during the source's transit. Similarly, for measurements at other radio telescopes, interplanetary flickering is also insignificant, given the longer measurement times for continuum flux density. 

The scintillations caused by the propagation of radio waves in the interstellar medium can have a diffractive and refractive character. Diffractive scintillation is usually associated with an extremely compact source, such as a pulsar or a fast radio burst. Additionally, observations must be made in a narrow frequency band in order to resolve the fine-scale structures responsible for the diffractive scintillation \citep{1992RSPTA.341..151N}. This is because the scintillation pattern can be highly frequency-dependent. The continuum radio measurements are carried out in a wide band (from hundreds of MHz to several GHz); therefore, this type of flickering can not be detected in our data.

Let's focus on estimating the potential contribution of refractive interstellar scintillations (RISS) to the variability at the \mbox{RATAN-600} frequencies (1.2, 2.3, 4.7, 8.2, 11.2, and 22.3~GHz) for the sources from the sample (in the declination range from $-34\degree$ to $+49\degree$). According to \cite{1998MNRAS.294..307W}, the transition frequency $\nu_{0}$ separating strong and weak scattering modes lies in the range from 5 to 15 GHz for most quasars. We will adopt these transition frequency values to estimate the modulation level of the flux density $m$ and its timescale $t$. We used the following equations in the weak mode \citep{1998MNRAS.294..307W}: $m=\left(\frac{\nu_{0}}{\nu}\right)^{17/12}\left(\frac{\theta_{F}}{\theta_{\rm s}}\right)^{7/6}$, $t=\frac{1}{12}\sqrt{\frac{\nu_{0}}{\nu}}\frac{\theta_{\rm s}}{\theta_{F}}$ (days), where $\theta_{F}=\theta_{F0}\sqrt{\frac{\nu_{0}}{\nu}}$ is the size of the first Fresnel zone at the frequency $\nu$, and $\theta_{\rm s}$ is the source's angular size. Equations in the strong mode are according to \citep{1998MNRAS.294..307W}: $m=\left(\frac{\nu}{\nu_{0}}\right)^{17/30}\left(\frac{\theta_{r}}{\theta_{\rm s}}\right)^{7/6}$, $t=\frac{1}{12}\left(\frac{\nu_{0}}{\nu}\right)^{\frac{11}{5}}\frac{\theta_{\rm s}}{\theta_{r}}$ (days), where $\theta_{r}=\theta_{F0}\left(\frac{\nu_{0}}{\nu}\right)^{\frac{11}{5}}$. We estimated an angular size as $\theta_{\rm s} = 5\times\theta_{\rm min}$, where $\theta_{\rm min}=0.6\sqrt{S}/\nu$ is the minimum angular size for the stationary source of synchrotron radiation \citep{1988gera.book..563K}, $S$ is the flux density (in Jy) at the observer's frequency (in GHz). We adopted $S = 100$~mJy, taking into account our sample selection criterion and characteristic flux densities measured with RATAN-600. As a result we calculate the flux density modulations and their typical timescales, as shown in Table~\ref{tab:RISS}. We can conclude that for quasars far from the Galactic plane, the modulation level is about 1~per~cent, which is significantly lower than the average values of the variability level for our sample. Therefore, RISS contribute insignificantly to the overall variability for most of these objects, except for a few individual cases where variability reaches $\leq 5$~per~cent. For objects located close to the Galactic plane (the transitional frequency around 15 GHz), RISS can make a significant contribution at low frequencies of 1--2 GHz. However, at higher frequencies ($> 5$ GHz), this contribution is considerably smaller (again, for individual objects with a level of variability of no more than 10\%, the contribution of RISS is significant but not decisive). Only three quasars in the sample are relatively close to the Galactic plane, within the Galactic latitude $-15^{\degree}< b < +15^{\degree}$: J0624$+$3856, J0733$+$0456, and J2050$+$3127. Their variability level at 2.3 GHz is in the range of 19--47\%, which may have a contribution from RISS but can not be explained only by it.

Thus, with the exception of a few specific objects, the contribution of interstellar scintillation effects to the total variability of most quasars in our sample is insignificant (see Table~\ref{tab:RISS}).

\begin{table}[H]
\caption{\label{tab:RISS} RISS modulation and its typical timescale estimates for RATAN-600 frequencies at the transitional frequencies of 5 and 15 GHz. The numbers of sources with the variability index and the fractional variability less than 5\% and 10\% are shown in the two last columns.}
\begin{adjustwidth}{-\extralength}{0cm}
\newcolumntype{C}{>{\centering\arraybackslash}X} 
\begin{tabularx}{\fulllength}{CCCCCCCC}
\toprule
\textbf{Frequency}  & \textbf{Source's size} & \textbf{m (5 GHz)}  & \textbf{t (5 GHz)} &  \textbf{m (15 GHz)} & \textbf{t (15 GHz)} & \textbf{N, $var\leq5\%$} & \textbf{N, $5\%<var\leq10\%$} \\
\textbf{(GHz)}  & \textbf{(mas)} & \textbf{(\%)} & \textbf{days} &  \textbf{(\%)}  & \textbf{days} & \textbf{$V_{S}/F_{S}$} & \textbf{$V_{S}/F_{S}$} \\
\midrule
2.3 & 0.41 & 2.1 & 8.6 & 8.5 & 17.2 & 4/1 & 10/4 \\
4.7 & 0.20 & 1.2 & 4.2 & 4.7 & 8.4 & 6/3 & 4/14 \\
8.2 & 0.12 & 0.7 & 2.4 & 2.9 & 4.8 & 6/3 & 6/10 \\
11.2 & 0.08 & 0.6 & 1.8 & 2.3 & 3.5 & 11/7 & 9/9 \\
22.3 & 0.04 & 0.3 & 0.9 & 1.3 & 1.8 & 4/1 & 4/1 \\
\bottomrule
\end{tabularx}
\end{adjustwidth}
\end{table}

\subsection{Variability characteristics}
\begin{figure}
\begin{minipage}{\linewidth}
\center{\includegraphics[width=0.5\linewidth]{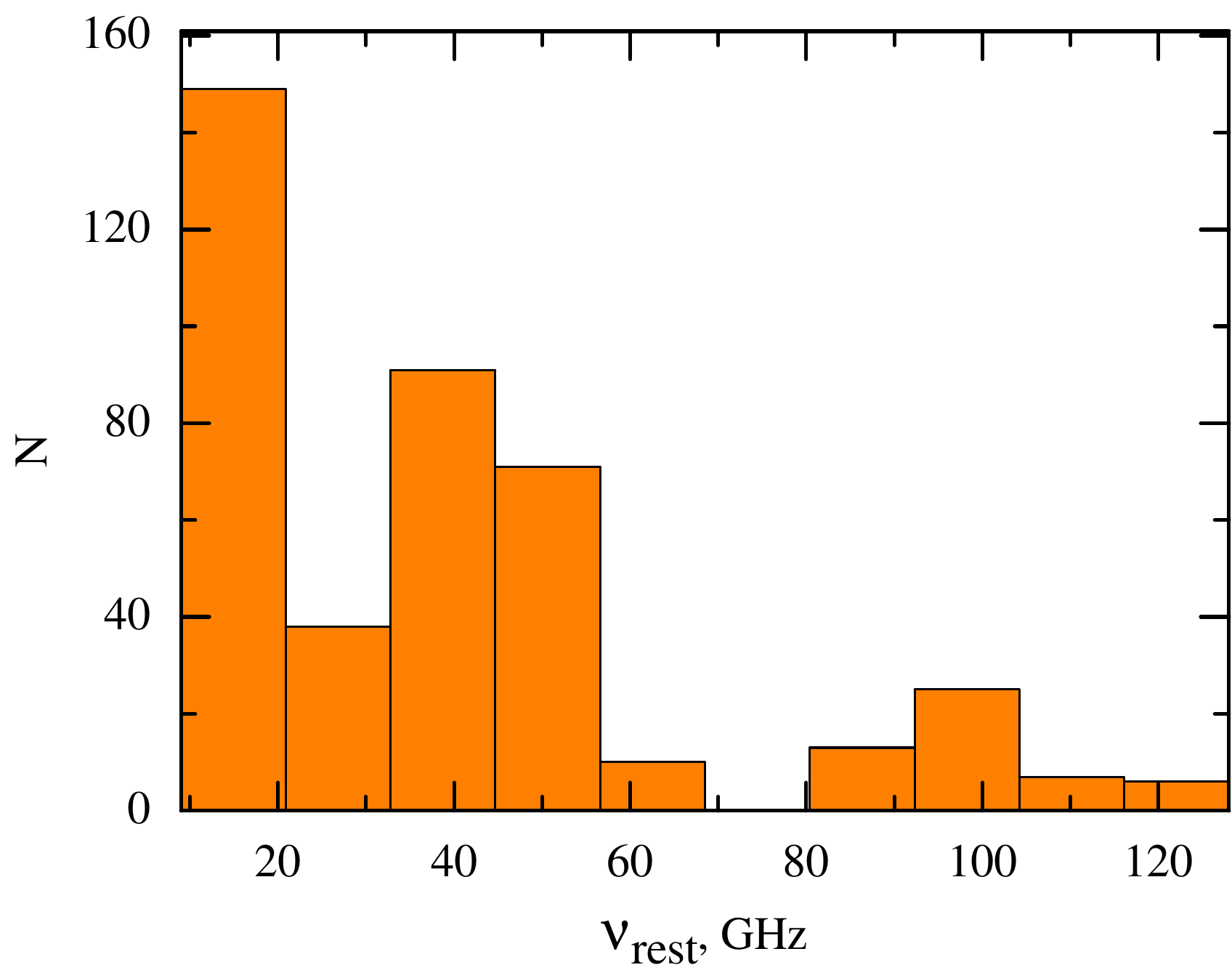}}
\end{minipage}
\hfill
\begin{minipage}{\linewidth}
\center{\includegraphics[width=0.5\linewidth]{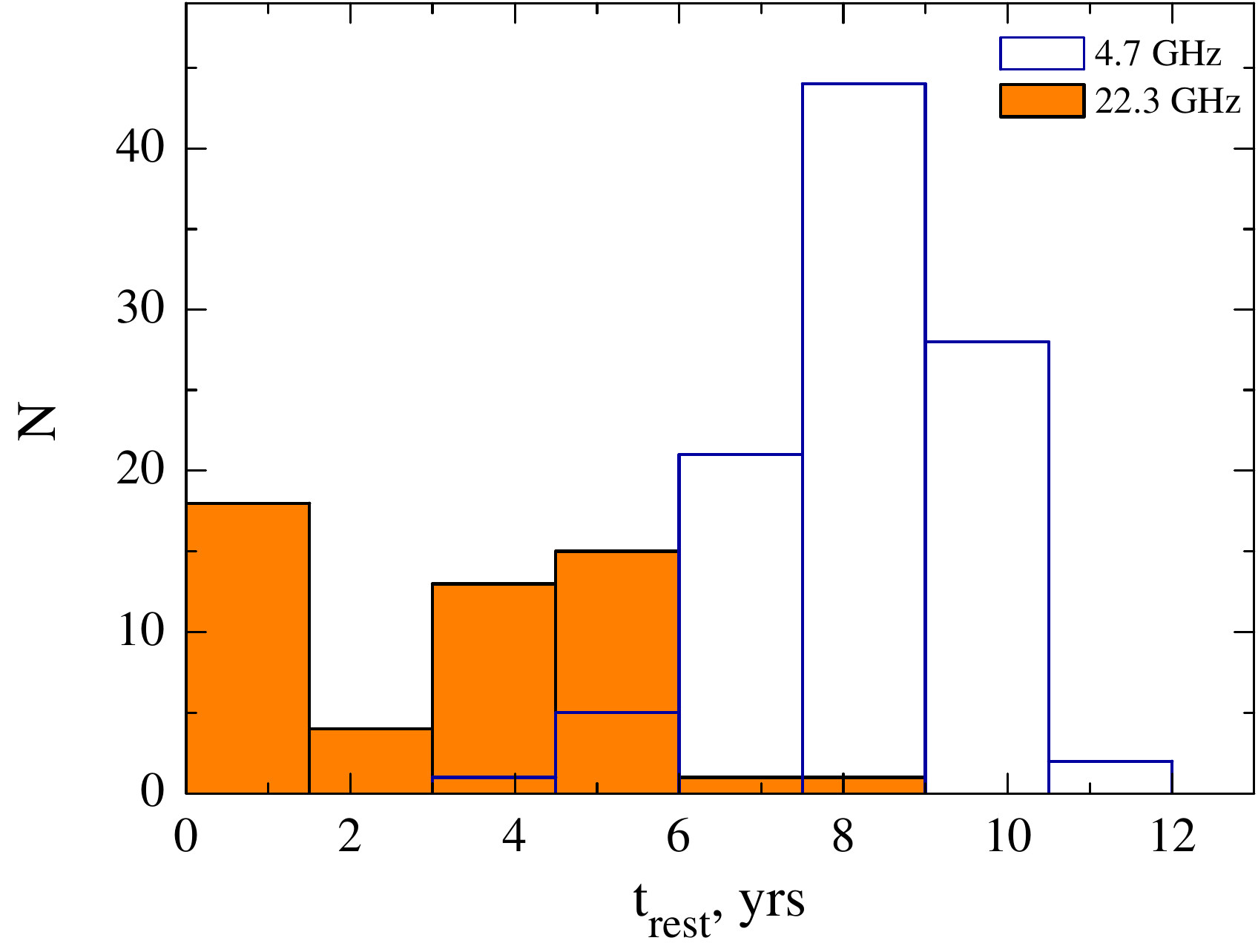}}
\end{minipage}
\caption{The distribution of the observational frequencies converted to the rest frame (top), and the distribution of the light curve durations in the rest frame, $t_{\mathrm{rest}}$ in years (bottom).}
\label{fig:nu_rest}
\end{figure}

\begin{figure}
\centerline{\includegraphics[width=0.75\columnwidth]{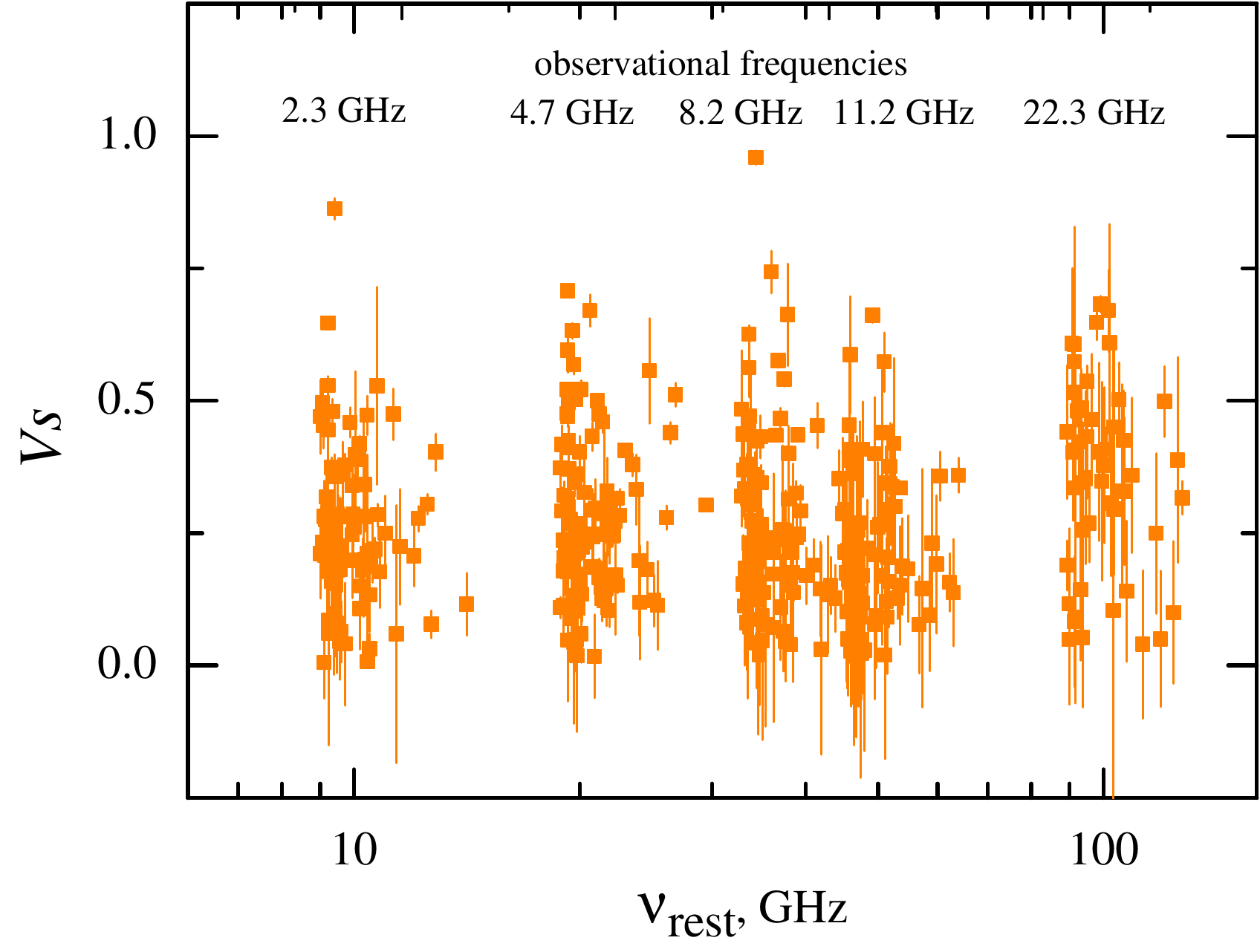}}
\caption{The variability indices plotted relatively to the rest frame frequencies. The outlier source is J0941$+$1145 with $V_S$ equal to 0.86 and 0.96.} 
\label{fig:Var-nu_rest}
\end{figure}

We computed variability amplitudes at certain observing frequencies. They correspond to different rest frame frequencies (Fig.~\ref{fig:nu_rest}, top), since the sources span a large redshift range. Furthermore, the timescale of the observations spans from about half a month to 7 years at 22.3 GHz and from 4 to 11 years at 4.7 GHz in the source's rest frame (Fig.~\ref{fig:nu_rest}, bottom), and consequently the median values of the variability timescale are 3 and 8 years at observing frequencies of 22.3 and 4.7 GHz respectively (Table~\ref{table:obs}). We found fairly wide scatter of the variability index throughout all the range of 9--130 GHz rest frame emission frequencies (Fig.~\ref{fig:Var-nu_rest}). Those emission frequencies are all associated with the dominating core component. There are no trends in the $V_S$ to $\nu_{\rm rest}$ relation, confirmed by Pearson's $r$=0.12 (p-val.=0.02).


We obtained the average variability level for the sample to be 0.20--0.30 at five frequencies (Table~\ref{table1}). For all spectral types, the average variability is higher at 22.3 GHz, up to $\sim0.40$ (Table~\ref{table3}). Maximum variability is observed in sources with flat, peaked, and complex spectra. Kolmogorov-Smirnov test confirms that the $V_S$ and $F_{S}$ distributions are not significantly different for the flat and peaked spectrum quasars.

We did not find a correlation between redshift and variability $V_{S}$ at 4.7 and 22.3 GHz. At redshifts greater than 3.5, we obtained the spread of variability indices at 4.7 GHz almost comparable to that at smaller redshifts (Fig.~\ref{fig55}). But there are scarce statistics on both the number of the highest-redshift quasars and the time scale of their radio monitoring. At 4.7 GHz ten extremely distant ($z>4$) sources exhibit a wide range of variability on a long time scale $t_{\rm rest} \sim 5$--10 yrs. The variability index varies from 0.12, which was obtained for the bright VLBI quasar \citep{2022ApJS..260...49K} J2134$-$0419 ($z=4.34$, N$_{\rm obs}$=18, $t_{\rm rest} \sim$ 6 yrs), to 0.71 for the peaked spectrum blazar J0525$-$3343 ($z=4.41$, N$_{\rm obs}$=12, at $t_{\rm rest} \sim$ 5.5 yrs) with an unresolved or slightly resolved core--jet structure in VLBI images \citep{2022ApJS..260...49K}. The most distant blazar in the sample ($z=5.28$) J1026+2542 with a one-sided jet morphology \citep{2015MNRAS.446.2921F} has a well determined peaked spectrum and a variability index of 0.30 over \mbox{$t_{\rm rest} \sim 3.9$}~yrs. However, even on a scale of 6--10 yrs in the rest frame there are about 10\% of quasars with $V_{S_{4.7}}< 0.10$ (Fig.~\ref{fig111}). Among them, about one third are PS quasars. 

\begin{table}[H]
\caption{\label{table3}The mean values of $F_{S}$ and $V_{S}$ for quasars of different radio spectrum types, N is the number of spectra at each frequency.}
\begin{adjustwidth}{-\extralength}{0cm}
\newcolumntype{C}{>{\centering\arraybackslash}X}
\begin{tabularx}{\fulllength}{LCCCCCCCCCC}
\toprule
 & \textbf{N$_{2.3}$} & \textbf{$F_{S_{2.3}}$} & \textbf{N$_{4.7}$} & \textbf{$F_{S_{4.7}}$} & \textbf{N$_{8.2}$} & \textbf{$F_{S_{8.2}}$} &\textbf{ N$_{11.2}$} & \textbf{$F_{S_{11.2}}$} & \textbf{N$_{22.3}$} & \textbf{$F_{S_{22.3}}$} \\
\midrule
peaked   & 41 & 0.19$\pm$0.08 & 43 & 0.19$\pm$0.12 & 41 & 0.18$\pm$0.10 & 31 & 0.18$\pm$0.09 & 23 & 0.28$\pm$0.13 \\
flat     & 20 & 0.22$\pm$0.09 & 22 & 0.20$\pm$0.11 & 22 & 0.21$\pm$0.19 & 20 & 0.14$\pm$0.09 & 10 & 0.38$\pm$0.17 \\
steep    &  9 & 0.20$\pm$0.04 & 10 & 0.21$\pm$0.12 &  1 & 0.50          &  4 & 0.35$\pm$0.19 &  7 & 0.17$\pm$0.13 \\
inverted &  5 & 0.32$\pm$0.17 &  8 & 0.20$\pm$0.13 &  8 & 0.28$\pm$0.05 &  6 & 0.11$\pm$0.04 &  6 & 0.24$\pm$0.12 \\
upturn   &  1 & 0.46          &  2 & 0.24$\pm$0.11 &  2 & 0.38$\pm$0.38 &  1 & 0.17          &  2 & 0.29$\pm$0.09 \\
complex  &  4 & 0.16$\pm$0.03 &  5 & 0.18$\pm$0.08 &  4 & 0.21$\pm$0.09 &  4 & 0.17$\pm$0.09 &  3 & 0.32$\pm$0.17 \\
\midrule
 & \textbf{N$_{2.3}$} & \textbf{$V_{S_{2.3}}$} & \textbf{N$_{4.7}$} & \textbf{$V_{S_{4.7}}$} & \textbf{N$_{8.2}$} & \textbf{$V_{S_{8.2}}$} & \textbf{N$_{11.2}$} & \textbf{$V_{S_{11.2}}$} & \textbf{N$_{22.3}$} & \textbf{$V_{S_{22.3}}$} \\
\midrule
peaked   & 43 & 0.26$\pm$0.14 & 47 & 0.28$\pm$0.16 & 44 & 0.28$\pm$0.17 & 38 & 0.23$\pm$0.16 & 25 & 0.37$\pm$0.16 \\
flat     & 20 & 0.27$\pm$0.18 & 24 & 0.30$\pm$0.16 & 24 & 0.25$\pm$0.19 & 21 & 0.18$\pm$0.11 & 12 & 0.34$\pm$0.21 \\
steep    & 12 & 0.20$\pm$0.14 & 15 & 0.23$\pm$0.14 & 12 & 0.17$\pm$0.14 &  6 & 0.27$\pm$0.15 &  1 & 0.61 \\
inverted & 7  & 0.26$\pm$0.16 &  8 & 0.26$\pm$0.13 &  8 & 0.36$\pm$0.08 &  6 & 0.18$\pm$0.13 &  8 & 0.23$\pm$0.17 \\
upturn   & 1  & 0.53          &  2 & 0.35$\pm$0.17 &  2 & 0.38$\pm$0.35 &  2 & 0.14$\pm$0.13 &  2 & 0.46$\pm$0.15 \\
complex  & 4  & 0.21$\pm$0.05 &  5 & 0.32$\pm$0.13 &  5 & 0.29$\pm$0.11 &  4 & 0.27$\pm$0.12 &  3 & 0.40$\pm$0.12 \\
\bottomrule
\end{tabularx}
\end{adjustwidth}
\end{table}

\begin{figure}
\centerline{\includegraphics[width=\columnwidth]{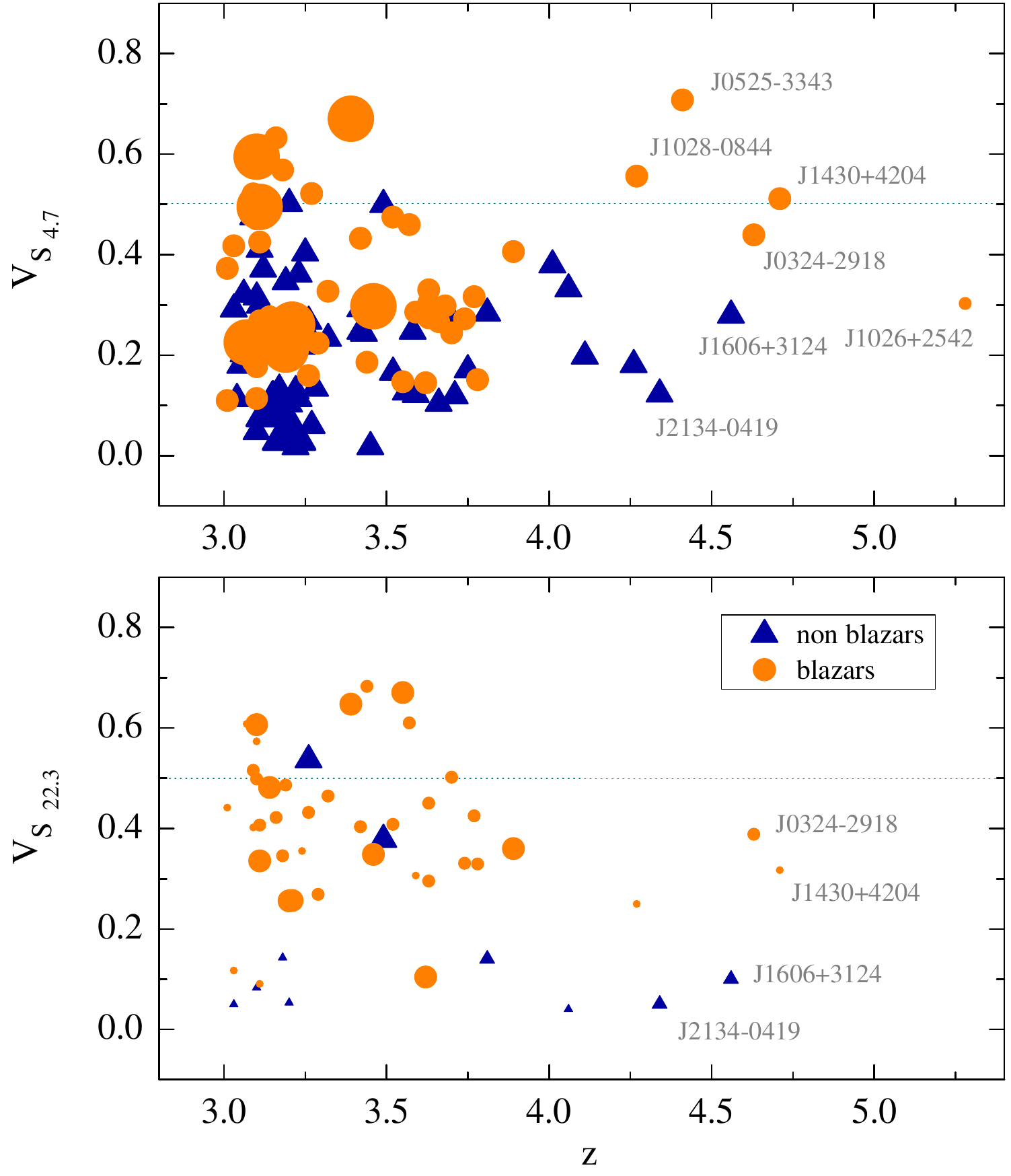}}
\caption{\label{fig55} The variability index $V_{S}$ at 4.7 (top) and 22.3 GHz (bottom) versus redshift $z$ for blazars (orange circles) and non-blazars (blue triangles). The size of the symbols is defined as in Fig.~\ref{fig:N}. The five well-known blazars are tagged: J1026$+$2542 at $z=5.25$ \citep{2012MNRAS.426L..91S,2015MNRAS.446.2921F}; J1430$+$4204 at $z=4.71$ \citep{2020SciBu..65..525Z}; J0324$-$2918 at $z=4.63$ \citep{2002A&A...391..509H}; J0525$-$3343 at $z=4.41$ \citep{1998MNRAS.294L...7H,2008MNRAS.385..283C}; J1028$-$0844 at $z=4.27$ \citep{1997A&A...323L..21Z}.}
\end{figure}

\begin{figure}
\begin{minipage}[h]{0.9\linewidth}
\center{\includegraphics[width=\linewidth]{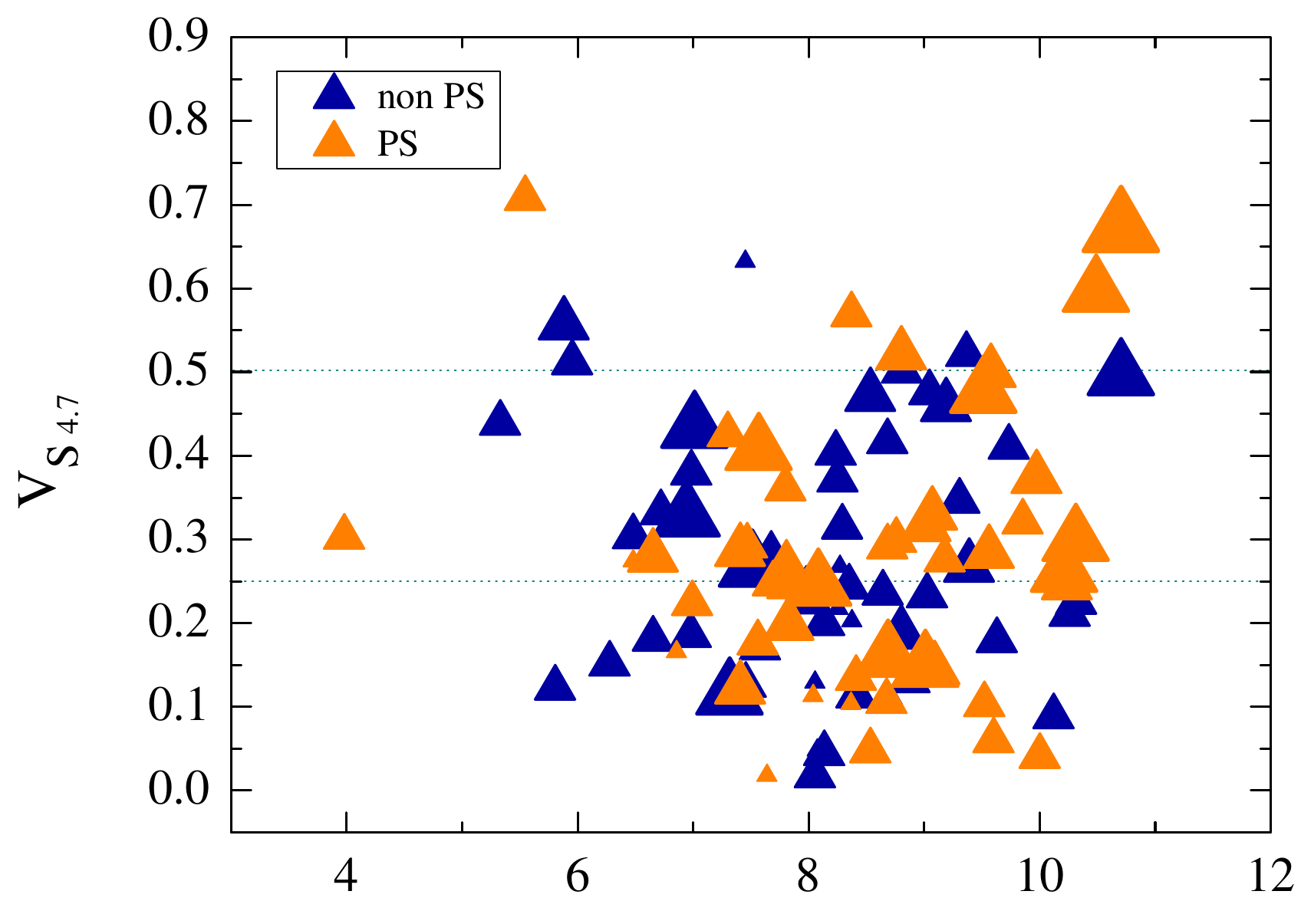}}
\end{minipage}
\vfill
\begin{minipage}[h]{0.9\linewidth}
\center{\includegraphics[width=\linewidth]{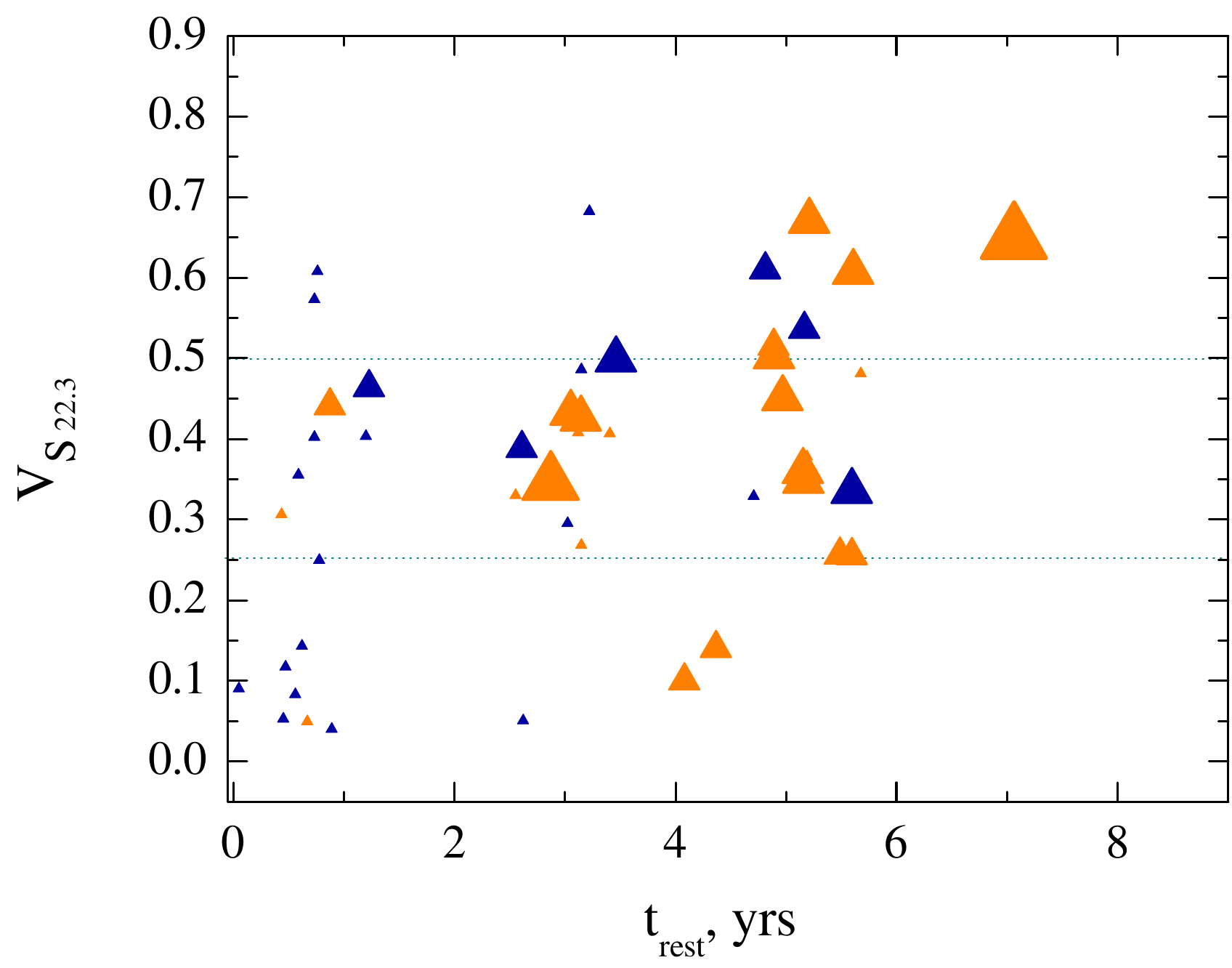}}
\end{minipage}
\caption{The $V_{S_{4.7}}$ and $V_{S_{22.3}}$ versus the rest frame timescale $t_{\rm rest}$ for the PS and non-PS quasars (orange and blue triangles). The symbol sizes are proportional to N$_{\rm obs}$, they correspond to 10, 20, 50, 100, and more number of observations. The dashed lines mark the 0.25 and 0.50 variability index levels.}
\label{fig111}
\end{figure}

\begin{figure}
\centerline{\includegraphics[width=0.95\columnwidth]{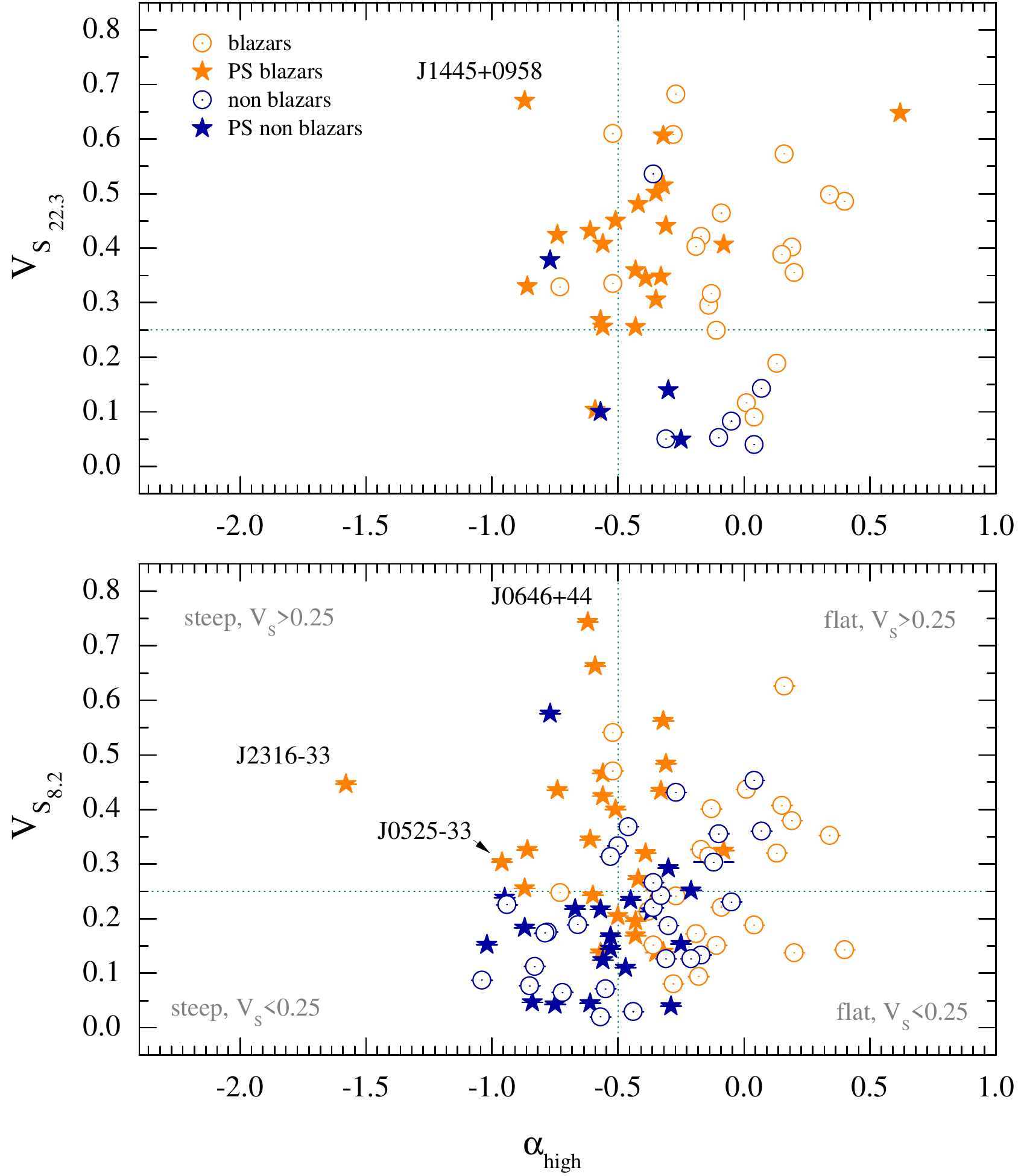}}
\caption{The variability indices $V_{S_{22.3}}$ (top) and $V_{S_{8.2}}$ (bottom) versus the high frequency spectral index $\alpha_{\rm high}$, measured from the averaged historical spectra of the PS blazars and PS non-blazars candidates (orange and blue stars) and the blazars/non-blazars with another spectral types (orange and blue circles). The vertical dotted line corresponds to the spectral index $\alpha=-0.5$, the horizontal line corresponds to $V_{S}=0.25$. Some variable quasars with steep averaged radio spectra are tagged.}
\label{fig5}
\end{figure}

For averaged spectra of the quasars measured over the entire historical time period, we did not find a correlation between the variability at 2.3, 4.7, 8.2, 11.2 and 22.3 GHz ($V_{S}$, $F_{S}$) and the spectral indices $\alpha_{\rm high}$ and $\alpha_{\rm low}$ in terms of Pearson and Spearman correlation coefficients. Such correlation would be expected for AGNs due to the fact that flat spectrum radio sources ($\alpha>-0.5$) are more variable than the steep ones. Indeed, a strong correlation between the spectral index and variability of bright AGNs has been found: for example, \cite{1987ApJS...65..319F,2001ApJS..136..265L}. This lack of correlation is possibly due to the prevalence of variable PS quasars with steep averaged spectra in the optically thin emission part. In addition, the blazars make up at least half of the sample and almost half of them have a peaked spectral shape \citep{2021MNRAS.508.2798S}. On a long time scale, these PS blazars demonstrate high radio variability (0.25--0.75) near the peak frequency. In Fig.~\ref{fig5} the ``$\alpha$--$V_{S}$'' plane is divided into four quadrants: steep-spectra with $V_{S}>0.25$ and $<0.25$, flat-spectra with $V_{S}>0.25$ and $<0.25$. It is clearly seen that PS quasars (orange and blue stars) make up the vast majority in the steep-spectrum area with $V_{S}>0.25$. During a flare, the radio spectra of some sources can exhibit characteristic spectrum evolution. In some cases, the spectra move up while maintaining a peaked shape, while in the others, they first move towards a flatter shape with increasing flux density at high frequencies before moving to the steep-spectrum area after the flare, when emission becomes optically thin.

\begin{figure}
\centerline{\includegraphics[width=0.85\columnwidth]{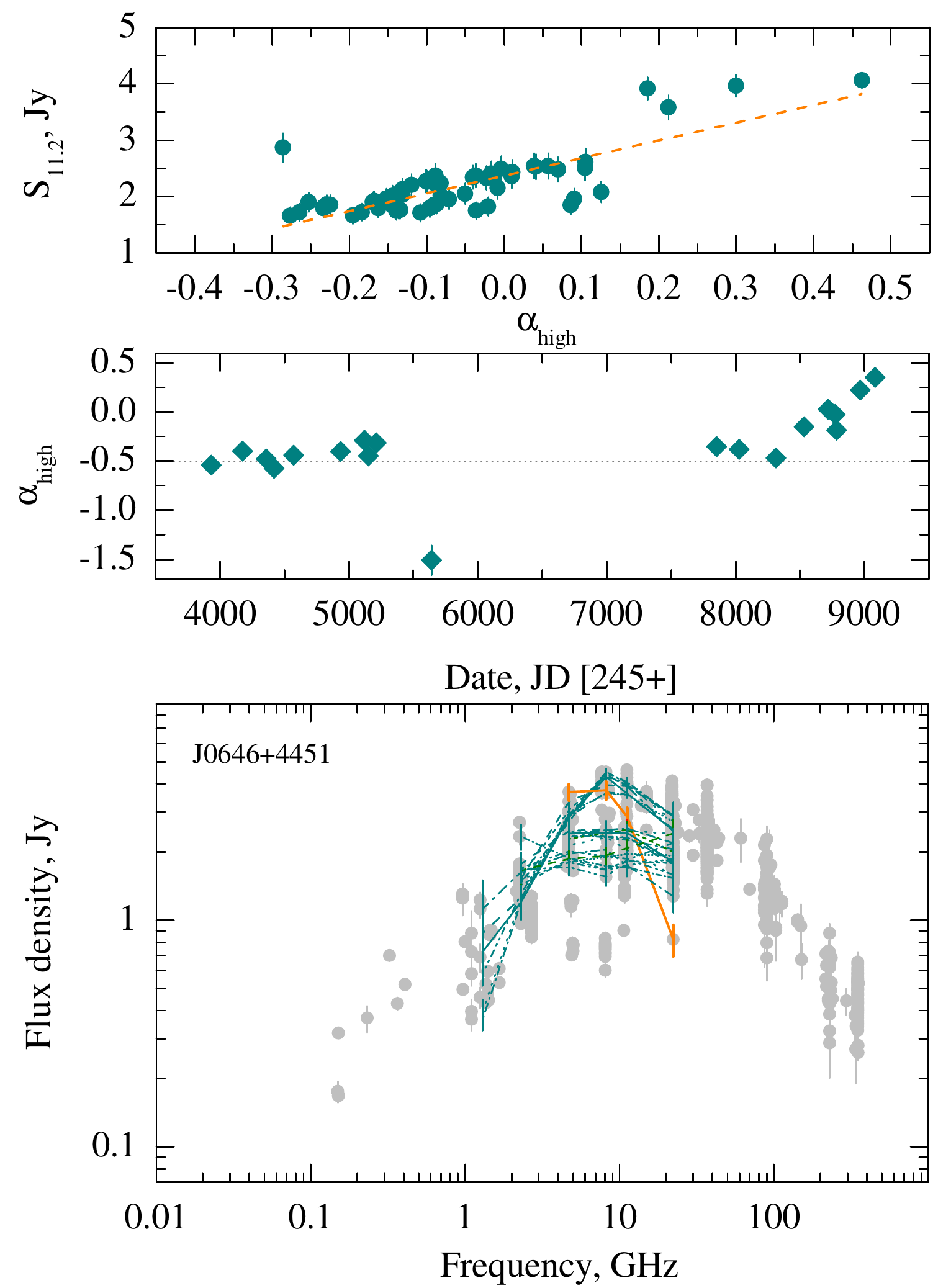}}
\caption{The relation $\alpha_{\rm high}-S_{11.2}$ for the blazar J0646$+$4451 is shown in the top panel. The spectral index $\alpha_{\rm high}$ evolution is presented in the middle panel. Corresponding RATAN-600 quasi-simultaneous spectra are shown in the bottom panel (green color) together with the literature data (grey color). The orange line corresponds to the ultra-steep spectrum measured in 2011.}
\label{fig7}
\end{figure}

An example illustrating this behaviour is the spectrum evolution of the PS blazar J0646$+$4451 with $V_{S_{8.2}}=0.74$ and $\alpha_{\rm high}$=-0.62. During the 17 years of RATAN-600 observations (Fig.~\ref{fig7}), its spectral index $\alpha_{\rm high}$ changed from $-1.51$ to $+0.35$, which corresponds to a spectral index variability of 0.60. We estimated the variability of the spectral index $V_{\alpha}$ using the same equation [\ref{eq:Var}] in terms of the $\alpha_{\nu}$ maximum and minimum values with their uncertainties. The blazar had an ultra-steep spectrum ($\alpha=-1.51$) in March 2011 at the beginning of its flux density decreasing at 2.3--22.3 GHz (orange spectrum), when radio emission had become optically thin at the frequencies above 5 GHz after several years of a relatively high state. There is a strong correlation between $\alpha_{\rm high}$ and the flux density at 11.2 GHz, Pearson's $r=0.81$, p-val.$<10^{-3}$ (Fig.~\ref{fig7}, top panel). Similar behavior is observed in other PS blazars such as J2316$-$3349 and J0525$-$3343. These spectral evolution patterns are probably caused by changes in the physical conditions in the source during a flare, such as variations in the magnetic field strength or electron density, which affect the observed flux densities at different frequencies.

We found significant correlation (Pearson's $r>0.5$, p-val.$\ll0.005$) between the spectral index $\alpha_{\rm high}$ and the flux density variability at 11.2 GHz for 15 of the most frequently observed (N~$>10$) quasars with $V_{S_{11.2}}>0.25$. This indicates that for these quasars, changes in the spectral index are predominantly driven by variation in the flux density at and above 11.2 GHz. Further studies are needed to investigate the underlying physical processes driving the observed variability in these quasars.  

The estimates of AGN variability at different redshifts can be biased due to the varying cosmological time dilation and different frequencies in the rest frame. The poorly sampled light curves also strongly impact variability estimates. Several factors, such as the time scale of monitoring $t_{\rm rest}$, the frequency $\nu_{\rm rest}$, and the number of measurements N$_{\rm obs}$ should be taken into account to compare variability at different $z$ \citep{2011A&A...536A..84V,2016A&A...593A..55V}. In order to search for patterns and tendencies by localizing groups of similar objects in a multidimensional parameter space (feature space) and investigating their statistical properties, we used two independent methods of cluster analysis of the quasars  (Appendix~\ref{sec:clusters}). The first uses PCA dimensionality reduction with k-means clustering, the second is based on the self-organizing maps \citep{1283494,2001som..book.....K}. We chose the variability parameters at an observed frequency of 4.7~GHz, which varies from 18 to 30 GHz in the rest frame. As a result, five quasar groups were determined and their parameters described in Table~\ref{tab:clusters}, which allowed us to compare the variability features within the clusters. An extremely highly variable quasar group with median $V_{S_{4.7}}=0.48$ was found at the highest redshifts, \mbox{$z=4.3$--$5.3$}, and they were observed as well at the shortest time scales, the median $t_{\rm rest}=5.7$~yrs. The clustering algorithm also determined a group of the most observed quasars with \mbox{$N_{\rm obs}=24$--$187$}. These quasars are at intermediate $z\leq3.9$ with long $t_{\rm rest}>7$ yrs. It is interesting that these most monitored quasars do not generally show the highest variability indices, which are nevertheless statistically higher than those for low-variable cluster 2 and intermediate-variable cluster 1. It is clearly seen that two factors impact the clustering result: the evolutionary state of the objects and observational constraints, such as $N_{\rm obs}$ and $t_{\rm rest}$.

\section{Peaked-spectrum sources at high redshifts}
\label{sec:PS}

Peaked spectrum quasars make up half of our sample \citet{2021MNRAS.508.2798S}. We consider the source as PS when the spectral index changes its sign from positive to negative in the average spectrum compiled with both the \mbox{RATAN-600} and literature data. It is well known that only several per cent of genuine GPS sources have been found in complete samples of bright AGNs spreading in a wide range of redshifts: see, e.g., \cite{2005A&A...435..839T,2007A&A...469..451T,2013AstBu..68..262M,2019AstBu..74..348S}. It is commonly believed that the mechanism which produces the absorption in the optically thick part in these compact luminous objects is due to synchrotron self-absorption caused by the high density of synchrotron-emitting electrons \citep{1998A&AS..131..435S}. Additionally, free-free absorption (FFA) has been suggested as a mechanism that can produce the GHz-frequency turnover as well as the observed anticorrelation between peak frequency and size in many of these sources \citep{1997ApJ...485..112B}. Several explanations have been put forth to account for the properties of the genuine GPS sources: they might be young radio galaxies in the process of evolving into larger ones \citep{1995A&A...302..317F,2012ApJ...760...77A}, or their compact nature could be a result of confinement through interaction with a dense environment \citep{1984AJ.....89....5V,2005JKAS...38..125B,2006A&A...447..481L,2007A&A...461..923O,2012ApJ...745..172D}, or the sources might be recurrent or intermittent sources \citep{1990A&A...232...19B,1997ApJ...487L.135R,2005A&A...443..891S,2012ApJ...760...77A, 2023arXiv230311361R}.

While most scenarios are well applicable to the known GPS galaxies, the nature of the GPS quasars has received limited attention in the literature.  GPS quasars are thought to be quite rare and challenging to identify due to strong contamination from blazars. Compact beamed sources can exhibit temporarily inverted radio spectra and may be easily disguised as GPS sources \citep{2008PhDT.......182T,2009AN....330..128T}. Study of the radio variability of PS quasars on historical timescales allows one to detect genuine GPS and MPS quasars, to estimate their fraction in the population of bright distant AGNs, and also to check how the MPS method is effective in searching for high-redshift AGNs  \citep{2015MNRAS.450.1477C,2016MNRAS.459.2455C}.

To search for GPS/MPS sources, we employ the following common criteria: the observed peak frequency $\nu_{\rm obs,peak}$ is less than 1 GHz for MPS sources and from 1 up to 5 GHz for GPS sources; spectral indices of optically thick $\alpha_{\rm thick}$ and thin $\alpha_{\rm thin}$ emission parts are close to $+0.5$ and $-0.7$, respectively (for the peaked spectra they match with $\alpha_{\rm low}$ and $\alpha_{\rm high}$, respectively); radio variability does not exceed 25\% on a long timescale, as GPS/MPS sources are known as the least variable class of compact extragalactic radio sources \citep{1990A&AS...82..261O,1991ApJ...380...66O,1997A&A...321..105D,1998PASP..110..493O,2004A&A...424...91E,2021A&ARv..29....3O}. Another PS type is high frequency peakers (HFPs), which are considered to be powerful and intrinsically compact (\textless 1 kpc) extragalactic radio sources with a spectral peak $\nu_{\rm peak,obs}$ above 5 GHz, representing the earliest stage of radio source evolution with typical ages of a few hundred years \citep{1995A&A...302..317F,2002NewAR..46..299D,2009AN....330..167O}. 
According to the criteria in \cite{2007A&A...475..813O,2008A&A...479..409O}, the HFPs lacking flux density variability and with low fractional polarization (\textless 1\%) and intrinsically small linear sizes (\textless 1 kpc) are considered good candidates for young radio sources.

Table~\ref{table5} presents the list of PS sources with their peak frequency, the width of the radio spectra at the half maximum level (FWHM), and spectral type based on \citep{2021MNRAS.508.2798S}, radio morphology from the existing literature, timescale of radio measurements, and the number of observing epochs. 

\begin{figure}
\centerline{\includegraphics[width=0.85\columnwidth]{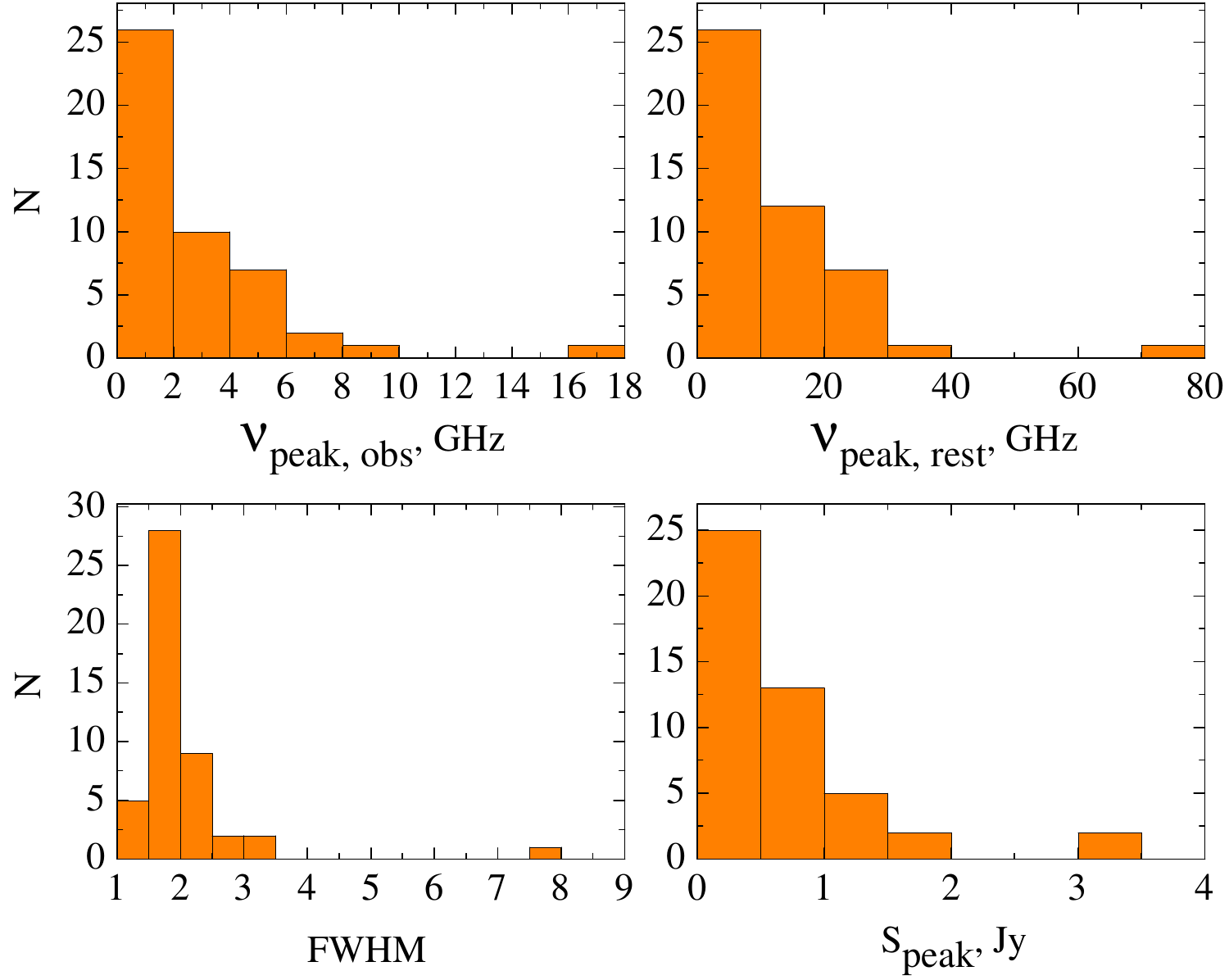}}
\caption{The distributions of the parameters for PS sources. The observed and rest frame peak frequencies are on top, and the flux densities at the peak frequency and FWHMs are at the bottom.}
\label{fig:param_distr}
\end{figure}

The distributions of the parameters calculated for PS sources in this section are shown in Fig.~\ref{fig:param_distr}, and their mean values are listed in Table~\ref{table6}. When PS blazars and non-blazars are considered separately, significant difference in the spectral and variability parameters emerge, emphasizing the importance of a careful approach in distinguishing intrinsically different sources that happen to have a spectral peak in the MHz--GHz frequency range. As shown in Table \ref{table6}, PS blazars have their spectral peaks at higher frequencies and have higher variations of flux densities. The  distributions of variability estimates at 2.3--22.3 GHz for  the PS sources and blazars of the sample are shown in Fig.~\ref{fig:var_distr} for comparison with the whole sample.

\begin{table}[H]
\caption{\label{table5} The peaked spectrum (PS) sources with their redshifts (Col.~2) and peak frequencies from \protect\citet{2021MNRAS.508.2798S} calculated in the observer's frame of reference (Col.~3), radio spectrum width at half maximum (Col.~4), timescale of radio measurements (in yrs) and the number of observing epochs at 4.7 GHz (Col.~5), variability index at 4.7 GHz (Col.~6), radio morphology and its reference (Col.~7), blazar type from BZCAT \protect\citep{2015Ap&SS.357...75M} (Col.~8), and PS type based on $\nu_{\rm peak,obs.}$ (Col.~9). Morphology designations are the following: c---core, c+j---core with a jet, cso---compact symmetric object, c+ext---compact structure with extension, E---core with small extension, T---triple morphology.}
\begin{adjustwidth}{-\extralength}{0cm}
\newcolumntype{C}{>{\centering\arraybackslash}X}
\begin{tabularx}{\fulllength}{p{1.5cm}CCCCCCCCC}
\toprule
\textbf{NVSS name} &	\textbf{$z$	}& \textbf{$\nu_{\rm peak,obs.}$}	& \textbf{FWHM} & \textbf{$t_{4.7}$/N$_{4.7}$} & \textbf{$V_{S_{4.7}}$} & \textbf{Morph.} & \textbf{Blazar}  & \textbf{Spectral}  \\ 
 & & \textbf{(GHz)} & \textbf{dex} & &  & & \textbf{type} & \textbf{type} \\
1 & 2 & 3 & 4 & 5 & 6 & 7 & 8 & 9\\
\midrule
000657$+$141546  &	3.20	&   1.5	& 7.7 & 40/14  	&		0.10 & c+j {\citep{1999A&A...344...51P}}  & & GPS \\ 
004858$+$064005	 &	3.58	&	3.9	& 1.5 & 36/25	&		0.25 & cso {\citep{2022ApJ...937...19Z}}  &	& GPS \\ 
010012$-$270851$^*$ &	3.52 & 1.4 & 1.7 & 31/10    &	    0.17 & c+j {\citep{2022ApJS..260....4P}}  & & GPS \\ 
012100$-$280623	 &	3.11	&	0.3	& 3.0 & 	30/19  	&	0.43 & c+j {\citep{2022ApJS..260....4P}}  &  FSRQ & MPS \\ 
020346$+$113445$^a$	 &	3.63	&	3.7	& 1.5 & 42/38	&	0.33 & c+j {\citep{2022ApJS..260....4P}}  &  FSRQ & GPS \\ 
021435$+$015703$^*$ &	3.28	&	0.5 & 1.9 & 46/19   &	0.13 & c+j {\citep{2022ApJS..260....4P}}  &  & MPS \\ 
023220$+$231756$^a$  &	3.42	&	0.5	& 2.0 & 33/13   &	0.29 & c+j {\citep{2022ApJS..260....4P}}  &  & MPS \\ 
024611$+$182330  &	3.59	&	1.8	& 1.8 & 34/44 &	0.29         & c+j  {\citep{1999A&A...344...51P}} & FSRQ & GPS \\ 
052506$-$334304  &	4.41	&	0.9	& 1.6 & 30/11  & 0.71        & c+j  {\citep{2022ApJS..260...49K}} & Uncert. & MPS \\ 
053954$-$283956$^a$ &	3.10	&	7.0	& 1.8 &  43/66	&	0.59 & c+j {\citep{2022ApJS..260....4P}}  & FSRQ &  HFP \\ 
062419$+$385648  &	3.46	&	0.3	& 2.5 &  46/68	&	0.30     & c+j {\citep{2022ApJS..260....4P}}  & FSRQ &  MPS \\ 
064632$+$445116  &	3.39	&	17.3 & 1.5 & 47/187	&	0.67     & c+j {\citep{2022ApJS..260....4P}}  & FSRQ &  HFP \\ 
073357$+$045614  &	3.10	&	5.8	& 1.8 & 40/35   &	0.37     & c+j {\citep{2022ApJS..260....4P}}  & FSRQ &  HFP \\ 
075141$+$271631	 &	3.20  	&	0.2 & 1.5 & 42/13  	&	0.04     &   &   		   &  MPS \\ 
075303$+$423130	 &	3.59	&	0.9 & 1.5 & 34/23	&	0.12     & cso  {\citep{2022ApJ...937...19Z}}  & & MPS \\ 
083910$+$200208$^*$ &	3.30	&	1.8 & 2.0 & 35/16   &	0.29 & c+j  {\citep{2022ApJS..260....4P}}  & & GPS \\ 
090549$+$041010$^*$ &	3.15	&	0.5	& 1.6 & 36/12	&	0.11 & c+j  {\citep{1999A&A...344...51P}}  & & MPS \\ 
093337$+$284532$^*$ &	3.42	&	1.7	& 1.5 & 34/17  	&	0.25 & E    {\citep{2014arXiv1406.4797G}}$^b$  & & GPS \\ 
102623$+$254259  &	5.28	&	0.3 & 1.7 & 25/13   &		0.30 & c+j  {\citep{2022ApJS..260...49K}}  & FSRQ &	MPS \\ 
104523$+$314232$^*$ &	3.23	&	0.9	& 2.3 & 34/10   &	0.11 & c    {\citep{2014arXiv1406.4797G}}$^b$  & & MPS \\ 
123055$-$113910	 &	3.52	&	1.1	& 2.1 & 43/51  &	0.47     & c+j  {\citep{2022ApJ...937...19Z}}  & FSRQ &	GPS \\ 
124209$+$372005  &	3.81	&	1.7	& 2.1 & 46/24   &	0.28     & T    {\citep{2014arXiv1406.4797G}}$^b$  & & GPS \\ 
130121$+$190421$^*$ &	3.10	& 1.7 & 1.9 & 35/13   &	0.05     & c+j  {\citep{2022ApJS..260....4P}}  & & GPS \\ 
134022$+$375443 &	3.11	&	2.7	& 1.4 & 36/12   &	0.30     & c+j  {\citep{1999A&A...344...51P}}  & & GPS \\ 
135406$-$020603 &	3.70	&	2.6	& 1.8 & 38/60   &	0.24     & c+j  {\citep{2022ApJS..260....4P}}  & FSRQ & GPS \\ 
135706$-$174402$^a$ &	3.14	&	1.8	& 1.7 & 38/19   &	0.28 & c+j  {\citep{2022ApJS..260....4P}}  & FSRQ & GPS \\ 
140135$+$151325$^*$ &	3.23	&	1.3	& 2.0 & 33/11   &	0.36 &   &     		& GPS \\ 
140501$+$041536$^a$ &	3.20	&	8.3	& 2.1 &  43/26  &	0.25 & c+j  {\citep{2014arXiv1406.4797G}}$^b$   & FSRQ & HFP \\ 
141821$+$425020$^*$ &	3.45	&	0.9	& 1.5 & 34/4    &	0.02 &   &     		& MPS \\ 
142438$+$225600 &	3.62	&	3.3	& 1.5 & 42/29  &	0.14     & c+j  {\citep{2022ApJS..260....4P}}   & FSRQ & GPS \\ 
144516$+$095835 &	3.55	&	1.3	& 1.6 & 41/76  &	0.15     & c+j  {\citep{2022ApJ...937...19Z}}   & FSRQ & GPS \\ 
150328$+$041949 &	3.66	&	5.6	& 1.5 & 39/8   &	0.10     & c+ext {\citep{1999A&A...344...51P}}  & & HFP \\ 
152117$+$175601$^*$ &	3.06	&	3.5	& 1.4 &	40/13  & 0.32    & c    {\citep{2014arXiv1406.4797G}}$^b$ & & GPS \\ 
153815$+$001905 &	3.49	&	2.9	& 1.6 & 43/20  &	0.50     & c+j  {\citep{2022ApJS..260....4P}}   & & GPS \\ 
155930$+$030447 &	3.89	&	4.5	& 1.7 & 37/63  &	0.41     & T    {\citep{2014arXiv1406.4797G}}$^b$ & FSRQ & GPS \\ 
160608$+$312447 &	4.56	&	2.5	& 1.5 & 37/26  &	0.28     & cso  {\citep{2022MNRAS.511.4572A}}   & & GPS \\ 
161637$+$045932 &	3.21	&	4.4	& 1.4 & 43/64  &	0.26 & c+j  {\citep{2014arXiv1406.4797G}}$^b$   & FSRQ & GPS \\ 
165844$-$073918$^a$ &	3.74	&	5.9	& 1.5 & 37/12 &	0.27 & c+j {\citep{2022ApJS..260....4P}} & FSRQ & HFP \\ 
184057$+$390046 &	3.1	&	4.5	& 1.8 & 36/23  &	0.52 & c+j {\citep{2022ApJS..260....4P}} & FSRQ & GPS \\ 
200324$-$325144	 &	3.77	&	5.6	& 1.4 & 43/39  &	0.32 & c+j {\citep{2022ApJS..260....4P}} & FSRQ & HFP \\ 
201918$+$112712$^*$ &	3.27	&	0.5 & 1.8 & 41/12   & 0.06 & c+ext  {\citep{1994MNRAS.269..902G}} &  & MPS \\ 
204257$-$222326 &	3.63	&	3.5	& 2.1 & 30/4    &	0.27 & c+j {\citep{2022ApJS..260....4P}}  & FSRQ & GPS \\ 
205051$+$312727 &	3.18	&	1.9	& 2.9 & 35/15   &	0.57 & c+j {\citep{2022ApJS..260....4P}}  & FSRQ & GPS \\ 
212912$-$153840	 &	3.26	&	6.8	& 1.5 & 37/54   &	0.16 & c+j {\citep{2023Galax..11...42C}}  & FSRQ & HFP \\ 
224800$-$054117	 &	3.29	&	0.4	& 3.3 & 30/13   &	0.22 & c+j {\citep{2022ApJS..260....4P}}  & Uncert. & MPS \\ 
231448$+$020150$^*$ &	4.11	&	1.7	& 2.1 &	40/20   &	0.20 & c+j  {\citep{2022ApJS..260...49K}}  &  & GPS \\ 
231643$-$334912	 &	3.10	&	3.9	& 1.3 & 31/18   &	0.18 & c+j {\citep{2022ApJS..260....4P}}  &  FSRQ & GPS \\ 
\bottomrule
\end{tabularx}
\end{adjustwidth}
\noindent{\footnotesize{$^a$ sources with two components in the radio spectrum.}}
\noindent{\footnotesize{$^b$ VLA measurements.}}
\noindent{\footnotesize{$^*$ new GPS/MPS candidates.}}
\end{table} 

\begin{table}[H]
\caption{\label{table6} 
The mean values of the parameters for the PS source subsamples.}
\begin{adjustwidth}{-\extralength}{0cm}
\newcolumntype{C}{>{\centering\arraybackslash}X}
\begin{tabularx}{\fulllength}{LCCCCCCC}
\toprule
     & \textbf{N} 	& \textbf{$\nu_{\rm peak,rest}$} & \textbf{$S_{\rm peak}$} & \textbf{FWHM} & \textbf{$V_{S_{11.2}}$} & \textbf{$V_{S_{8.2}}$} & \textbf{$V_{S_{4.7}}$} \\
     &      & \textbf{(GHz)} & \textbf{(Jy)} & \textbf{(dex)} & & & \\ 
\midrule
PS blazars     & 25 & $17.11\pm15.82$ & $0.94\pm0.78$ & $1.88\pm0.53$  & $0.32\pm0.15$ & $0.35\pm0.16$ & $0.35\pm0.16$ \\
PS non-blazars & 22 & $7.91\pm6.12$   & $0.41\pm0.42$ & $2.0\pm1.3$ & $0.12\pm0.07$ & $0.18\pm0.12$ & $0.20\pm0.12$ \\
\midrule
all        & 47   & $12.8\pm13.1$ & $0.69\pm0.68$ & $1.93\pm0.96$ & $0.23\pm0.16$ & $0.28\pm0.17$ & $0.28\pm0.16$ \\
\bottomrule
\end{tabularx}
\end{adjustwidth}
\end{table}

\subsection{Rest frame peak frequency and FWHM}
\begin{figure}
\centerline{\includegraphics[width=\columnwidth]{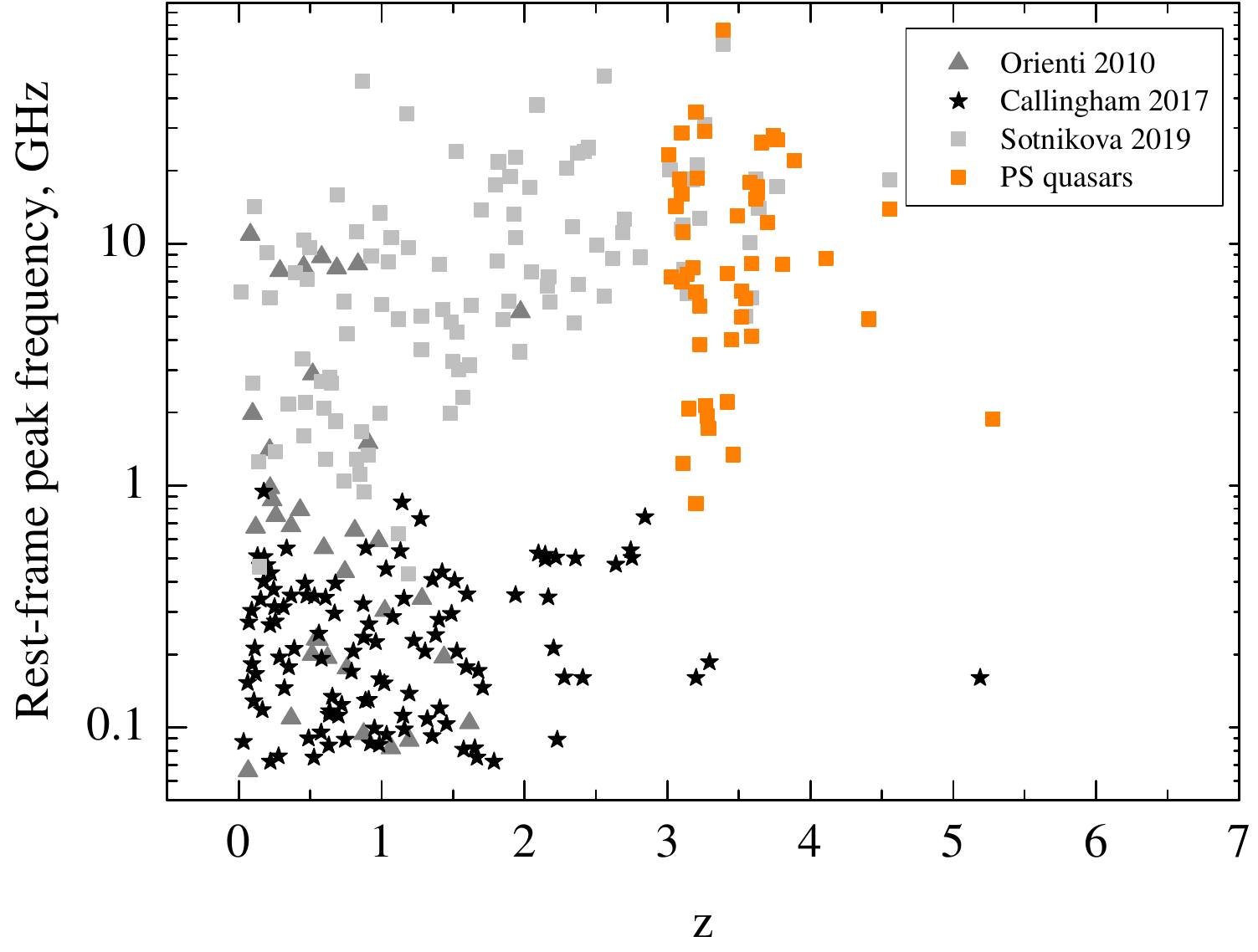}}
\caption{Redshift versus rest frame peak frequency for 47 PS quasars from the sample under investigation. The PS sources from \protect\cite{2014MNRAS.438..463O} are marked by grey triangles, from \protect\cite{2017ApJ...836..174C} by star symbols, and from \protect\cite{2019AstBu..74..348S} by grey squares.}
\label{fig3}
\end{figure}

For the sources with a peaked radio spectrum, we estimated the rest frame peak frequency $\nu_{\rm peak,rest} = \nu_{\rm peak,obs}(1+z)$, which ranges from 0.8 to 76 GHz. The redshift versus $\nu_{\rm peak,rest}$ relationship is shown in Fig.~\ref{fig3}, where we additionally plot PS sources at different redshifts from other studies: the sample from \cite{2014MNRAS.438..463O} marked by cross symbols, sources from \cite{2017ApJ...836..174C} marked by star symbols, and from \cite{2019AstBu..74..348S} by grey circles. \cite{1997A&A...321..105D} found a trend towards higher turnover frequencies at higher redshifts, but Fig.~\ref{fig3} shows a wide spread in the peak frequencies in our sample at redshifts $z$=3--5 and no evidence for correlation between them. Thus, sources with synchrotron self-absorbed components around and even below 1 GHz in the rest frame occur at high redshifts and at high radio luminosities.

We calculated the Full Width at Half Maximum (FWHM) for the PS sources in units of decades of frequency (dex). For classical GPS sources, the parameter FWHM~$\sim1$--1.5 dex, as has been found in \cite{1990A&AS...84..549O,1991ApJ...380...66O} and \cite{1997A&A...321..105D} for a sample mostly represented by $z < 3$ GPS sources (with only 6 of 74 sources having $z > 3$ in the latter study). \cite{2013AstBu..68..262M} compared the samples of GPS sources classified as galaxies and quasars and found that the radio spectrum of GPS quasars is on average wider (FWHM~$\sim1.6$) than that of the GPS galaxies (FWHM~$\sim1.4$). In our PS sample sources, the FWHM varies from 1.3 to 7.7 dex with a mean value of $1.93\pm0.96$ (Table~\ref{table6}).

\subsection{Spectral variability of identified HFP/GPS/MPS quasars}
\label{sec:knownPS}

Blazars have the potential to contaminate GPS source samples since they exhibit peaked radio spectra during their flaring states \citep{1986AJ.....92.1262O,2005BaltA..14..413K,2011A&A...536A..14P,2021AN....342.1195S}. In this section, we focus on the broadband long-term radio spectra of 23 quasars previously classified as HFP, GPS, or MPS sources \citep{1990MNRAS.245P..20O,1991ApJ...380...66O,1998PASP..110..493O,1999A&AS..135..273M,2000A&A...363..887D,2005A&A...432...31T,2013AstBu..68..262M,2017ApJ...836..174C,2021MNRAS.503.4662M}. Their radio spectra and RATAN-600 light curves are presented in Figs.~\ref{fig:B1}--\ref{fig:B2}.
The longest-term measurements, spanning up to 30--43 years, are available at frequencies at 4.7, 11.2, and 22.3 GHz. They are listed in Table~\ref{table8} with the information about their variability index at 11.2 and 22.3 GHz and the historical period of observations. 

Among the known HFP/GPS/MPS sources, six are not classified as blazars. Three of these six sources (J0048$+$0640, J0751$+$2716, J0753$+$4231) have relatively small variability indices $V_{S_{4.7}}=0.04$-$0.25$ and $V_{S_{11.2}}=0.05$-$0.12$ over the timescales of 36--43 and 3--9 years, respectively. Two of them have a compact symmetric morphology \citep{2022ApJ...937...19Z,2022MNRAS.511.4572A}. J1606$+$3124 has been classified as a galaxy with a radio structure similar to that of CSOs, and it may be the highest redshift ($z=4.56$) CSO galaxy discovered so far \citep{2022MNRAS.511.4572A}. We measured the variability indices as \mbox{0.10--0.16} at 11.2 and 22.3 GHz and 0.28 at 4.7 GHz for this source at a long timescale (37--43 yrs). The other two objects, J1340$+$3754 and J1538$+$0019, have the variability indices \mbox{$V_{S_{4.7}}=0.30$--$0.50$} ($t=36$ and 43 yrs) and $V_{S_{11.2}}=0.03$--$0.27$ ($t=5$ and 23 yrs). 

Among seventeen GPS quasars in Table~\ref{table8}, which belong to the blazar population, thirteen demonstrate high variability indices $V_{S_{22.3}}$ up to 0.67. Only four blazars, namely J1424$+$2256, J1616$+$0459, J2248$-$0541, and J2316$-$3349, have relatively low variability of less than 0.26 at 4.7 GHz and less than 0.27 at 11.2 GHz. However, for J2248$-$0541 and J2316$-$3349, there are no
enough radio data to establish their variability reliably. 

Thus, among previously reported 23 GPS, MPS, and HFP sources, seven quasars demonstrate low radio variability ($V_{S} < 0.27$) at 4.7--22.3 GHz, with four of them being blazars. Only two GPS (J0048+0640, J1606+3124) and two MPS sources (J0751+2716, J0753+4231) have variability attributes of young compact radio sources. The description of 17 individual sources from this section that have significant radio variability is presented in Appendix~\ref{sec:identifiedGPS}.

\begin{table}[H]
\caption{\label{table8} The variability indices at 22.3 and 11.2 GHz of the earlier determined high-redshift HFP/GPS/MPS sources. The columns are the following: source name (Col.~1), timescale of radio measurements (in years) and the number of observing epochs at 22.3 and 11.2 GHz (Cols.~2--5), $V_{S_{22.3}}$ and $V_{S_{11.2}}$ (Cols.~6--7), blazar type according to BZCAT (Col.~8), references for the spectral classification (Col.~9), where [1]---\protect\citet{2013AstBu..68..262M}; [2]---\protect\citet{1991ApJ...380...66O}; [3]---\protect\citet{2017MNRAS.467.2039C}; [4]---\protect\citet{2000A&A...363..887D}; [5]---\protect\citet{2017ApJ...836..174C}; [6]---\protect\citet{1990MNRAS.245P..20O}; [7]---\protect\cite{1999A&AS..135..273M}.}
\newcolumntype{C}{>{\centering\arraybackslash}X}
\begin{tabularx}{\textwidth}{CCrCCCCrr}
\toprule
\textbf{Name }& \textbf{$t_{22.3}$} & \textbf{N$_{22.3}$} & \textbf{$t_{11.2}$} & \textbf{N$_{11.2}$ }& \textbf{$V_{S_{22.3}}$} & \textbf{$V_{S_{11.2}}$} & \textbf{Blazar}  &  \textbf{Spectral} \\
      &  \textbf{(yrs)}   & & \textbf{(yrs)} & & &  & \textbf{type} & \textbf{type} \\
\midrule  
  1          & 2 & 3 & 4 & 5 & 6 & 7 & 8 & 9  \\
\midrule
J0048$+$0640 & -- & --  & 7  & 17 & --   & 0.12 & --   & GPS [1] \\
J0121$-$2806 & 7  & 18  & 5  & 14 & 0.41 & 0.20 & FSRQ & MPS [1] \\
J0203$+$1134 & 22 & 31  & 13 & 22 & 0.45 & 0.38 & FSRQ & GPS [2] \\
J0324$-$2918 & 15 & 15  & 5  & 16 & 0.39 & 0.14 & FSRQ & GPS [3] \\
J0646$+$4451 & 30 & 319 & 14 & 68 & 0.65 & 0.66 & FSRQ & HFP [2, 4]\\
J0751$+$2716 & -- & --  & 3  & 9  & --   & 0.05 &  -- & MPS [5] \\
J0753$+$4231 & -- & --  & 9  & 10 & --   & 0.09 &  --  & MPS [1] \\
J1230$-$1139 & 14 & 7   & 23 & 44 & 0.41 & 0.44 & FSRQ & GPS [1] \\
J1340$+$3754 & -- & --  & 5  & 3  & --   & 0.03 &  -- & GPS [1] \\
J1354$-$0206 & 23 & 33  & 23 & 44 & 0.50 & 0.30 &  FSRQ  & GPS [2, 6] \\
J1357$-$1744 & 23 & 3   & 25 & 3  & 0.48 & 0.36 & FSRQ  & GPS [2] \\
J1424$+$2256 & 22 & 18  & 14 & 14 & 0.10 & 0.16 & FSRQ & GPS [2, 4] \\
J1445$+$0958 & 23 & 27  & 42 & 33 & 0.67 & 0.57 & FSRQ & GPS [2, 6] \\
J1538$+$0019 & 23 & 10  & 23 & 9  & 0.38 & 0.27 &  --   & GPS [1] \\
J1559$+$0304 & 8  & 29  & 26 & 51 & 0.36 & 0.18 & FSRQ & GPS [5] \\
J1606$+$3124 & 43 & 22  & 41 & 15 & 0.10 & 0.16 &  --   & GPS [2] \\
J1616$+$0459 & 23 & 22  & 27 & 19 & 0.26 & 0.17 & FSRQ & HFP [2, 4, 6] \\
J1658$-$0739 & 10 & 9   & 12 & 6  & 0.33 & 0.13 & FSQR & GPS [2] \\
J1840$+$3900 & 20 & 13  & 13 & 10 & 0.52 & 0.45 & FSRQ & GPS [4, 7] \\
J2003$-$3251 & 15 & 36  & 13 & 8  & 0.42 & 0.15 & FSRQ & HFP [2] \\
J2129$-$1538 & 20 & 24 & 41 & 21 & 0.43 & 0.41 & FSRQ & HFP [2, 6] \\
J2248$-$0541 & 2 & 2 & 1 & 6  & 0.27 & -- &  Uncert. & MPS [5] \\
J2316$-$3349 & -- & -- & 7 & 13 & -- & 0.26 & FSRQ & GPS [1] \\
\bottomrule
\end{tabularx}
\end{table}

\subsection{New GPS/MPS candidates}
\label{sec:GPS/MPS/HFP candidates}

We found 12 new GPS/MPS candidates, which are marked with an asterisk symbol in Table~\ref{table5} and described in Appendix~\ref{sec:newGPS}. Seven of them, J0100$-$2708, J0839$+$2002, J0933$+$2845, J1301$+$1904, J1401$+$1513, J1521$+$1756, and J2314$+$0201, we consider as new GPS candidates. Five quasars, J0214$+$0157, J0905$+$0410, J1045$+$3142, J1418$+$4250, and J2019$+$1127, we suggest as new MPS candidates.

\subsection{HFP candidates}
There are 6 PS sources previously identified as HFP candidates (Table~\ref{table5}), with a turnover peak at about 5 GHz or higher. If we consider the source's rest frame frequency, it is above 20 GHz. All of them are classified as blazars of the FSRQ type in BZCAT and have high variability indices from 0.26 to 0.65 at 22.3 GHz according to our estimates. Four quasars, J0646$+$4451, J1616$+$0459, J2003$-$3251, and J2129$-$1538, have been previously classified as HFPs (Appendix~\ref{sec:identifiedGPS}). For the remaining two quasars, J0733$+$0456 and J1503$+$0419, we obtained the turnover frequency $\nu_{\rm peak,obs}$ equal to 5.8 and 5.6 GHz respectively \citep{2021MNRAS.508.2798S}. 

J0733$+$0456 was not previously considered as an HFP candidate. The multifrequency data from the GLEAM survey at 72--231 MHz in 2013--2014 \citep{2017MNRAS.464.1146H} and the RATAN-600 measurements in 2016--2020 made it possible to classify the quasar as a PS source \citep{2021MNRAS.508.2798S} with a high turnover frequency $\nu_{\rm peak,obs}=5.8$ GHz. The wide spectral peak FWHM~$=1.8$ reflects the high radio variability of the quasar at all the considered frequencies ($V_{S} > 0.30$).

J1503$+$0419, initially identified as an FSRQ in \cite{2007ApJS..171...61H}, exhibits a core and some extension in the VLA 5 GHz image. Later  authors \cite{2011AJ....141..182K} confirmed the presence of a lobe in the 1.4~GHz FIRST survey image. However, when observed using VLBI, no discernible structure was detected, as reported in \cite{1999A&A...344...51P}. Unfortunately, there are very little radio data. We measured a low variability index $V_{S}=0.10$ at 4.7 GHz spanning a time period of $t_{4.7}=39$ years with a total of 8 observations.  The variability at 11.2 and 22.3 GHz can not be evaluated because the flux density errors are larger than the observed scatter of the data, according to the only two RATAN-600 observing epochs. The quasar was classified as a PS source for the first time with an estimated peak frequency $\nu_{\rm peak,obs}=5.6$ GHz and FWHM~$=1.5$ \cite{2021MNRAS.508.2798S}.

Despite the presence of a peak above 5 GHz, we do not consider three quasars as HFP candidates: J0539$-$2839, J1405$+$0415, and J1658$-$0739, since they have two-component spectra or a strongly variable spectral shape with a wide peak (Table~\ref{table5}, Table~\ref{TableA2}). 

Note that both the quasar J0257$+$4338 with a core--jet morphology \citep{2016AJ....151..154G} and the BL\,Lac blazar \citep{2014ApJS..215...14D} or blazar of an uncertain type in BZCAT with an unresolved core in VLBA images \citep{VCS4}. J0525$-$2338 could hypothetically be considered as HFP candidates due to their spectral shape, which has a hint of a possible peak near 8 and 20 GHz \citep{2021MNRAS.508.2798S}. The blazar J0525$-$2338 exhibits a rising spectrum ($\alpha_{\rm high}$=+0.60) and has the variability indices $V_{S_{22.3}}=0.50$ and $V_{S_{11.2}}=0.36$. J0257$+$4338 is one of the most distant quasars in the sample at $z=4.06$ \citep{2006Ap.....49..184A} with a variability level $V_{S_{8.2}}=0.45$. Its turnover frequency showed a remarkable change from 2 GHz up to 8 GHz between 1990 and 2017, with a strong increase in flux density from 0.13 Jy \citep{1992MNRAS.254..655P} up to 0.36 Jy. 

\section{Discussion}
\label{sec:Discussion}
Studying the variability of quasars in the radio band is an approach to understanding the astrophysical processes occurring in the sources. Long-term (months, years, and decades) variability provides insights into the properties of the jet, flaring activity, and interaction between the jet and the surrounding medium, while short-term (from several hours to several days) variability could provide information about the interstellar medium based on the measurements of scintillations \citep{2008ApJ...689..108L}. In our study, we consider flux density variations at GHz frequencies over long periods of time of a sample of high-redshift AGNs, which represent a superposition of the aforementioned short-term events and long-term processes. 

The redshift values range from 3.0 to 5.3 for the quasars in our sample, which corresponds to a wide range of emission frequencies in the source's reference frame, from 9 to 130 GHz, and the rest frame time scales of monitoring from several weeks to 11 years. Consequently, we can get the most general picture of the dependence of the radio emission variability on the redshift and frequency and on the time scale of monitoring.

The quasars that have showed a high level of variability but are not listed as blazars could be considered as blazar candidates after checking their morphology and other multi-band properties, thus usefully supplementing the current lists of the known blazars in the ``Open Universe for Blazars~--~Reference list V2.0'' and the Roma-BZCAT catalogue.

\subsection{Variability properties}

Estimates of the ﬂux density variability show that the averaged variability is higher at higher frequencies ($M_{22.3}=0.28$, $F_{S_{22.3}}=0.31$, and $V_{S_{22.3}}=0.36$) with typical values of about 0.20--0.25 at other frequencies. These values are comparable to those observed in blazars at any redshift. 

The distributions of $V_S$ and $F_S$ at 22.3 GHz are significantly different from those at 2.3--11.2 GHz, which may be attributed to intrinsic physical differences. At the source's rest frequencies greater than 45 GHz (i.e., $11.2\times (1+z)$, where $z\geq 3$), the radio emission from the compact AGN core becomes more dominant, leading to a different variability behavior (the $V_S$ distribution peaks around 0.4) compared to the lower frequencies (the $V_S$ distribution peaks around 0.1--0.2). At frequencies lower than 45 GHz, the variability may be mixed with variations from multiple jet components, which can lead to smearing of the variability index.

Only the brightest peaked-spectrum sources can be detected at high redshifts due to their steep radio spectrum above the peak frequencies. Compared to the other sources in the sample, the PS sources are about two times brighter, while the variability level is of the same order at all frequencies. The PS sources are also found to be more variable than the AGNs with power-law spectra at MHz frequencies in \cite{2018MNRAS.474.4937C} and \cite{2021MNRAS.501.6139R}. Some PS sources even demonstrate high amplitude variability of 100--2500\% over 24 years at 1.5 GHz \citep{2020ApJ...905...74N}. More than half of PS quasars in the sample are blazars (25 out of 47), which affects the variability level of the PS subsample. 

We have not found a correlation between radio variability and redshift or the spectral index; apparently, dense and long-term time series are required to study such relationships more robustly. Comparing the variability of high-$z$ AGNs, the contribution of observational restrictions and different source conditions in such a wide redshift range can not be easily distinguished for our sample. We found several quasar groups with distinctive variability features. The most interesting is a small highly variable group of quasars located at redshifts of 4.3--5.3, notable by a shorter time scale of radio monitoring, $t_{\rm rest}=4.0$--$6.7$ yrs, compared to the rest of the quasars in the sample. One of the simple explanations is that the youngest quasars are more compact, and as a consequence the variability time scale is shorter. This can be confirmed by the recently reported relationship between variability timescales and core sizes in \cite{2023arXiv230809626C}. On the other hand the most bright sources can be detected at such high redshifts, and these bright sources may be more variable on average. We should be careful making far-reaching conclusions because only future long-term measurements can provide evidence of the different properties of intrinsic variability.

\subsection{Population of peaked spectrum sources at high~redshifts}

A comparison of radio spectra features and flux density variations of quasars reveals that the peaked spectrum is a typical feature of blazars at redshifts $z>3$, distinguishing them from the flat spectrum low-redshift blazars. Obviously, only a small fraction of quasars can be associated with the so-called classical GPS/MPS sources. The long-term radio data showed a significant variability at GHz frequencies, up to 70~per~cent, in the most known GPS quasars (Section~\ref{sec:knownPS}). Previous studies have mentioned the scarcity of genuine GPS quasars \citep{2023Galax..11...42C,2005A&A...435..839T,1990A&AS...82..261O}, and we need long-term multifrequency observations to determine their spectra reliably. 

Because of the young age of HFPs, one could expect a larger proportion of them at high redshifts compared to the intermediate ones. Previously, in \cite{2013AstBu..68..262M}, we identified about 2\% objects in the flux-density-complete sample of GPS galaxies and quasars at lower redshifts (69\% sources are at $z\leq 2$) as HFP candidates. In our sample of high-redshift quasars, all 6 HFP candidates are FSRQ blazars, which exhibit high variability levels and  show a peaked spectrum shape occasionally, during a flare. These six quasars are, on average, more variable, with the mean value of $V_{S_{22.3}}=0.43\pm0.14$, compared to the whole sample ($V_{S_{22.3}}=0.35\pm0.18$). Authors \cite{2000A&A...363..887D} found that only 40\% of the objects displayed all the typical characteristics of young radio sources. Furthermore, all quasars in the sample were excluded. So, to prevent contamination from selection and boosting effects, authors \cite{2014MNRAS.438..463O} considered only galaxies in their statistical analysis of HFP sources.

Three quasars with peaked radio spectra and a CSO-type morphology showed relatively low variability levels at all the five frequencies we considered, providing evidence in favor of them being bona-fide CSOs. These findings are consistent with the estimates in \cite{2023arXiv230311359K,2023arXiv230311357K}, where they correspond to $\sim 3$\% of the sample.

It should be noted that for 12\% of the objects there are no multifrequency and long-term measurements, and our analysis is based on non-uniform sparse time series. Only for about 10 quasars there are flux density measurements with high cadence (N$_{\rm obs}=40$--60) over a long time period. The strong correlation between the number of observations and variability level confirms the influence of the observational selection effect on the result.

\section{Summary}
In this study, we present the results of a multi-frequency radio variability analysis of 101 bright radio quasars at $z \geq 3$ based on the RATAN-600 observations at 2.3--22.3 GHz conducted in 2017--2020 \citep{2021MNRAS.508.2798S}, as well as the information from the literature covering a time period of 30--40 years, which corresponds to a timescale of about ten years in the source's frame of reference. The main results are summarised as follows.

\begin{enumerate}

\item The quasars show a wide spread of radio variability on any time scale of the measurements.  About half of the objects in the sample show a variability level (25--50\%) comparable to that of the blazars at low redshifts. Some quasars demonstrate flux density variations of up to 60--90\% at 8.2--22.3 GHz. The median values of modulation index $M$, fractional variability $F_{S}$, and variability index $V_{S}$ are about 0.20 at 2.3--8.2 GHz and are higher at 22.3 GHz (0.28--0.36). The proportion of sources with low variability of $V_{S}< 0.10$ varies from 8 to 20\% at different frequencies.

\item The distribution of the variability index is significantly different at 22.3 GHz compared to lower frequencies, demonstrating the peak at higher levels of variability. This suggests that the emission from the compact core dominates at higher frequencies, while at lower frequencies, the emission from jet components could blend with the core emission, shifting the distribution peak to lower values. In general, the variability we obtained at the range from dozens to about a hundred GHz in the rest frame is attributed to the dominating core component in quasars. 

\item Several quasar groups with distinctive variability characteristics were found. Six highest-redshift quasars demonstrate average variability of about 0.50 and at a shorter time scale than the others.

\item Among the 23 GPS and MPS sources reported previously, only six objects show low radio variability ($V_{S} < 0.25$). We propose seven new candidates for GPS sources (J0100$-$2708, J0839$+$2002, J0933$+$2845, J1301$+$1904, J1401$+$1513, J1521$+$1756, and J2314$+$0201) and five new MPS candidates (J0214$+$0157, J0905$+$0410, J1045$+$3142, J1418$+$4250, and J2019$+$1127) based on the analysis of their spectral parameters and radio variability. We found that about 18 and 9\% of the sample of bright quasars at $z\geq3$ are GPS and MPS candidates, respectively. 

\item A peaked spectrum is a typical feature of blazars at high redshifts, and this distinguishes them from the flat spectrum low-redshift blazars. The radio variability exceeding more than 30\% in high redshift PS sources is common for blazars.  

\end{enumerate}

\authorcontributions{Conceptualization, Yu.S.; observations, Yu.S., A.M., T.M., T. S.; data reduction, Yu.S., A.M., R.U., T. S.; methodology, Yu.S., A.M., T.M., D.K., V.S.; software, D.K., V.S.; writing---original draft preparation, Yu.S., A.M., T.M., D.K., A.K.; writing---review and editing, T.A., M.M.; funding acquisition, Yu.S. All authors have read and agreed to the published version of the manuscript.}

\institutionalreview{Not applicable.}

\informedconsent{Not applicable.}

\funding{This work is supported in the framework of the State project ``Science'' by the Ministry of Science and Higher Education of the Russian Federation under the contract 075-15-2024-541, and the National SKA Program of China (2022SKA0120102, 2022SKA0130103).} 

\dataavailability{The data underlying this article are available in the article. \autoref{TableA2} is distributed in the VizieR database.\!\footnote{\url{https://cdsarc.cds.unistra.fr/viz-bin/cat/J/other/Galax/12.25}}} 

\acknowledgments{The observations were carried out with the RATAN-600 scientific facility. This research has made use of the NASA/IPAC Extragalactic Database (NED), which is operated by the Jet Propulsion Laboratory, California Institute of Technology, under contract with the National Aeronautics and Space Administration; the CATS database, available on the Special Astrophysical Observatory website; the SIMBAD database, operated at CDS, Strasbourg, France. This research has made use of the VizieR catalogue access tool, CDS, Strasbourg, France. We thank the anonymous referees for their comments and review of the manuscript.}

\conflictsofinterest{The authors declare no conflict of interest.} 

\abbreviations{Abbreviations}{
The following abbreviations are used in this manuscript:\\

\noindent 
\begin{tabular}{@{}ll}
AGN & Active galactic nuclei \\
CSO & Compact symmetric object \\
FWHM & Full width at half maximum \\
GPS & Gigahertz peaked-spectrum \\
HFP & High-frequency peaker \\
MPS & Megahertz peaked-spectrum \\
PS & peaked radio spectrum \\
RISS & Refractive interstellar scintillation 
\end{tabular}
}


\appendixtitles{yes} 
\appendixstart
\appendix
\section[\appendixname~\thesection]{Sample parameters}

\begin{landscape}
\begin{longtable}{|l|c|c|c|c|c|c|c|c|c|c|c|}
\caption{\label{TableA1}The sample parameters: source name (Col.~1), redshift (Col.~2), average flux density at 4.7 GHz (Col.~3), radio structure (Col.~4) with references to the literature (Col.~5), spectral indices $\alpha_{\rm low}$ and $\alpha_{\rm high}$ (Cols.~6--7), radio spectrum type (Col.~8), blazar type from BZCAT (Col.~9), and number of cluster (Col.~10). Columns 3, 6--9 are adopted from \protect\cite{2021MNRAS.508.2798S}. Morphology designations are the following: c -- core, c+j -- core with a jet, cso -- compact symmetric object, c+ext -- compact structure with extension, E -- core with small extension, T -- triple morphology.}\\
\hline
\textbf{NVSS name} &  \textbf{$z$} & \textbf{$S_{4.7}$}  & \textbf{Morph.}        &  \textbf{Morph.} &\textbf{ $\alpha_{\rm low}$} & \textbf{$\alpha_{\rm high}$} & \textbf{Spectral}  & \textbf{Blazar} & \textbf{Cluster} \\
	  &      &        \textbf{(Jy)}& \textbf{type}          &  \textbf{ref.}  &  & &          \textbf{type} &   \textbf{type} &   \\
 1 & 2 & 3 & 4 & 5 & 6 & 7 & 8 & 9 & 10 \\
\hline
\endfirsthead
\caption[]{continued.}\\
\hline
NVSS name &  $z$ & $S_{4.7}$ & Morph.       &  Morph. & $\alpha_{\rm low}$ & $\alpha_{\rm high}$ & Spectral & Blazar & Cluster \\
	  &      &  (Jy)    & type          & ref.   &  & &           type &   type & \\
 1 & 2 & 3 & 4 & 5 & 6 & 7 & 8 & 9 & 10 \\
\hline
\endhead
\hline
\endfoot
000108$+$191434     &   3.10 & $0.15\pm0.01$    &  c+j  &    10     & $-0.87\pm0.11$ & $+0.16\pm0.01$ & upturn  & FSRQ          &  3   \\
000657$+$141546     &   3.20 & $0.14\pm0.01$    &  c+j  &    1      & $+0.02\pm0.02$ & $-0.21\pm0.01$ & peaked  &               &  2   \\
004858$+$064005     &   3.58 & $0.17\pm0.01$    &  cso  &    8      & $+0.05\pm0.03$ & $-0.95\pm0.01$ & peaked  &               &  1   \\
010012$-$270852     &   3.52 & $0.12\pm0.01$    &  c+j  &    10     & $+0.41\pm0.04$ & $-0.47\pm0.01$ & peaked  &               &  1   \\
012100$-$280623     &   3.11 & $0.17\pm0.01$    &  c+j  &    10     & $+1.45\pm0.15$ & $-0.08\pm0.01$ & peaked & FSRQ           &  3   \\
013113$+$435812$^{a}$ &  3.12 & $0.03\pm0.01$   &  c+j  &    2      & $-1.16\pm0.02$ & $-0.12\pm0.08$ & complex  &              &  3   \\
014844$+$421518     &   3.24 & $0.11\pm0.01$    &       &           & $-0.27\pm0.02$ & $+0.20\pm0.01$ & upturn & FSRQ           &  2   \\
015106$+$251729     &   3.10 & $0.13\pm0.01$    &  c+j  &    10     & $-0.05\pm0.01$ & $-0.05\pm0.01$ & flat &                  &  3   \\
020346$+$113445     &   3.63 & $0.65\pm0.03$    &  c+j  &    10     & $+0.29\pm0.01$ & $-0.51\pm0.01$ & peaked$^*$ & FSRQ       &  4   \\
021435$+$015702     &   3.28 & $0.08\pm0.01$    &  c+j  &    10     & $+0.34\pm0.09$ & $-0.45\pm0.01$ & peaked   &              &  2   \\
023220$+$231756     &   3.42 & $0.31\pm0.02$    &  c+j  &    10     & $+0.87\pm0.08$ & $-0.37\pm0.01$ & peaked$^*$ &            &  1   \\
024611$+$182330     &   3.59 & $0.17\pm0.01$    &  c+j  &    1      & $+0.49\pm0.02$ & $-0.35\pm0.01$ & peaked & FSRQ           &  4   \\
025759$+$433836     &   4.06 & $0.31\pm0.02$    &  c+j  &    5      & $+0.36\pm0.01$ & $+0.04\pm0.01$ & inverted &              &  1   \\
032444$-$291821     &   4.63 & $0.15\pm0.01$    &  c+j  &    5      & $+0.15\pm0.01$ & $+0.15\pm0.01$ & inverted &  FSRQ        &  0   \\
033755$-$120404     &   3.44 & $0.29\pm0.02$    &  c+j  &    10     & $-0.32\pm0.01$ & $-0.27\pm0.01$ & flat & FSRQ             &  1   \\
033900$-$013317     &   3.19 & $0.37\pm0.02$    &  c+j  &    10     & $-0.03\pm0.01$  & $-0.40\pm0.01$ & flat & FSRQ            &  2   \\
035424$+$044107     &   3.26 & $0.39\pm0.02$    &  c+j  &    10     & $+0.04\pm0.02$  & $-0.36\pm0.01$ & complex  &             &  4   \\
042457$+$080517     &   3.09 & $0.19\pm0.01$    &  c+j  &    10     & $+0.11\pm0.02$  & $+0.19\pm0.01$ & inverted & FSRQ        &  2   \\
042835$+$173223     &   3.32 & $0.15\pm0.01$    &  c+j  &    10     & $-0.09\pm0.01$  & $-0.09\pm0.01$ & flat & FSRQ            &  4   \\
052506$-$233810     &   3.1  & $0.74\pm0.03$    &  c+j  &    10     & $+0.60\pm0.01$  & $+0.34\pm0.01$ & inverted & FSRQ        &  4   \\
052506$-$334304     &   4.41 & $0.40\pm0.01$    &  c+j  &    5      & $+0.43\pm0.02$  & $-0.96\pm0.02$ & peaked & Uncert.       &  0   \\
053954$-$283956     &   3.10 & $0.99\pm0.10$    &  c+j  &    10     & $+0.32\pm0.01$  & $-0.32\pm0.01$ & peaked$^{*}$ & FSRQ    &  4   \\
062419$+$385648     &   3.46 & $0.61\pm0.03$    &  c+j  &    10     & $+0.40\pm0.01$  & $-0.33\pm0.01$ & peaked & FSRQ          &  4   \\
064632$+$445116     &   3.39 & $2.27\pm0.10$    &  c+j  &    10     & $+0.52\pm0.01$  & $-0.62\pm0.01$ & peaked & FSRQ          &  4   \\
073357$+$045614     &   3.01 & $0.53\pm0.03$    &  c+j  &    10     & $+0.39\pm0.01$  & $-0.31\pm0.01$ & peaked & FSRQ          &  4   \\
075141$+$271631$^{a}$ &  3.20 & $0.20\pm0.01$   &       &           & $+1.02\pm0.03$  & $-0.75\pm0.01$ & peaked &               &  2   \\ 
075303$+$423130     &   3.59 & $0.41\pm0.02$    &  cso  &    8      & $+0.87\pm0.04$  & $-0.61\pm0.01$ & peaked &               &  1   \\
083322$+$095941     &   3.75 & $0.09\pm0.01$    &  c+j  &    1      & $-0.33\pm0.01$  & $-0.33\pm0.01$ & flat &                 &  1   \\
083910$+$200208     &   3.03 & $0.12\pm0.01$    &  c+j  &    10     & $+0.36\pm0.01$  & $-0.25\pm0.01$ & peaked &               &  3   \\
084715$+$383110     &   3.18 & $0.11\pm0.01$    &  c    &    4$^{c}$      & $+0.07\pm0.01$ & $+0.07\pm0.01$ & inverted &        &  2   \\
090549$+$041010     &   3.15 & $0.09\pm0.01$    &  c+j  &    1      & $+0.57\pm0.06$  & $-0.53\pm0.01$ & peaked &               &  2   \\
090915$+$035443     &   3.20 & $0.13\pm0.01$    &  c+j  &    4,9    & $-0.35\pm0.01$   & $-0.10\pm0.01$ &  flat &               &  3   \\
091551$+$000712     &   3.07 & $0.21\pm0.01$    &  T    &    4$^{c}$      & $-0.28\pm0.01$   & $-0.28\pm0.01$ &  flat & FSRQ    &  2   \\
093337$+$284532     &   3.42 & $0.05\pm0.01$    &  E    &    4$^{c}$      & $+0.17\pm0.02$   & $-0.67\pm0.01$ &  peaked &       &  1   \\
094113$+$114533     &   3.19 & $0.12\pm0.01$    &  c+j  &    1      & $-0.51\pm0.01$   & $-0.51\pm0.01$ &  flat &               &  3   \\
101644$+$203746     &   3.11 & $0.55\pm0.03$    &  c+j  &    1      & $-0.07\pm0.01$ & $-0.52\pm0.01$ &  complex & FSRQ         &  4   \\
102010$+$104003     &   3.15 & $0.15\pm0.01$    &       &           & $-0.63\pm0.01$   & $-0.94\pm0.01$ &  steep &              &  2   \\
102107$+$220921     &   4.26 & $0.12\pm0.01$    &  c+j  &    5      & $-0.51\pm0.01$   & $-0.17\pm0.01$ &  flat &               &  1   \\
102623$+$254259     &   5.28 & $0.11\pm0.01$    &  c+j  &    5      & $+0.11\pm0.03$   & $-0.50\pm0.01$ &  peaked & FSRQ        &  0   \\
102645$+$365825     &   3.25 & $0.19\pm0.01$    &  c    &    4$^{c}$      & $-0.08\pm0.01$   & $-0.27\pm0.01$ &  flat  &        &  3   \\
102838$-$084438     &   4.27 & $0.10\pm0.01$    &  c+j  &    5      & $-0.11\pm0.01$   & $-0.11\pm0.01$ &  flat & FSRQ          &  0   \\
103626$+$132654     &   3.09 & $0.04\pm0.01$    &  T    &    3      & $-0.89\pm0.01$   & $-0.53\pm0.01$ &  steep  &             &  3   \\
104523$+$314232     &   3.23 & $0.08\pm0.01$    &  c    &    4$^{c}$      & $+0.25\pm0.03$   & $-0.56\pm0.01$ & peaked  &       &  2   \\
110147$+$001038     &   3.69 & $0.06\pm0.01$    &  c    &    4$^{c}$      & $-0.74\pm0.01$   & $-0.78\pm0.01$ & steep &         &  1   \\
112500$+$333858     &   3.43 & $0.08\pm0.01$    &       &           & $-0.78\pm0.01$ & $-0.55\pm0.01$ & steep &                 &  2   \\
112851$+$232616     &   3.04 & $0.10\pm0.01$    &  E    &    4$^{c}$      & $-0.46\pm0.01$   & $-0.46\pm0.01$ & flat  &         &  2   \\
115016$+$433206     &   3.03 & $0.11\pm0.01$    &  c    &    4$^{c}$      & $+0.01\pm0.01$   & $+0.01\pm0.01$ & inverted & FSRQ &  3   \\
123055$-$113910     &   3.52 & $0.18\pm0.01$    &  c+j  &    8      & $+0.34\pm0.04$   & $-0.56\pm0.01$ & peaked & FSRQ         &  4   \\ 
124209$+$372005     &   3.81 & $0.56\pm0.02$    &  T    &    4$^{c}$      & $+0.26\pm0.01$   & $-0.30\pm0.01$ & peaked$^*$ &    &  4   \\
130122$+$190421     &   3.10 & $0.08\pm0.01$    &  c+j  &    10     & $+0.13\pm0.01$   & $-0.68\pm0.01$ & peaked &              &  2   \\ 
134022$+$375443     &   3.11 & $0.30\pm0.02$    &  c+j  &    1      & $+1.07\pm0.01$ & $-1.02\pm0.01$ & peaked &                &  3   \\ 
135406$-$020603     &   3.70 & $0.70\pm0.04$    &  c+j  &    10     & $+0.46\pm0.01$ & $-0.35\pm0.01$ & peaked & FSRQ           &  4   \\ 
135646$-$110130     &   3.01 & $0.23\pm0.01$    &  c+j  &    10     & $+0.13\pm0.01$ & $+0.13\pm0.01$ & inverted & FSRQ         &  2   \\
135652$+$291817     &   3.24 & $0.05\pm0.01$    &  T    &    4$^{c}$      & $-0.57\pm0.01$ & $-0.57\pm0.01$ & steep  &          &  2   \\
135706$-$174402     &   3.14 & $0.65\pm0.04$    &  c+j  &    10     & $+0.28\pm0.01$ & $-0.42\pm0.01$ & peaked$^{*}$  & FSRQ    &  4   \\ 
140135$+$151325$^{a}$ &   3.23 & $0.04\pm0.01$  &       &           & $+0.05\pm0.01$ & $-0.78\pm0.02$ & peaked &                &  3   \\ 
140501$+$041536     &   3.20 & $0.73\pm0.03$    &  c+j  &    4$^{c}$      & $+0.01\pm0.01$ & $-0.43\pm0.01$ & peaked$^*$ & FSRQ &  4   \\
141152$+$430024     &   3.21 & $0.09\pm0.01$    &  c+j  &    4$^{c}$      & $-0.21\pm0.01$ & $-0.21\pm0.01$ & flat  &           &  2   \\
141300$+$394745     &   3.71 & $0.05\pm0.01$    &       &           & $-0.55\pm0.01$ & $-0.44\pm0.01$ & flat  &                 &  1   \\
141318$+$450523     &   3.11 & $0.16\pm0.01$    &  c    &    4$^{c}$      & $+0.04\pm0.01$ & $+0.04\pm0.01$ & inverted  & FSRQ  &  2   \\
141821$+$425020     &   3.45 & $0.05\pm0.01$    &       &           & $+0.32\pm0.01$ & $-0.77\pm0.01$ & peaked &                &  2   \\ 
142107$-$064355     &   3.68 & $0.22\pm0.01$    &  c+j  &    8      & $-0.65\pm0.01$ & $-0.38\pm0.01$ & complex  & FSRQ         &  1   \\
142438$+$225600$^{a}$ &  3.62 & $0.61\pm0.03$   &  c+j  &    10     & $+0.62\pm0.01$ & $-0.59\pm0.01$ & peaked  & FSRQ          &  4   \\ 
143023$+$420436     &   4.71 & $0.12\pm0.01$    &  c+j  &    5      & $-0.13\pm0.01$ & $-0.13\pm0.01$ & flat & FSRQ             &  0   \\
144516$+$095835     &   3.55 & $0.98\pm0.04$    &  c+j  &    8      & $+0.68\pm0.01$ & $-0.87\pm0.01$ & peaked & FSRQ           &  4   \\
145722$+$051922     &   3.17 & $0.07\pm0.01$    &  c+j  &    10     & $-0.30\pm0.01$ & $-0.30\pm0.01$ & flat   &                &  2   \\
145805$+$085529     &   3.06 & $0.06\pm0.01$    &       &           & $-0.53\pm0.01$ & $-0.87\pm0.01$ & steep  &                &  2   \\
145927$+$325358     &   3.32 & $0.05\pm0.01$    &  complex & 4$^{c}$      & $-0.67\pm0.01$ & $-0.85\pm0.01$ & steep  &          &  2   \\
150328$+$041949     &   3.66 & $0.19\pm0.01$    &  c+ext$^{b}$ &  1 & $+0.21\pm0.01$ & $-0.29\pm0.01$ & peaked  &               &  2   \\
152117$+$175601     &   3.06 & $0.13\pm0.01$    &  c    &    4$^{c}$      & $+0.40\pm0.03$ & $-0.87\pm0.01$ & peaked  &         &  3   \\
152219$+$211957     &   3.22 & $0.08\pm0.01$    &       &           & $-0.99\pm0.01$ & $-1.04\pm0.01$ & steep     &             &  2   \\
153815$+$001905     &   3.49 & $0.32\pm0.02$    &  c+j  &    10     & $+0.38\pm0.01$ & $-0.77\pm0.01$ & peaked &                &  3   \\ 
155930$+$030447     &   3.89 & $0.43\pm0.02$    &   T   &    4$^{c}$      & $+0.11\pm0.01$ & $-0.43\pm0.01$ & peaked & FSRQ     &  4   \\ 
160608$+$312447     &   4.56 & $0.64\pm0.03$    &  cso  &    6      & $+1.10\pm0.02$ & $-0.57\pm0.01$ & peaked &                &  0   \\ 
161005$+$181143     &   3.11 & $0.07\pm0.01$    &   T   &    4$^{c}$      & $-0.25\pm0.01$ & $-0.79\pm0.02$ & steep &           &  3   \\
161637$+$045932     &   3.21 & $1.03\pm0.04$    &  c+j  &    4$^{c}$      & $+0.64\pm0.01$ & $-0.56\pm0.01$ & peaked$^*$ & FSRQ &  4   \\
163257$-$003321$^{a}$ &  3.42 & $0.15\pm0.01$   &  c+j  &    10     & $-0.19\pm0.01$ & $-0.19\pm0.01$ & flat  & FSRQ            &  4   \\ 
165519$+$324241     &   3.18 & $0.06\pm0.01$    &  c+j  &    1      & $-0.80\pm0.01$ & $-0.80\pm0.01$ & steep  &                &  2   \\
165543$+$194847     &   3.26 & $0.11\pm0.01$    &  E    &    4$^{c}$      & $-0.10\pm0.01$ & $-0.36\pm0.01$ & flat  &           &  2   \\
165844$-$073918     &   3.74 & $0.83\pm0.03$    &  c+j  &    10     & $+0.16\pm0.01$ & $-0.86\pm0.01$ & peaked$^*$ & FSRQ       &  1   \\
171521$+$214534     &   4.01 & $0.16\pm0.01$    &  T    &    5      & $-0.66\pm0.01$ & $-0.66\pm0.01$ & steep &                 &  1   \\
174020$+$350048     &   3.22 & $0.06\pm0.01$    &       &           & $-0.69\pm0.01$ & $-0.69\pm0.01$ & steep  &                &  2   \\
184057$+$390046     &   3.1 & $0.18\pm0.01$    &  c+j  &    10     & $+0.27\pm0.01$ & $-0.32\pm0.01$ & peaked & FSRQ            &  3   \\ 
193957$-$100241     &   3.78 & $0.59\pm0.02$    &  c+j  &    8      & $-0.17\pm0.01$ & $-0.73\pm0.01$ & complex   & FSRQ        &  1   \\
200324$-$325144     &   3.77 & $0.68\pm0.03$    &  c+j  &    10     & $+0.40\pm0.01$ & $-0.74\pm0.01$ & peaked  & FSRQ          &  4   \\
201918$+$112712$^{a}$  &   3.27 & $0.08\pm0.01$ &  c+ext&    7      & $+0.71\pm0.01$ & $-0.84\pm0.01$ & peaked  &               &  2   \\ 
204124$+$185502     &   3.05 & $0.17\pm0.01$    &       &           & $-0.83\pm0.01$ & $-0.83\pm0.01$ & steep   &               &  2   \\
204257$-$222326     &   3.63 & $0.09\pm0.01$    &  c+j  &    10     & $+0.13\pm0.01$ & $-0.60\pm0.02$ & peaked  & FSRQ          &  1   \\
204310$+$125513     &   3.27 & $0.14\pm0.02$    &  c+j  &    10     & $-0.18\pm0.01$ & $-0.18\pm0.01$ & flat  & FSRQ            &  3   \\
205051$+$312727     &   3.18 & $0.57\pm0.03$    &  c+j  &    10     & $+0.15\pm0.01$ & $-0.39\pm0.01$ & peaked  & FSRQ          &  3   \\ 
212912$-$153841     &   3.26 & $1.46\pm0.10$    &  c+j  &    11     & $+0.56\pm0.01$ & $-0.61\pm0.01$ & peaked$^{*}$ & FSRQ     &  4   \\ 
213412$-$041909     &   4.34 & $0.22\pm0.01$    &  c+j  &    5      & $-0.31\pm0.01$ & $-0.31\pm0.01$ & flat &                  &  1   \\
221748$+$022011     &   3.57 & $0.32\pm0.02$    &  c+j  &    10     & $-0.65\pm0.01$ & $-0.52\pm0.01$ & steep & FSRQ            &  3   \\
221935$-$271902     &   3.63 & $0.21\pm0.01$    &  c+j  &    10     & $-0.14\pm0.01$ & $-0.14\pm0.01$ & flat  & FSRQ            &  1   \\
222536$+$204014     &   3.56 & $0.13\pm0.01$    &  c+j  &    10     & $-0.72\pm0.01$ & $-0.72\pm0.01$ & steep  &                &  1   \\
224800$-$054117     &   3.29 & $0.20\pm0.01$    &  c+j  &    10     & $+0.02\pm0.01$ & $-0.57\pm0.01$ & peaked & Uncert.        &  1   \\
225153$+$221737     &   3.66 & $0.11\pm0.01$    &  c+j  &    10     & $-0.36\pm0.01$ & $-0.36\pm0.01$ & flat   & FSRQ           &  4   \\
231448$+$020150     &   4.11 & $0.07\pm0.01$    &  c+j  &    5      & $+0.06\pm0.01$ & $-0.53\pm0.01$ & peaked  &               &  1   \\
231643$-$334912     &   3.1  & $0.48\pm0.03$    &  c+j  &    10     & $+0.75\pm0.02$ & $-1.58\pm0.01$ & peaked  & FSRQ          &  2   \\ 
232118$-$082721     &   3.16 & $0.18\pm0.01$    &  c+j  &    10     & $-0.46\pm0.01$ & $-0.17\pm0.01$ & flat & FSRQ             &  3   \\
234451$+$343348     &   3.05 & $0.09\pm0.01$    &       &           & $-0.50\pm0.01$ & $-0.50\pm0.01$ & flat  &                 &  2   \\
\hline
\multicolumn{9}{l}{\footnotesize$^a$ gravitationally lensed system.}\\
\multicolumn{9}{l}{\footnotesize$^b$ also classified as core dominated with a lobe in \cite{2011AJ....141..182K}.}\\
\multicolumn{9}{l}{\footnotesize$^c$ VLA measurements.}\\
\multicolumn{9}{l}{\footnotesize$^*$ sources with two components in the radio spectrum.}\\
\multicolumn{9}{l}{\footnotesize 5---\cite{2022ApJS..260...49K}, 6---\cite{2022MNRAS.511.4572A}, 7---\cite{1994MNRAS.269..902G}, 8---\cite{2022ApJ...937...19Z}, 9---\cite{2021ApJ...915...98P},}\\
\multicolumn{9}{l}{\footnotesize 10---\cite{2022ApJS..260....4P}, 11---\cite{2023Galax..11...42C}.}\\
\end{longtable}
\end{landscape}

\begin{landscape}
\begin{longtable}{|l|c|c|c|c|c|c|c|c|c|c|c|c|c|c|c|c|c|c|c|c|}
\caption{\label{TableA2} The variability, modulation and fractional variability indices for the sample. Columns 3--6 show the timescale (in year) at which we have estimated the variability and the number of measurements at 4.7 and 22.3 GHz. The table is available in the VizieR database at \url{https://cdsarc.cds.unistra.fr/viz-bin/cat/J/other/Galax/12.25}}\\
\hline
Name &  $z$ & $t_{\rm 4.7}$ & N$_{4.7}$ & $t_{\rm 22.3}$ & N$_{22.3}$ & $V_{S_{2.3}}$ & $M_{2.3}$ & $F_{2.3}$ & $V_{S_{4.7}}$ & $M_{4.7}$ & $F_{4.7}$ & $V_{S_{8.2}}$ & $M_{8.2}$ & $F_{8.2}$ & $V_{S_{11.2}}$ & $M_{11.2}$ &  $F_{11.2}$ & $V_{S_{22.3}}$ & $M_{22.3}$ & $F_{22.3}$ \\
 1 & 2 & 3 & 4 & 5 & 6 & 7 & 8 & 9 & 10 & 11 & 12 & 13 & 14 & 15 & 16 & 17 & 18 & 19 & 20 & 21 \\
\hline
\endfirsthead
\caption[]{continued.}\\
\hline
Name & $z$ & $t_{\rm 4.7}$ & N$_{4.7}$ & $t_{\rm 22.3}$ & N$_{22.3}$ & $V_{S_{2.3}}$ & $M_{2.3}$ & $F_{2.3}$ & $V_{S_{4.7}}$ & $M_{4.7}$ & $F_{4.7}$ & $V_{S_{8.2}}$ & $M_{8.2}$ & $F_{8.2}$ & $V_{S_{11.2}}$ & $M_{11.2}$ & $F_{11.2}$ &  $V_{S_{22.3}}$ & $M_{22.3}$ & $F_{22.3}$ \\
 1 & 2 & 3 & 4 & 5 & 6 & 7 & 8 & 9 & 10 & 11 & 12 & 13 &14 & 15 & 16 & 17 & 18 & 19 & 20 & 21\\
\hline
\endhead
\hline
\endfoot
J0001$+$1914 &  3.10      &  35  &   21    & 3 & 9 &   0.53    &       0.42   &    0.46    &       0.47    &       0.31   &    0.31    &       0.63    &       0.64   &    0.65    &       0.23    &       0.17   &   0.17     &       0.57    &       0.34   &    0.35   \\
J0006$+$1415 &       3.20      &  40  &   14    & -- & -- &    0.08    &       0.1    &    0.09    &       0.1     &       0.13   &    0.12    &       0.25    &       0.17   &    0.16    &       0.03    &       0.09   &   --        &       --      &       --     &    --      \\
J0048$+$0640 &       3.58      &  36  &   25    & -- & -- &    0.28    &       0.18   &    0.18    &       0.25    &       0.14   &    0.13    &       0.24    &       0.15   &    0.15    &       0.12    &       0.16   &   0.1      &       --      &       --     &    --      \\
J0100$-$2708 &       3.52      &  31  &   10    & -- & -- &    0.11    &       0.14   &    0.13    &       0.17    &       0.14   &    0.14    &       0.11    &       0.13   &    0.1     &       0.17    &       0.2    &   0.17     &       --      &       --     &    --      \\
J0121$-$2806 &       3.11      &  30  &   19    & 7 & 18 &    0.06    &       0.2    &    0.15    &       0.43    &       0.37   &    0.4     &       0.32    &       0.24   &    0.24    &       0.2     &       0.17   &   0.15     &       0.41    &       0.33   &    0.34   \\
J0131$+$4358 &       3.12      &  34  &   12    & -- & -- &    --      &       --     &    --       &       0.37    &       0.22   &    0.22    &       0.3     &       0.27   &    --       &       0.17    &       0.28   &   0.1      &       --      &       --     &    --      \\
J0148$+$4215 &       3.24      &  34  &   14    & 3 & 8 &    --      &       0.07   &    --       &       0.23    &       0.17   &    0.16    &       0.14    &       0.12   &    0.11    &       0.05    &       0.1    &   --        &       0.36    &       0.25   &    0.22   \\
J0151$+$2517 &       3.10      &  34  &   17    & 2 & 8 &    0.25    &       0.2    &    0.2     &       0.32    &       0.29   &    0.29    &       0.23    &       0.16   &    0.15    &       0.1     &       0.11   &   0.06     &       0.08    &       0.12   &    --      \\
J0203$+$1134 &       3.63      &  42  &   38    & 22 & 31 &    0.21    &       0.09   &    0.08    &       0.33    &       0.22   &    0.21    &       0.4     &       0.18   &    0.17    &       0.38    &       0.26   &   0.26     &       0.45    &       0.33   &    0.33   \\
J0214$+$0157 &       3.28      &  36  &   19    & -- & -- &    0.23    &       0.21   &    0.2     &       0.13    &       0.12   &    0.08    &       0.23    &       0.21   &    0.21    &       0.03    &       0.16   &   --        &       --      &       --     &    --      \\
J0232$+$2317 &       3.42      &  33  &   13    & -- & -- &   0.2     &       0.2    &    0.22    &       0.29    &       0.19   &    0.19    &       0.21    &       0.15   &    0.14    &       0.08    &       0.17   &   0.12     &       --      &       0.11   &    --      \\
J0246$+$1823 &       3.59      &  34  &   44    & 2 & 5 &    0.34    &       0.26   &    0.27    &       0.29    &       0.17   &    0.08    &       0.25    &       0.15   &    0.15    &       0.29    &       0.2    &   --        &       0.31    &       0.33   &    0.34   \\
J0257$+$4338 &       4.06       &  34  &   16    & 5 & 10 &    0.06    &       0.26   &    0.18    &       0.33    &       0.31   &    0.32    &       0.45    &       0.37   &    0.37    &       0.08    &       0.1    &   0.05     &       0.04    &       0.13   &    0.05   \\
J0324$-$2918 &       4.63      &  30  &   19    & 15 & 15 &    0.08    &       0.08   &    --       &       0.44    &       0.43   &    0.45    &       0.41    &       0.25   &    0.25    &       0.14    &       0.16   &   0.13     &       0.39    &       0.4    &    0.39   \\
J0337$-$1204 &       3.44      &  31  &   12    & 14 & 6 &    0.26    &       0.18   &    0.18    &       0.19    &       0.13   &    0.12    &       0.24    &       0.13   &    0.12    &       0.09    &       0.11   &   0.07     &       0.68    &       0.64   &    0.28   \\
J0339$-$0133 &       3.19   &  $>$43  &   17    & 13 & 6 &    0.2     &       0.13   &    0.12    &       0.21    &       0.12   &    0.11    &       0.14    &       0.1    &    0.08    &       0.25    &       0.16   &   0.13     &       0.49    &       0.46   &    0.5    \\
J0354$+$0441 &       3.26      &  40  &   31    & 22 & 13 &    0.18    &       0.15   &    0.14    &       0.27    &       0.13   &    0.11    &       0.22    &       0.18   &    0.17    &       0.17    &       0.15   &   0.12     &       0.54    &       0.47   &    0.47   \\
J0424$+$0805 &       3.8       &  36  &   21    & 3 & 9 &   0.32    &       0.23   &    0.23    &       0.19    &       0.14   &    0.13    &       0.38    &       0.29   &    0.3     &       0.16    &       0.12   &   0.07     &       0.4     &       0.36   &    0.35   \\
J0428$+$1732 &       3.32      &  30  &   56    & 5 & 17 &    0.38    &       0.27   &    0.28    &       0.33    &       0.16   &    0.14    &       0.22    &       0.19   &    0.19    &       0.22    &       0.13   &   0.07     &       0.46    &       0.34   &    0.32   \\
J0525$-$2338 &       3.1      &  30  &   55    & 14 & 50 &    0.28    &       0.19   &    --       &       0.11    &       0.08   &    0.06    &       0.35    &       0.24   &    0.23    &       0.36    &       0.15   &   0.13     &       0.5     &       0.25   &    0.23   \\
J0525$-$3343 &       4.41     &   30  &   18    & -- & -- &    0.29    &       0.3    &    0.36    &       0.71    &       0.62   &    0.66    &       0.3     &       0.25   &    0.18    &       --      &       --     &   --        &       --      &       --     &    --      \\
J0539$-$2839 &       3.10   &  $>$43  &   66    & 23 & 47 &    0.44    &       0.33   &    0.3     &       0.59    &       0.35   &    0.34    &       0.56    &       0.35   &    0.35    &       0.59    &       0.4    &   0.39     &       0.61    &       0.45   &    0.43   \\
J0624$+$3856 &       3.46      &  46  &   68    & 23 & 47 &    0.4     &       0.35   &    0.35    &       0.3     &       0.18   &    0.18    &       0.43    &       0.2    &    0.19    &       0.26    &       0.21   &   0.18     &       0.35    &       0.27   &    0.2    \\
J0646$+$4451 &       3.39      &  47  &  187    & 30 & 319 &    0.46    &       0.23   &    0.21    &       0.67    &       0.31   &    0.31    &       0.74    &       0.49   &    0.5     &       0.66    &       0.37   &   0.36     &       0.65    &       0.25   &    0.24   \\
J0733$+$0456 &       3.1       &  40  &   35    & 4 & 15 &    0.47    &       0.21   &    0.21    &       0.37    &       0.24   &    0.24    &       0.48    &       0.36   &    0.37    &       0.29    &       0.21   &   0.18     &       0.44    &       0.27   &    0.24   \\
J0751$+$2716 &       3.20      &  42  &   13    & -- & -- &    0.17    &       0.16   &    0.16    &       0.04    &       0.05   &    --       &       0.04    &       0.1    &    0.05    &       0.05    &       0.12   &   --        &       --      &       --     &    --      \\
J0753$+$4231 &       3.59      &  34  &   23    & 18 & 2 &    0.2     &       0.17   &    0.17    &       0.12    &       0.06   &    0.05    &       0.04    &       0.08   &    0.05    &       0.09    &       0.1    &   0.05     &       --      &       0.17   &    --      \\
J0833$+$0959 &       3.75      &  36  &   17    & -- & -- &    --      &       0.05   &    --       &       0.17    &       0.14   &    0.11    &       0.24    &       0.17   &    0.13    &       --      &       0.07   &   --        &       --      &       --     &    --      \\
J0839$+$2002 &       3.3       &  35  &   16    & 3 & 2 &    0.23    &       0.2    &    0.21    &       0.29    &       0.21   &    0.2     &       0.15    &       0.15   &    0.13    &       0.21    &       0.21   &   0.19     &       0.05    &       0.17   &    --      \\
J0847$+$3831 &       3.18      &  34  &   16    & 3 & 5 &    0.36    &       0.35   &    0.37    &       0.2     &       0.14   &    0.11    &       0.36    &       0.24   &    0.24    &       --      &       0.05   &   --        &       0.14    &       0.24   &    0.23   \\
J0905$+$0410 &       3.15      &  36  &   12    & -- & -- &    0.17    &       0.19   &    0.23    &       0.11    &       0.12   &    0.06    &       0.17    &       0.2    &    0.21    &       0.04    &       0.15   &   --        &       --      &       --     &    --      \\
J0909$+$0354 &       3.20      &  37  &   18    & 2 & 2 &    0.15    &       0.18   &    0.19    &       0.5     &       0.29   &    0.29    &       0.36    &       0.23   &    0.23    &       0.26    &       0.22   &   0.22     &       0.05    &       0.18   &    --      \\
J0915$+$0007 &       3.7       &  42  &   19    & 3 & 7 &    0.27    &       0.23   &    0.19    &       0.23    &       0.15   &    0.14    &       0.08    &       0.11   &    --       &       0.05    &       0.1    &   0.05     &       0.61    &       0.68   &    0.72   \\
J0933$+$2845 &       3.42      &  34  &   17    & -- & -- &    0.34    &       0.25   &    0.24    &       0.25    &       0.19   &    0.14    &       0.22    &       0.21   &    0.09    &       0.21    &       0.23   &   --        &       --      &       --     &    --      \\
J0941$+$1145 &       3.19      &  39  &   16    & 1 & 2 &    0.86    &       0.38   &    0.42    &       0.35    &       0.33   &    0.33    &       0.96    &       1      &    0.97    &       0.07    &       0.13   &   0.05     &       --      &       0.09   &    --      \\
J1016$+$2037 &       3.11      &  44  &   91    & 23 & 44 &   0.18    &       0.15   &    0.14    &       0.5     &       0.23   &    0.22    &       0.47    &       0.33   &    0.34    &       0.41    &       0.18   &   0.15     &       0.34    &       0.21   &    0.14   \\
J1020$+$1040 &       3.15      &  42  &   11    & -- & -- &    0.18    &       0.17   &    0.17    &       0.09    &       0.08   &    0.06    &       0.23    &       0.17   &    0.15    &       --      &       0.06   &   --        &       --      &       --     &    --      \\
J1021$+$2209 &       4.23      &  35  &   12    & -- & -- &    --      &       0.06   &    --       &       0.18    &       0.13   &    0.11    &       0.13    &       0.16   &    0.15    &       0.09    &       0.13   &   0.1      &       --      &       --     &    --      \\
J1026$+$2542 &       5.28      &  25  &   13    & -- & -- &    0.12    &       0.12   &    0.12    &       0.3     &       0.15   &    0.14    &       0.2     &       0.17   &    0.17    &       --      &       0.08   &   --        &       --      &       --     &    --      \\
J1026$+$3658 &       3.25      &  35  &   11    & -- & -- &    0.05    &       0.08   &    0.07    &       0.4     &       0.3    &    0.31    &       0.43    &       0.3    &    0.32    &       0.02    &       0.08   &   0.05     &       --      &       --     &    --      \\
J1028$-$0844 &       4.27      &  31  &   49    & 4 & 5 &    --      &       --     &    --       &       0.56    &       0.35   &    0.31    &       0.15    &       0.15   &    0.14    &       0.23    &       0.31   &   0.33     &       0.25    &       0.3    &    0.31   \\
J1036$+$1326 &       3.9       &  37  &   13    & -- & -- &    --      &       --     &    --       &       0.48    &       0.4    &    0.39    &       0.31    &       0.3    &    0.26    &       0.39    &       0.35   &   0.25     &       --      &       --     &    --      \\
J1045$+$3142 &       3.23      &  34  &   10    & -- & --&    0.2     &       0.23   &    0.24    &       0.11    &       0.13   &    0.06    &       0.12    &       0.17   &    0.09    &       --      &       0.14   &   --        &       --      &       --     &    --      \\
J1101$+$0010 &       3.69      &  36  &   14    & -- & -- &   0       &       0.1    &    --       &       0.28    &       0.23   &    0.18    &       0.18    &       0.25   &    0.15    &       0.42    &       0.57   &   0.63     &       --      &       --     &    --      \\
J1125$+$3338   &       3.43      &  37  &   14    & -- & -- &    0.29    &       0.22   &    0.22    &       0.24    &       0.15   &    0.11    &       0.07    &       0.14   &    --       &       --      &       0.09   &   --        &       --      &       --     &    --      \\
J1128$+$2326   &       3.4       &  34  &   14    & -- & -- &    0.45    &       0.25   &    0.26    &       0.11    &       0.09   &    0.01    &       0.37    &       0.19   &    0.15    &       --      &       0.09   &   --        &       --      &       --     &    --      \\
J1150$+$4332   &       3.3       &  35  &   14    & 2 & 6 &    0.5     &       0.53   &    0.59    &       0.42    &       0.28   &    0.28    &       0.44    &       0.27   &    0.28    &       0.29    &       0.18   &   0.14     &       0.12    &       0.19   &    0.16   \\
J1230$-$1139 &       3.52   &  $>$43  &  51     & 14 & 7 &   0.42    &       0.33   &    0.36    &       0.47    &       0.35   &    0.38    &       0.47    &       0.41   &    0.43    &       0.44    &       0.24   &   0.25     &       0.41    &       0.36   &    0.37   \\
J1242$+$3720 &       3.81      &  46  &   24    & 21 & 13 &   0.18    &       0.15   &    0.12    &       0.28    &       0.21   &    0.21    &       0.29    &       0.18   &    0.19    &       0.19    &       0.14   &   0.12     &       0.14    &       0.17   &    0.11   \\
J1301$+$1904 &       3.10      &  35  &   13    & -- & -- &    0.25    &       0.21   &    0.21    &       0.05    &       0.12   &    -       &       --      &       0.06   &    --       &       --      &       0.15   &   --        &       --      &       --     &    --      \\
J1340$+$3754 &       3.11      &  36  &   12    & -- & -- &    0.08    &       0.11   &    0.12    &       0.3     &       0.17   &    0.17    &       0.15    &       0.1    &    0.09    &       0.03    &       0.11   &   --        &       --      &       --     &    --      \\
J1354$-$0206 &       3.70      &  38  &   60    & 23 & 33 &    0.22    &       0.16   &    0.16    &       0.24    &       0.13   &    0.12    &       0.14    &       0.09   &    0.09    &       0.3     &       0.16   &   0.1      &       0.50    &       0.24   &    --      \\
J1356$-$1101 &       3.1       &  35  &    9    & -- & -- &    0.21    &       0.2    &    0.22    &       0.11    &       0.1    &    0.11    &       0.32    &       0.29   &    0.32    &       --      &       --     &   --        &       0.19    &       0.25   &    --      \\
J1356$+$2918 &       3.24      &  35  &    7    & -- & -- &    --      &       --     &    --       &       0.22    &       0.19   &    0.19    &       0.02    &       0.1    &    0.04    &       --      &       --     &   --        &       --      &       --     &    --      \\
J1357$-$1744 &       3.14      &  38  &   19    & 23 & 3 &    0.37    &       0.26   &    0.25    &       0.28    &       0.12   &    0.12    &       0.27    &       0.21   &    0.22    &       0.36    &       0.39   &   0.44     &       0.48    &       0.55   &    0.63   \\
J1401$+$1513 &       3.23      &  33  &   11    & -- & -- &    --      &       0.03   &    --       &       0.36    &       0.28   &    0.28    &       --      &       0.14   &    --       &       0.09    &       0.27   &   0.18     &       --      &       --     &    --      \\
J1405$+$0415 &       3.20   &  $>$43  &   26    & 24 & 12 &    0.23    &       0.18   &    0.18    &       0.25    &       0.16   &    0.16    &       0.2     &       0.12   &    0.11    &       0.19    &       0.17   &   0.14     &       0.26    &       0.3    &    0.25   \\
J1411$+$4300 &       3.21      &  34  &   10    & -- & -- &    0.08    &       0.16   &    0.17    &       0.05    &       0.06   &    --       &       0.13    &       0.15   &    0.12    &       0.13    &       0.17   &   0.14     &       --      &       --     &    --      \\
J1413$+$3947 &       4.12      &  34  &   10    & -- & -- &    0.22    &       0.27   &    0.29    &       0.12    &       0.14   &    --       &       0.03    &       0.15   &    --       &       0.15    &       0.26   &   0.2      &       --      &       --     &    --      \\
J1413$+$4505 &       3.11      &  34  &    8    & $<1$ & 2 &    --      &       --     &    --       &       0.27    &       0.16   &    0.16    &       0.19    &       0.19   &    0.21    &       0.03    &       0.13   &   0.11     &       0.09    &       0.24   &    --      \\
J1418$+$4250 &       3.45      &  34  &    4    & -- & -- &    --      &       --     &    --       &       0.02    &       0.08   &    --       &       --      &       --     &    --       &     --      &       --     &   --        &       --      &       --     &    --      \\
J1421$-$0643 &       3.68      &  30  &    6    & -- & -- &    0.19    &       0.18   &    0.2     &       0.3     &       0.23   &    0.26    &       0.21    &       0.16   &    0.17    &       --      &       --     &   --        &       --      &       --     &    --      \\
J1424$+$2256 &       3.62      &  42  &   29    & 22 & 18 &    0.47    &       0.14   &    0.14    &       0.14    &       0.12   &    0.11    &       0.66    &       0.21   &    0.17    &       0.16    &       0.16   &   0.14     &       0.1     &       0.28   &    0.17   \\
J1430$+$4204 &       4.71      &  34  &   16    & 2 & 4 &    0.4     &       0.31   &    0.32    &       0.51    &       0.43   &    0.44    &       0.4     &       0.29   &    0.3     &       0.36    &       0.3    &   0.31     &       0.32    &       0.29   &    0.32   \\
J1445$+$0958 &       3.55      &  41  &   76    & 23 & 27 &    0.15    &       0.1    &    0.08    &       0.15    &       0.1    &    0.09    &       0.26    &       0.17   &    0.15    &       0.57    &       0.26   &   0.26     &       0.67    &       0.4    &    0.38   \\
J1457$+$0519 &       3.17      &  37  &   12    & -- & -- &    0.1     &       0.17   &    0.16    &       0.13    &       0.13   &    0.08    &       0.19    &       0.2    &    0.17    &       0.05    &       0.17   &   0.05     &       --      &       --     &    --      \\
J1458$+$0855 &       3.6       &  34  &    7    & -- & -- &    --      &       0.35   &    --       &       0.2     &       0.21   &    --       &       --      &       0.06   &    --       &       --      &       --     &   --        &       --      &       --     &    --      \\
J1459$+$3253 &       3.32      &  39  &   11    & -- & -- &    0.04    &       0.16   &    --       &       0.23    &       0.17   &    0.07    &       0.08    &       0.19   &    --       &       --      &       0.22   &   --        &       --      &       --     &    --      \\
J1503$+$0419 &       3.66      &  39  &    8    & -- & -- &    0.03    &       0.04   &    0.03    &       0.1     &       0.08   &    0.07    &       0.04    &       0.03   &    --       &       --      &       0.01   &   --        &       --      &       --     &    --      \\
J1521$+$1756 &       3.6       &  40  &   13    & -- & -- &    0.21    &       0.17   &    --       &       0.32    &       0.23   &    0.23    &       0.18    &       0.17   &    --       &       0.1     &       0.15   &   0.11     &       --      &       --     &    --      \\
J1522$+$2119 &       3.22      &  34  &   10    & -- & -- &    0.18    &       0.17   &    0.18    &       0.13    &       0.1    &    --       &       0.09    &       0.18   &    --       &       --      &       0.18   &   --        &       --      &       --     &    --      \\
J1538$+$0019 &       3.49   &  $>$43  &   20    & 23 & 10 &    0.39    &       0.34   &    0.35    &       0.5     &       0.36   &    0.36    &       0.58    &       0.37   &    0.38    &       0.27    &       0.2    &   0.19     &       0.38    &       0.28   &    0.12   \\
J1559$+$0304 &       3.89      &  37  &   63    & 8 & 29 &    0.25    &       0.19   &    0.21    &       0.41    &       0.16   &    0.16    &       0.17    &       0.13   &    0.12    &       0.18    &       0.15   &   0.13     &       0.36    &       0.22   &    0.19   \\
J1606$+$3124 &       4.56      &  37  &   26    & 43 & 22 &    0.3     &       0.16   &    0.15    &       0.28    &       0.18   &    0.19    &       0.22    &       0.15   &    0.15    &       0.16    &       0.13   &   0.12     &       0.1     &       0.15   &    0.07   \\
J1610$+$1811 &       3.11      &  40  &   12    & -- & -- &    0.23    &       0.18   &    0.18    &       0.41    &       0.31   &    0.31    &       0.17    &       0.18   &    0.08    &       --      &       0.23   &   --        &       --      &       --     &    --      \\
J1616$+$0459 &       3.21    &  $>$43 &   64    & 23 & 22 &    --      &       0.13   &    0.12    &       0.26    &       0.13   &    0.12    &       0.42    &       0.15   &    0.13    &       0.17    &       0.1    &   0.09     &       0.26    &       0.25   &    0.21   \\
J1632$-$0033 &       3.42      &  31  &   57    & 5 & 6 &    0.25    &       0.2    &    0.2     &       0.43    &       0.22   &    0.2     &       0.17    &       0.12   &    0.11    &       0.4     &       0.2    &   0.12     &       0.4     &       0.43   &    0.44   \\
J1655$+$3242 &       3.18      &  34  &   12    & -- & -- &    0.24    &       0.22   &    0.21    &       0.04    &       0.12   &    --       &       --      &       0.09   &    --       &       --      &       0.12   &   --        &       --      &       --     &    --      \\
J1655$+$1948 &       3.26      &  34  &   15    & -- & -- &    0.28    &       0.26   &    0.26    &       0.24    &       0.19   &    0.18    &       0.27    &       0.16   &    0.14    &       0.12    &       0.17   &   0.11     &       --      &       --     &    --      \\
J1658$-$0739 &       3.74      &  37  &   12     & 10 & 9 &    0.22    &       0.16   &    0.16    &       0.27    &       0.21   &    0.21    &       0.33    &       0.27   &    0.28    &       0.13    &       0.12   &   0.05     &       0.33    &       0.35   &    0.32   \\
J1715$+$2145 &       4.1       &  35  &   16    & -- & -- &    0.47    &       0.24   &    0.24    &       0.38    &       0.3    &    0.31    &       0.19    &       0.14   &    0.11    &       --      &       0.08   &   --        &       --      &       0.18   &    --      \\
J1740$+$3500 &       3.22      &  34  &   18    & -- & -- &    0.29    &       0.22   &    0.21    &       0.02    &       0.08   &    --       &       --      &       0.11   &    --       &       0.27    &       0.19   &   --        &       --      &       --     &    --      \\
J1840$+$3900 &       3.1       &  36  &   23    & 20 & 13 & 0.24    &       0.22   &    0.18    &       0.52    &       0.41   &    0.42    &       0.14    &       0.16   &    0.15    &       0.45    &       0.34   &   0.33     &       0.52    &       0.51   &    0.48   \\
J1939$-$1002 &       3.78      &  30  &   17    & 23 & 4 &   0.28    &       0.19   &    0.15    &       0.15    &       0.09   &    0.09    &       0.25    &       0.14   &    0.15    &       0.33    &       0.28   &   0.29     &       0.33    &       0.33   &    0.36   \\
J2003$-$3251 &       3.77   &  $>$43  &   39    & 15 & 36 &     0.53    &       0.32   &    0.18    &       0.32    &       0.2    &    0.2     &       0.44    &       0.15   &    0.14    &       0.15    &       0.16   &   0.13     &       0.42    &       0.31   &    0.31   \\
J2019$+$1127 &       3.27      &  41  &   12    & -- & -- &     0.06    &       0.12   &    --       &       0.06    &       0.09   &    0.04    &       0.05    &       0.23   &    --       &       --      &       --     &   --        &       --      &      --     &    --      \\
J2041$+$1855 &       3.5       &  39  &   12    & -- & -- &     0.01    &       0.07   &    --       &       0.18    &       0.16   &    0.15    &       0.11    &       0.22   &    --       &       0.17    &       0.24   &   0.27     &       --      &       --     &    --      \\
J2042$-$2223 &       3.63      &  30  &    4    & -- & -- &     0.01    &       0.01   &    --       &       0.27    &       0.25   &    --       &       0.24    &       0.19   &    0.22    &       --      &       --     &   --        &       --      &       --     &    --      \\
J2043$+$1255 &       3.27      &  40  &   12    & -- & -- &     --      &       --     &    --       &       0.52    &       0.42   &    0.43    &       0.09    &       0.11   &    0.1     &       --      &       0.08   &   --        &       --      &       --     &    --      \\
J2050$+$3127 &       3.18      &  35  &   15    & 12 & 60 &     0.19    &       0.14   &    0.11    &       0.57    &       0.22   &    0.21    &       0.32    &       0.16   &    0.15    &       0.27    &       0.17   &   0.13     &       0.34    &       0.23   &    0.12   \\
J2129$-$1538 &       3.26      &  37  &   54    & 20 & 24 &     0.19    &       0.15   &    0.12    &       0.16    &       0.11   &    0.1     &       0.34    &       0.26   &    0.25    &       0.41    &       0.22   &   0.21     &       0.43    &       0.27   &    0.25   \\
J2134$-$0419 &       4.34      &  31  &   12    & 14 & 3 &     0.21    &       0.15   &    0.14    &       0.12    &       0.09   &    0.08    &       0.13    &       0.12   &    0.11    &       0.19    &       0.21   &   0.2      &       0.05    &       0.14   &    0.11   \\
J2217$+$0220 &       3.57      &  42  &   25    & 22 & 14 &     0.28    &       0.29   &    0.28    &       0.46    &       0.34   &    0.34    &       0.54    &       0.42   &    0.43    &       0.34    &       0.24   &   0.23     &       0.61    &       0.52   &    0.5    \\
J2219$-$2719 &       3.63      &  30  &   12    & 14 & 5 &     0.19    &       0.17   &    0.16    &       0.3     &       0.19   &    0.19    &       0.31    &       0.22   &    0.22    &       0.34    &       0.22   &   0.21     &       0.3     &       0.28   &    0.28   \\
J2225$+$2040 &       3.56      &  34  &   12    & -- & -- &     0.18    &       0.13   &    0.13    &       0.13    &       0.07   &    --       &       0.06    &       0.1    &    --       &       0.02    &       0.18   &   --        &       --      &       --     &    --      \\
J2248$-$0541 &       3.29      &  30  &   13    & 2 & 2 &     0.36    &       0.27   &    0.28    &       0.22    &       0.14   &    0.14    &       0.14    &       0.1    &    0.08    &       --      &       --     &   --        &       0.27    &       0.35   &    0.42   \\
J2251$+$2217 &       3.66      &  35  &   56    & -- & -- &     0.13    &       0.14   &    0.14    &       0.27    &       0.2    &    0.17    &       0.15    &       0.14   &    0.12    &       0.34    &       0.17   &   0.13     &       --      &       --     &    --      \\
J2314$+$0201 &       4.11      &  40  &   20    & -- & -- &     --      &       0.1    &    --       &       0.2     &       0.17   &    0.14    &       0.14    &       0.16   &    0.08    &       0.15    &       0.24   &   0.14     &       --      &       --     &    --      \\
J2316$-$3349 &       3.10      &  31  &   18    & -- & -- &     0.65    &       0.29   &    0.22    &       0.18    &       0.1    &    0.1     &       0.45    &       0.28   &    0.28    &       0.26    &       0.2    &   0.19     &       --      &       --     &    --      \\
J2321$-$0827 &       3.16      &  31  &   10    & 13 & 4 &     0.48    &       0.34   &    0.36    &       0.63    &       0.55   &    0.25    &       0.33    &       0.28   &    0.29    &       0.13    &       0.2    &   0.22     &       0.42    &       0.45   &    0.51   \\
J2344$+$3433 &       3.5       &  35  &   15    & -- & -- &     0.28    &       0.23   &    0.23    &       0.24    &       0.22   &    0.2     &       0.33    &       0.36   &    0.36    &       0.18    &       0.19   &   --        &       --      &       0.2    &    --      \\
\hline
\end{longtable}
\end{landscape}

\section[\appendixname~\thesection]{Cluster analysis of the quasars with PCA+k-means and Self Organizing~Maps}
\label{sec:clusters}

We have a number of parameters that can potentially influence the obtained variability properties. They concern both the actual properties of the source, like the emission frequency $\nu_{\rm rest}$ or redshift $z$, and the characteristics connected with the observing process: the number of observations N$_{\rm obs}$ and monitoring duration for the rest frame $t_{\rm rest}$.

Thus, the dependencies may be multiparametric. To search for possible tendencies in this case, a technique of cluster analysis can be implemented. The main idea is to localize groups of similar objects in a multidimensional parameter space (feature space) and investigate their statistical properties. The measure of similarity in our case is the feature space Euclidean distance.

\begin{figure}
\centerline{\includegraphics[width=\columnwidth]{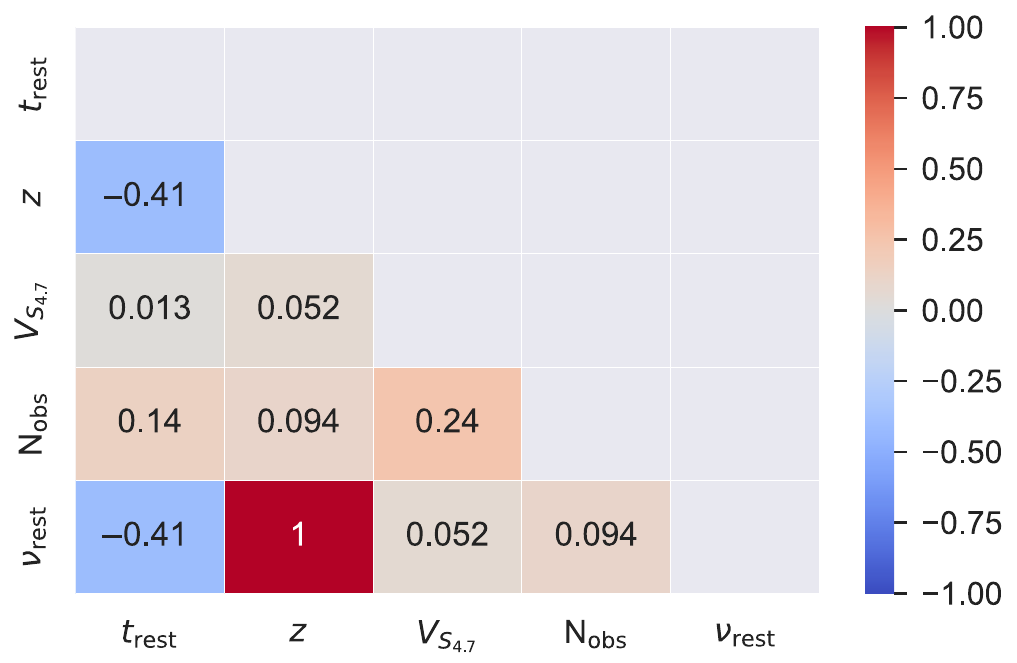}}
\caption{Correlation matrix with Kendall correlation coefficients}
\label{fig:corr_matrix}
\end{figure}

Our data are most complete for the observed frequency of 4.7~GHz, therefore we took the corresponding characteristics. The Kendall correlation matrix for the considered parameters is shown in Fig.~\ref{fig:corr_matrix}. We can see an analytical dependence between $\nu_{\rm rest}$ and $z$ (${\rm corr}=1$), some correlation between $\nu_{\rm rest}$ and $t_{\rm rest}$ (${\rm corr}=-0.41$), caused by their mutual dependence on $z$, and a weak dependence between N$_{\rm obs}$ and $V_{S_{4.7}}$ (\mbox{${\rm corr}=0.24$}).

As $\nu_{\rm rest}$ and redshift $z$ are analytically connected, they actually provide the same information about an object, so one of the characteristics can be omitted, we chose $\nu_{\rm rest}$ for our analysis.  As a result, we end up with a quite simple 4D feature space of $V_{S_{4.7}}$, $t_{\rm rest}$, $\nu_{\rm rest}$, and N$_{\rm obs}$. For modelling we took $\log{\rm N}_{\rm obs}$ due to very uneven distribution of this parameter.

The next step is dimensionality reduction using the standard primary component analysis (PCA). In this way we reduce ``the curse of dimensionality,\!'' get rid of correlation between feature space axes, and simplify visualization of the results. Hereafter, we used the scikit-learn \citep{JMLR:v12:pedregosa11a} library for our calculations. The explained variance along the primary components after the scaling and PCA transformation is as follows: $\sim\!52$\%, 29\%, 13\%, 6\%, thus we can drop the last component, saving 94\% of information and transforming our problem to the clustering in a 3D space of the first three primary components.

\begin{figure*}
\centerline{
\includegraphics[width=0.6\columnwidth]{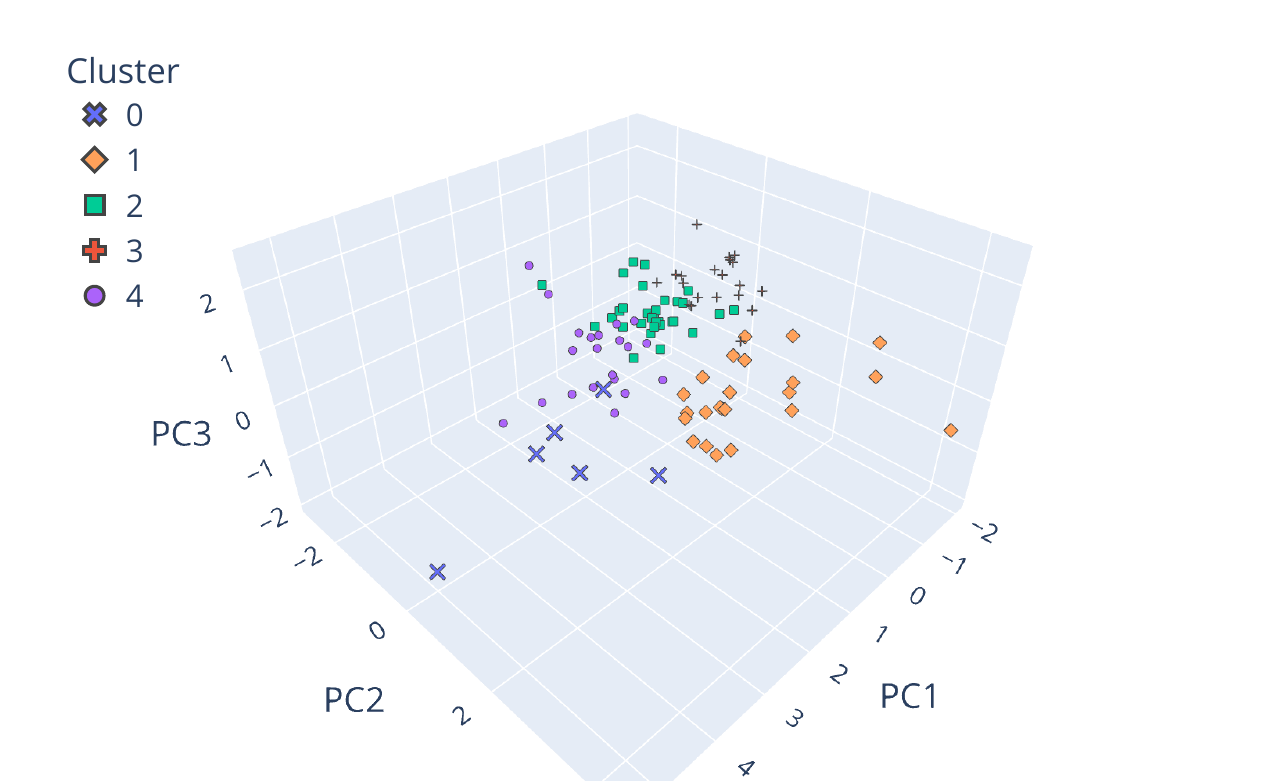}
\includegraphics[width=0.6\columnwidth]{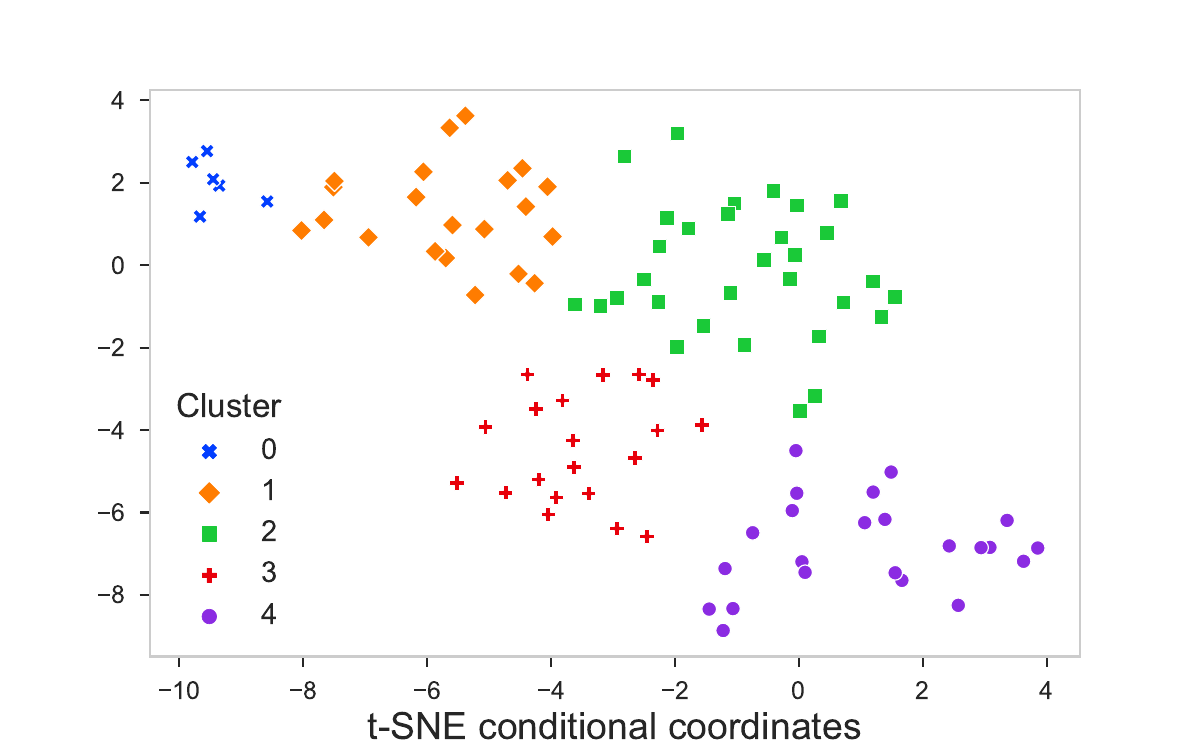}
}
\caption{Left: clusters in the primary component (PC) coordinates; right: 2D t-SNE representation of the left panel.}
\label{fig:clusters}
\end{figure*}

The clustering has been performed using the k-means algorithm \citep{1283494}. The optimal number of clusters, five, was estimated based on the ``elbow method'' implemented in the Yellowbrick library \citep{Bengfort2019}. The results are presented in Fig.~\ref{fig:clusters}. In the left panel one can see the clusters found by the k-means algorithm directly in the primary component (PC) space. For clarity, at right we show a 2D representation of the same figure. It has been obtained using t-distributed stochastic neighbor embedding (t-SNE) \citep{vanDerMaaten2008}. Although the t-SNE algorithm is strongly dependent on its ``perplexity'' hyperparameter, here we have a fairly good separation of the clusters at its default value ($=\!30$). Notice that we did not perform a separate clustering in the 2D coordinates, the cluster labels in the two panels are exactly the same, so the visualization demonstrates good accordance between the PC clustering and its t-SNE representation. The groups look more localized in the t-SNE coordinates.

\begin{figure*}
\centerline{
\includegraphics[width=\textwidth]{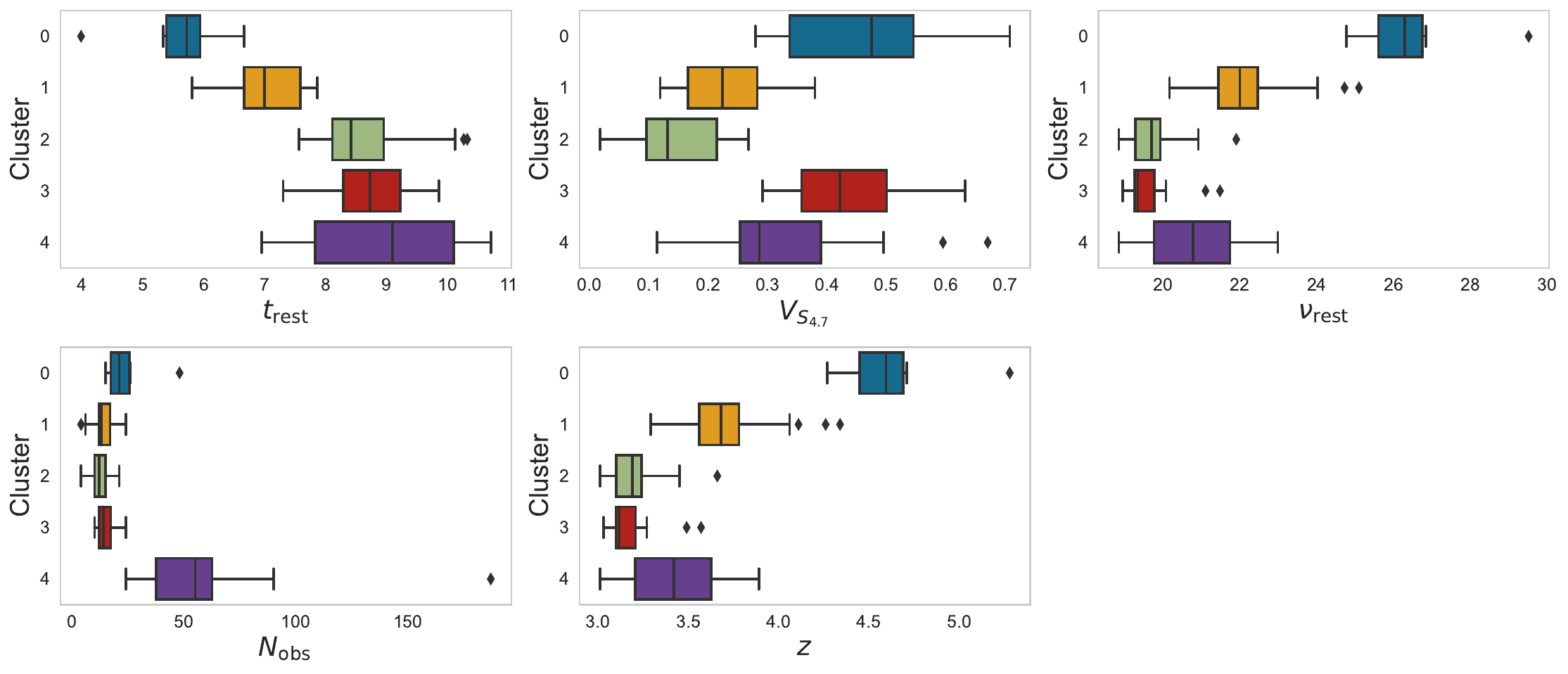}
}
\caption{Distributions of the considered characteristics for different clusters. Each panel corresponds to a certain characteristic, the $y$-axes show the cluster ordinal numbers. The distributions are shown as box plots, see description in the text. 
}
\label{fig:distribs}
\end{figure*}

Having the quasars marked with cluster labels, we can consider the difference in the distributions of our characteristics within different clusters. The distributions are shown in Fig.~\ref{fig:distribs} as box plots. The box is the interquartile range (IQR, 25th to 75th percentile, or Q1 to Q3), the median is shown as a vertical line inside the box, the ``whiskers'' extend to show the rest of the distribution, except for points that are determined as ``outliers,\!'' locating beyond the median~$\pm$~1.5~IQR range, these outliers are shown by dots. The numerical values are given in Table~\ref{tab:clusters}.

\begin{table}
\caption{Cluster characteristics}
\label{tab:clusters}
\newcolumntype{C}{>{\centering\arraybackslash}X}
\begin{tabularx}{\textwidth}{CCCCCCCCCCCCCC}
\toprule
\textbf{Cluster} & \textbf{N} && \multicolumn{5}{c}{\textbf{Medians}} && \multicolumn{5}{c}{\textbf{Min--max values}} \\
 &  && \textbf{$t_{\rm rest}$ (yr)} & \textbf{$V_{S_{4.7}}$} & \textbf{$\nu_{\rm rest}$ (GHz)} & \textbf{N$_{\rm obs}$} & \textbf{$z$} &&
\textbf{ $t_{\rm rest}$ (yr)} & \textbf{$V_{S_{4.7}}$} & \textbf{$\nu_{\rm rest}$ (GHz)} & \textbf{N$_{\rm obs}$} &\textbf{ $z$} \\
\midrule
 0 &  6 && 5.7 & 0.48 & 26.3 & 21 & 4.6 && 
     4.0--6.7 & 0.28--0.71 & 24.8--29.5 & 15--48 & 4.3--5.3 \\
 1 & 21 && 7.0 & 0.22 & 22.0 & 13 & 3.7 &&
     5.8--7.9 & 0.12--0.38 & 20.2--25.1 & 4--24 & 3.3--4.3 \\
 2 & 31 && 8.4 & 0.13 & 19.7 & 12 & 3.2 &&
     7.6--10.3 & 0.02--0.27 & 18.8--21.9 & 4--21 & 3.0--3.7 \\
 3 & 20 && 8.7 & 0.42 & 19.3 & 14 & 3.1 &&
     7.3--9.9 & 0.29-0.63 & 18.9--21.5 & 10--24 & 3.0--3.6 \\
 4 & 23 && 9.1 & 0.29 & 20.8 & 55 & 3.4 &&
     6.9--10.7 & 0.11--0.67 & 18.8--23.0 & 24--187 & 3.0--3.9 \\
\bottomrule
\end{tabularx}
\end{table}

Considering the $V_{S_{4.7}}$ panel in Fig.~\ref{fig:distribs}, we can  separate out clusters~0 and 3 as the quasar groups with highest variability. Remarkably, these are very different groups if we look at their other characteristics. 

The six quasars of cluster~0 are the most distant objects ($z\geq4.3$), respectively with the highest emission frequencies $\nu_{\rm rest}\geq24.8$~GHz and the shortest time of monitoring in the rest frame $t_{\rm rest}<7$~years, which, together with an ordinary number of observations N$_{\rm obs}=15$--$48$, does not prevent them to have prominent variability $V_{S_{4.7}}=0.28$--$0.71$.

At the same time, quasars of cluster~3 with similar variability indices $V_{S_{4.7}}=0.29$--$0.63$ are at the lowest redshifts in our sample, mostly within $z=3.0$--$3.3$, and, correspondingly, with low $\nu_{\rm rest}$, mostly less than 21~GHz. They have standard N$_{\rm obs}=10$--$24$, but a sufficiently long $t_{\rm rest}$ of 7--10~years.

Quasars of cluster~2 are very similar to those of cluster~3, except they have the ``opposite'' lowest variability indices within the sample, $V_{S_{4.7}}=0.02$--$0.27$. 

Quasars of the ``moderate'' cluster~1 at intermediate redshifts $z=3.3$--$4.3$ show characteristics in between the values for clusters~0--3.

Ultimately, the clustering algorithm separated out a special group of the most observed quasars with \mbox{N$_{\rm obs}=24$--$187$}. These quasars are at close or intermediate redshifts \mbox{$z\leq3.9$} with long $t_{\rm rest}$ greater 7~years and low-to-intermediate \mbox{$\nu_{\rm rest}\leq23$}~GHz. Noteworthy that these most monitored quasars do not generally show the highest variability indices, which are nevertheless statistically higher than those for low-variable cluster~2 and intermediate-variable cluster~1.

These results were double-checked with another method which used the Self Organizing Maps (SOM) to reduce the dimensionality of the input data set with a subsequent cluster analysis with k-means.

A Self Organizing Map \citep{2001som..book.....K} is a type of the artificial neural network that is trained using unsupervised learning. The aim of a SOM is to transform the multi-dimensional input data set to a low-dimensional output space (usually two-dimensional map), performing the dimensionality reduction. This transformation is made adaptively in a topologically ordered fashion.

The SOM neurons are placed in the nodes of a 2D lattice and, after their weights are randomly initialised, tuned to the input data vectors during epochs of competitive learning. The number of the learning epochs depends on the input data set, the size of the output map, the manner of updating the neighboring neuron weights and usually varies between 10 and 100.

After such tuning, the weights of the output neurons on the lattice become ordered, and it is possible to create a coordinate system on this lattice for the input data that we analyze. Thus, the SOM makes a topological map of the input vectors. We can consider these as non-linear generalisation of the PCA method mentioned above.

\begin{figure*}
\centerline{\includegraphics[width=\textwidth]{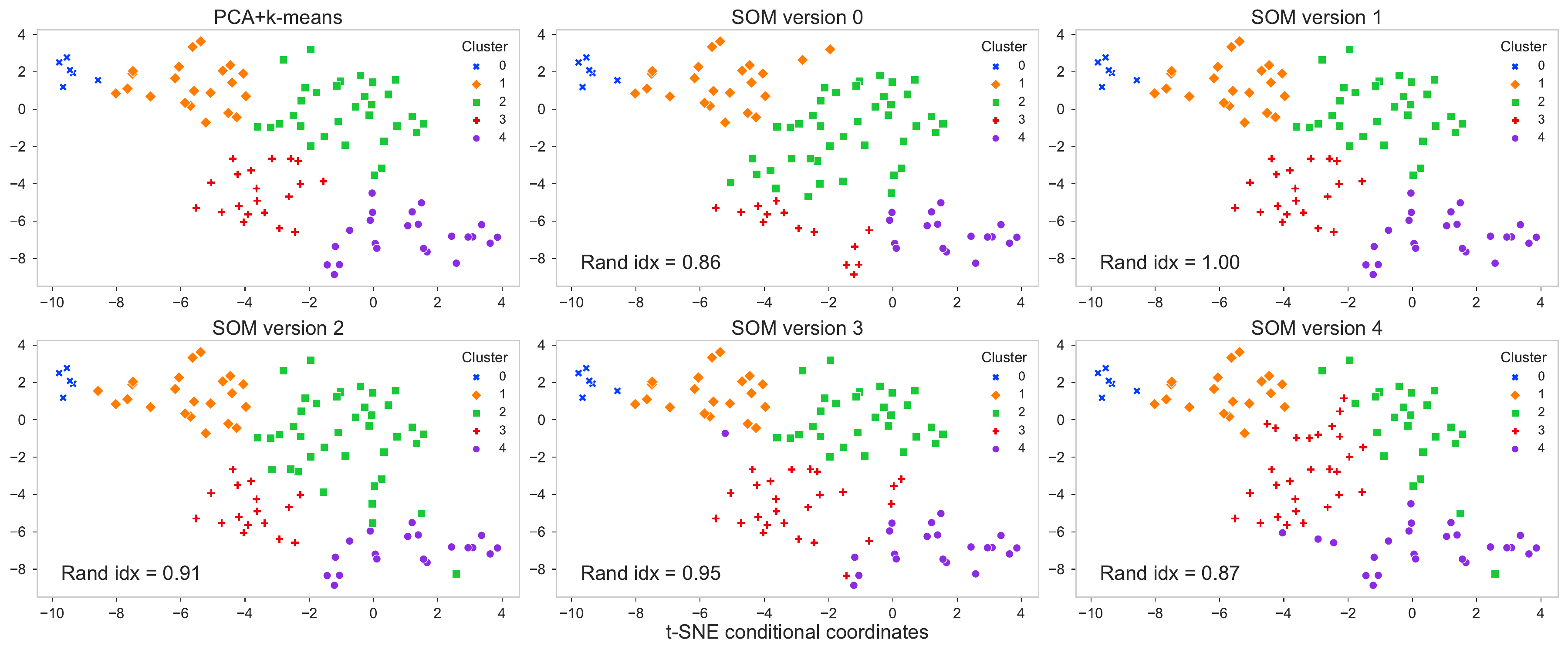}}
\caption{Comparison of the PCA+k-means clustering (upper left panel) with 5 runs of the SOM training (other panels).}
\label{fig:comparison}
\end{figure*}

There are several SOM libraries available, and in the current analysis we used the Python module \textit{somoclu}\footnote{ \url{https://github.com/peterwittek/somoclu}} \citep{JSSv078i09}. 
We chose a SOM map size of $200\times320$ neurons, and the number of learning epochs was 50. The same set of parameters ($V_{S_{4.7}}$, $t_{\rm rest}$, $\nu_{\rm rest}$, and 
$\log{\rm N}_{\rm obs}$) was included into the input vectors, forming a 4D-parameter space.

After the SOM algorithm had finished the learning stage, the k-means method was used to perform the clustering, assuming 5 clusters. We can assign a particular cluster number to the input data vector based on the nearest neuron in the lattice, looking for the ``winner'' neuron, minimizing the squared Euclidean distance between the input vector and the weight vector of the lattice neuron. The input vector will belong to the same cluster as the ``winner'' neuron.

A comparison of the SOM results with those of the PCA+k-means method, given in Fig.~\ref{fig:comparison} using the t-SNE conditional coordinates,  shows that different runs of SOM training can give somewhat different borders of the clusters, but their general structure sustains. The Rand index, which is commonly used to compare two clusterings, varies for our five SOM training runs from 0.86 to 1, showing in one case a perfect match between the two methods. 

Thus, we can state that the SOM results generally correspond to  PCA+k-means (with an accuracy of about 85\% based on the Rand indices range). For our analysis we chose the latter, as a more straightforward, unambiguous, and interpretable approach, also well corresponding to t-SNE.

\section[\appendixname~\thesection]{Comments on individual sources}
\label{sec:individ}

\subsection[\appendixname~\thesubsection]{Identified HFP/GPS/MPS candidates which revealed significant radio variability}
\label{sec:identifiedGPS}

{\bf J0121$-$2806} is an FSRQ classified as PS in \cite{2013AstBu..68..262M}. We calculated that the spectral peak is at 0.3 GHz \citep{2021MNRAS.508.2798S}. Due to the absence of radio data at frequencies from 0.2 to 1.4 GHz, the spectral maximum can not be determined reliably. The quasar demonstrates high flux density variability with $V_{S_{4.7}}=0.43$ and $V_{S_{8.2}}=0.32$. The spectral index $\alpha_{\rm thin}$ was changing from -0.70 to -0.08, remaining almost flat in 2014--2020.

{\bf J0203$+$1134} is an FSRQ known as a GPS source with a peak frequency of about 17 GHz in the rest frame \citep{1991ApJ...380...66O}. The RATAN-600 observations confirm the peaked spectral shape in the time period of 2006--2020. However, occasionally it shows an inverted spectrum (e.g., in 2001) and has been classified as having chaotic variability \citep{2012ARep...56..345G}. The variability indices $V_{S_{4.7}}$ and $V_{S_{22.3}}$, which have been measured at a historical scale of 20--40 years, are 0.33 and 0.45 respectively. The broadband spectra are well described by two components, linear at the observer frequencies of 0.07--0.5 GHz and convex at higher frequencies. 

{\bf J0324$-$2918} was considered as a PS source in \citet{2017MNRAS.467.2039C}, where authors estimated a possible spectral turnover between 1.4 GHz and 7 GHz. The uncertainty of positioning the peak was caused by the scatter of the observed flux densities around 5 GHz, where the AT20G and PMN surveys contributed with different resolutions and at different epochs of observations. In our study we clarified the 2.3--22.3 GHz part of the spectrum as being inverted using quasi-simultaneous RATAN-600 measurements. Thus we see that the radio spectrum of J0324$-$2918 has a complex structure: below 1 GHz it has a peaked shape, but above it is inverted. We obtained significant variability (up to 45\%) for the source during our monitoring, which is in agreement with the conclusion in \citet{2016MNRAS.463.3260C} that its VLBI emission is Doppler-boosted.

{\bf J0646$+$4451} is one of the brightest high-redshift PS blazars. The object is known as a high-frequency peaker \citep{2000A&A...363..887D,2005A&A...432...31T} and has for the first time been considered as a GPS source in \cite{1990MNRAS.245P..20O}. The quasar has a spectral peak around 17--18 GHz in the observer's frame of reference \citep{2005A&A...435..839T,2021MNRAS.508.2798S}. Flux variations are observed on both sides of the peak (Fig.~\ref{fig:B1}), and the variability index reaches 0.70 at 5--8 GHz. We observed variable peaked spectra for the blazar during 2006--2011. In 2017--2021 the quasar had flat and even rising spectra, i.e., the spectrum became optically thick at the high frequencies more than 10 GHz, with $\alpha_{\rm thick} =+0.22$ and $+0.35$ in April and August 2020. Its RATAN light curves show a slow flux density decrease about 1.5--2 times, from 4 to 1.5--2 Jy, at 4.7--22.3 GHz in 2012--2020. The steepest radio spectrum $\alpha_{\rm thin}=-1.58$ was observed in 2011, when the flux density started decreasing at 2.3--22.3 GHz, and the spectrum became optically thin at frequencies above 5 GHz after several years of a relatively high state.

{\bf J1230$-$1139} is an FSRQ and has been classified as a GPS candidate in \citet{2013AstBu..68..262M}. Low frequency data are well represented by the GLEAM and TGSS surveys. The VLBA observations show a large jet bending ($\sim90\degree$) within 2 mas from one bright radio component at 8.4 GHz \citep{2022ApJ...937...19Z}. We have found the variability indices at 4.7 and 22.3 GHz equal to 0.47 and 0.41 on timescales of $>$43 and 14~years respectively. The RATAN-600 observational data covers the period of 2015--2020. The overall number of observations N$_{\rm obs}=51$ at 4.7 GHz.

{\bf J1340$+$3754} is an FSRQ \citep{2007ApJS..171...61H} at $z=3.11$ with a core--jet morphology \citep{1999A&A...344...51P}. It has been classified as a GPS candidate in \citet{2013AstBu..68..262M}. Extended monitoring revealed variability up to 30\% at our frequencies.

{\bf J1354$-$0206} is a known GPS source and a blazar \citep{1990MNRAS.245P..20O,1991ApJ...380...66O} with a core--jet morphology \citep{2022ApJS..260....4P}. It demonstrated the maximum variability at 22.3~GHz $V_{S_{22.3}}=0.50$ on a time scale of 23 yrs.

{\bf J1357$-$1744} is a known GPS \citep{1991ApJ...380...66O} with a core--jet morphology \citep{2022ApJS..260....4P}. It shows quite high flux density variation both at high and low frequencies ($V_{S_{2.3}}=0.37$ and $V_{S_{22.3}}=0.48$).

{\bf J1445$+$0958} at $z=3.55$ has been classified as a GPS source by \citet{1991ApJ...380...66O} and is also an FSRQ with the core--jet type of radio morphology \citep{2014arXiv1406.4797G,2022ApJ...937...19Z}. The compiled spectrum has a well-constrained absorbed part with $\alpha_{\rm thick}=+0.68$ ($0.01$). A high variability index $V_{S_{22.3}}=0.67$ was calculated using 27 observing epochs during a time period of about 20 years. At lower frequencies the variability has a smaller amplitude, $V_{S}=0.15$ at 2.3 and 4.7 GHz.

The quasar {\bf J1538$+$0019} was classified as a GPS source in \cite{2013AstBu..68..262M}. It has flux density variation at 2.3--22.3 GHz in the range $V_{S}=0.27$--0.58 on a timescale of up to 40 years  \citep{1997MNRAS.284...85D,1986ApJS...61....1B}. Its observed peak frequency shifted from 3 to 1 GHz during 2015--2020.

{\bf J1559$+$0304} at $z=3.89$ with a compact core and a two-sided jet at 5 and 8.4 GHz \citep{2014arXiv1406.4797G} was classified as a PS source with the peak between 72 MHz and 1.4 GHz by \citet{2017ApJ...836..174C}. We have obtained a greater value of the spectral peak $\nu_{\rm peak,obs}=4.5$~GHz and the variability index $V_{S_{22.3}}=0.36$ on a timescale of eight years. 
The highest variability is observed near the spectral peak, historical 37-year data show $V_{S_{4.7}}=0.41$ (N~$=63$). The spectral shape seems to remain convex during the RATAN observations in 2015--2020. But since the beginning of 2019, the flux density has started to increase slowly at 22.3--4.7~GHz.    

{\bf J1606$+$3124} is a radio galaxy at $z=4.56$ with a peak in the radio spectrum at $\nu_{\rm peak, obs}=2.5$ GHz. 
The radio properties of J1606$+$3124 studied in \cite{2022MNRAS.511.4572A,2022ApJ...937...19Z} suggest that it is a young radio source, the highest-redshift GPS galaxy known to date. The peak frequency in the source's rest frame corresponds to 13.9 GHz. A two-sided jet structure is revealed by VLBI at 8.4 GHz. 

{\bf J1616$+$0459}, an FSRQ at $z=3.21$, was classified as a GPS and HFP source in \cite{1990MNRAS.245P..20O,1991ApJ...380...66O,2000A&A...363..887D}. Analysing its compiled radio spectrum, we estimated $\nu_{\rm peak, obs}=4.4$~GHz.
The variability index $V_{S_{4.7}}=0.26$ on a timescale of more than 40~years (N~$=64$), and $V_{S_{22.3}}=0.26$ on the 23-year timescale (N~$=22$). The highest variability was found at 8.2 GHz, $V_{S_{8.2}}=0.42$.

{\bf J1658$-$0739} is an FSRQ \citep{2007ApJS..171...61H} at $z=3.74$ with a core--jet morphology \citep{2022ApJS..260....4P}. It has been introduced as a GPS candidate with $\nu_{\rm obs,peak} = 4.8$~GHz in \citep{2004A&A...424...91E} based on the VLA and ATCA data, but our study revealed variability of 13--33\%, which casts doubt that J1658$-$0739 is a canonical GPS source. Also its radio spectrum suggests the presence of another extended emission at lower frequencies, turning up below 200 MHz. 

The FSRQ blazar {\bf J1840$+$3900} at $z=3.1$ was classified as an HFP source in several papers \citep{1999A&AS..135..273M,2000A&A...363..887D,2008A&A...482..483T}, but the compiled radio spectra 
demonstrate a high level of variability ($V_{S}\sim0.52$) at 4.7 and 22.3 GHz on a timescale of 36 years. The RATAN radio spectra had rising shapes in July 2006 and July 2014 at 4.7--22.3 GHz ($\alpha_{\rm thin}=+0.22$ and $+0.34$). We conclude that J1840$+$3900 was misclassified as a GPS/HFP source.

The core--jet FSRQ blazar {\bf J2003$-$3251} at $z=3.77$ was classified as a GPS and HFP candidate \citep{1991ApJ...380...66O,1997A&A...321..105D}. We determined its spectral peak $\nu_{\rm peak, obs}=5.6$ GHz. The quasar had been demonstrating moderate variability $V_{S_{22.3}}=0.25$ up to 2019. Since 2019 its flux density has started to increase from 0.42 to 0.85 mJy at 22.3 GHz and from 0.50 to 0.61 at 11.2 GHz, 
and radio variability has reached 42\% at 22.3 GHz. Further long-term radio observations are required to determine its variability properties.
   
The quasar {\bf J2129$-$1538} at $z=3.26$ is a bright FSRQ blazar, one of the first known high-redshift GPS sources \citep{1990MNRAS.245P..20O,1991ApJ...380...66O}. The quasar was studied as an HFP candidate in \cite{2016AN....337..130J}, where its variability was estimated over 1.5 years of observations (\mbox{4--6}~epochs) with modulation indices of 9 and 22.5\% at 22 and 43 GHz respectively; its radio spectrum was reported as steep ($\alpha_{22-43\rm GHz}=-0.52$). \cite{2023Galax..11...42C} reported typical well-aligned core--jet structure for J2129$-$1538 at 8, 15 and 43 GHz radio images. The RATAN-600 observations were made in 2012--2020 and showed slow flux density rising at 4.7--22.3 GHz since 2017. The spectral index $\alpha_{\rm thin}$ was changing from $-0.04$ up to $-0.83$ (by 88\%). We estimated its variability index to be 43\% at 22.3 GHz using data of 20-years-long observations. 

Thus, as we have shown above, longer time periods of monitoring of blazars are likely to reveal higher flux density variations.

\subsection[\appendixname~\thesubsection]{New GPS/MPS candidates}
\label{sec:newGPS}

The quasar {\bf J0006$+$1415} has a structure with an extended component of 10 mas from the core \citep{1999A&A...344...51P}. Its radio spectrum at hundreds of MHz is presented by the TGSS and mostly by the GLEAM measurements, which demonstrate quite a scatter of flux densities, e.g., at \mbox{130--170}~MHz the variations of flux density reaches 100\% (\mbox{110--220}~mJy). Considering that, in some states its radio spectrum could be flat or steep. We obtained the low level of variability indices, 0.03--0.25 at 2.3--11.2 GHz. However, the spectral indices $\alpha_{\rm thick}=+0.02$ and $\alpha_{\rm thin}=-0.21$ make the spectrum close to flat, with FWHM~$=7.7$, and barely corresponding to the canonical GPS spectrum.

The quasar {\bf J0100$-$2708} with a core--jet angular structure \citep{2022ApJS..260....4P} has not much information about its radio continuum properties in the literature. It has a clear peaked shape (FWHM~$=1.7$) in the radio spectra, where the optically thin part, with $\alpha_{\rm thin}=-0.47\pm0.01$, is represented by seven epochs of RATAN-600 observations. The optically thick part mostly consists of the GLEAM data ($\alpha_{\rm thick}=+0.41\pm0.01$). Both modulation and variability indices do not exceed a value of 0.20 at 2.3--11.2 GHz.

For {\bf J0214$+$0157} the GLEAM and TGSS data for the optically thick part together with the RATAN-600 data for the optically thin part show a peaked spectrum (FWHM~$=1.9$) with  $\nu_{\rm peak,obs}=0.5$ GHz. We note a significant scatter in the measured flux densities (0.01--0.26 Jy) below 100 MHz, and $V_{S}= 0.23$ at 2.3 GHz. At the frequencies 4.7--11.2 the variability index equals 0.03--0.23. Due to moderate radio variability and 
the spectral shape, we suggest J0214$+$0157 as a new MPS candidate.

The core--jet \citep{2022ApJS..260...49K} PS blazar of an uncertain type {\bf J0525$-$3343} at $z=4.41$ has a spectral peak $\nu_{\rm peak,obs}=0.9$ GHz and could be considered as an MPS candidate with the spectral indices $\alpha_{\rm thick}=+0.43\pm0.02$ and $\alpha_{\rm thin}=-0.96\pm0.02$. But its radio variability reaches \mbox{29--71}~per~cent at 2.3, 4.7, and 8.2 GHz on a time scale of 30~years. The highest variability is found at 4.7 GHz, the maximum flux density S$_{\rm max}=0.21$ Jy was measured in 1990 in the Parkes-MIT-NRAO (PMN) survey at 4.85 GHz \citep{1996ApJS..103..145W}. In the NVSS survey at 1.4 GHz, carried out several years later \citep{1998AJ....115.1693C}, the flux density is 0.188~mJy, which gives a rising non-simultaneous spectrum.

The megahertz-peaked spectrum core--jet blazar {\bf J0624$+$3856} \citep{2022ApJS..260....4P} has a spectral peak $\nu_{\rm peak,obs}=0.3$ GHz and FWHM~$= 2.5$ dex. The highest variability $V_{S}= 0.43$ is found at 8.2 GHz on a timescale of 30 years. 

We consider {\bf J0839$+$2002} as a new GPS candidate despite some deviation from the criteria of genuine GPS sources. The low-frequency spectral part is poorly represented by a single TGSS measurement at 150 MHz \citep{2017A&A...598A..78I}. The averaged spectral indices of the optically thick and thin parts are $+0.36\pm0.01$ and $-0.25\pm0.01$. The variability indices are 0.05, 0.21, 0.15, and 0.29 at 22.3, 11.2, 8.2, and 4.7 GHz respectively.

The core--jet quasar {\bf J0905$+$0410} was detected in X-rays \citep{1995A&AS..109..147B}, its radio structure has an extension of 10 mas \citep{1999A&A...344...51P}. Its radio spectrum has a clear peak around $\nu_{\rm obs} \sim 0.5$ GHz with the spectral indices \mbox{$\alpha_{\rm thick}=+0.57\pm0.06$} and $\alpha_{\rm thin}=-0.53\pm0.01$. The variation of flux densities is $V_{S}=0.04$--$0.17$ at 2.3--11.2 GHz (Table~\ref{TableA2}). We suggest J0905$+$0410 as the second new MPS candidate in the sample.

The next new GPS candidate {\bf J0933$+$2845} with narrow absorption lines \citep{2003ApJ...594..684M} is detected in X-rays, but its structure is unresolved \citep{2021ApJ...914..130S}; in the radio band it has an extended structure \citep{2014arXiv1406.4797G}. Its radio spectrum has a clear peak at $\nu_{\rm peak,obs}\sim 1.7$ GHz. There is a lack of measurements for the optically thick part (\mbox{$\alpha_{\rm thick}=+0.17\pm0.02$}), we rely on only one measurement point at 150 MHz from the TGSS. At frequencies higher than 2.3 GHz we calculated flux densities variability in the range $V_{S}=0.21$--$0.34$.

The source {\bf J1045$+$3142} is classified as having an unresolved core morphology \citep{2014arXiv1406.4797G} and features a clear peak in its radio spectrum at 0.9 GHz with moderate variability at 2.3 and 8.2 GHz. We suggest the quasar as a new MPS candidate, although the optically thick spectral part ($\alpha_{\rm thick}=+0.25\pm0.03$) is represented by only three radio measurements  \citep{1997A&AS..124..259R,1999MNRAS.306...31R,2017A&A...598A..78I}.

For the core--jet quasar {\bf J1301$+$1904} we calculated moderate and negligible variabilities at frequencies of 2.3, 4.7, 8.2, and 11.2 GHz (Table~\ref{TableA2}) and a peaked spectrum with $\nu_{\rm peak,obs}\sim 1.7$~GHz. The low-frequency spectral part is represented by only single measurement at 150 MHz (TGSS) with $\alpha_{\rm thin}=+0.13\pm0.01$. 

A new GPS candidate {\bf J1401$+$1513} has the optically thick spectral part close to flat, $\alpha_{\rm thick}=+0.05\pm0.01$, and the variability indices are 0.09 and 0.36 at 11.2 and 4.7 GHz ($t_{4.7}=33$ years). The low-frequency spectral part, poorly represented by a single TGSS measurement, should be determined more precisely.

The PS blazar \citep{2007ApJS..171...61H} {\bf J1405$+$0415} has a compact core and a resolved jet extending to 18 mas \citep{2002A&A...381..757L,2006evn..confE..86Y,2008ApJS..175..314D}. Its radio spectrum is composed of several components: steep below 1~GHz, curved with a peak between 1 and 10 GHz, steep from 10 to 110 GHz, and inverted above 110 GHz. The quasar has moderate variability indices ranging from 0.19 to 0.26 at the radio frequencies.

A new MPS candidate {\bf J1418$+$4250} has $\nu_{\rm peak,obs}\sim 0.9$~GHz, 
two measurements below the peak (325 and 408~MHz), and measurements at five frequencies above it. More long-term observations would be complementary, both for peak frequency clarification and variability estimation. To estimate the variability for the time being, we used a frequency of 4.7 GHz and obtained $V_{S_{4.7}}=0.02$ on a timescale of 34 years, based however only on three measurements of Green Bank in April and October 1987 (4.85 GHz) and RATAN-600 in July 2020 (4.7 GHz). The flux density varies insignificantly from 0.047 and 0.048 Jy to 0.054 Jy, which is within the measurement uncertainties (0.006--0.009 Jy).

The new GPS candidate {\bf J1521$+$1756} has a spectral peak at 3.5 GHz, the variability indices are 0.10, 0.18, and 0.32 at 11.2, 8.2, and 4.7 GHz on a timescale $t_{4.7}=40$ years (Table~\ref{TableA2}). The VLA images at 1.4 GHz show an unresolved core morphology \citep{2014arXiv1406.4797G}.

The quasar {\bf J2019$+$1127} should be treated carefully because it is associated with the gravitationally lensed system MG B2016$+$112 consisting of three components \citep{1984Sci...223...46L}: a doubly imaged core and a counter--jet quadruply imaged at higher resolution \citep{2009MNRAS.394..174M}. The lensed radio images of the object can not be resolved by \mbox{RATAN-600}. Its radio spectrum has a complex structure below the peak frequency, caused either by the several emission components or by the variability of the source.

{\bf J2314$+$0201} is a new GPS source of X-ray emission \citep{2013ApJ...763..109W,2019MNRAS.482.2016Z}. We estimated the variability indices $V_{S_{4.7-11.2}}=0.14$--$0.20$. Its radio spectrum has a peak, however there is a lack of measurements in the optically thick spectral part, and we rely only on one measurement point at 150 MHz from the TGSS.

\subsection[\appendixname~\thesubsection]{Gravitationally lensed quasars}
\label{sec:gravlensed}

The lensing galaxy at $z=1.145$ \citep{2010ApJ...716L.185L} creates four images of the quasar {\bf J0131$+$4358} with a maximum image separation of $0.54^{\second}$ \citep{2000MNRAS.319L...7P}. The quasar {\bf J0751$+$2716} is lensed by a galaxy at $z=0.35$ \citep{1999AJ....117.2034T} into four images separated by $0.8^{\second}$ \citep{1997AJ....114...48L}. {\bf J1401$+$1513} is a unique sextuple system with a maximum angular separation between the quasar images of about $1.7^{\second}$, the lensing group of three galaxies are located at $z\sim1$ \citep{2001ApJ...557..594R}. The quadruple images of {\bf J1424$+$2256}, created by a lensing galaxy at $z=0.34$ \citep{1998AJ....115....1T}, have a maximum separation of $1.3^{\second}$ \citep{1992MNRAS.259P...1P,1998MNRAS.295..587M}. The lensing galaxy located at $z=1.165$ \citep{2006ApJ...641...70O} creates a triple image of {\bf J1632$-$0033} separated by about $1.5^{\second}$, according to the VLA observations at 43 GHz \citep{2002AJ....123...10W,2004Natur.427..613W}. {\bf J2019$+$1127} is a narrow-line Seyfert 1 galaxy, reviewed in Section~\ref{sec:GPS/MPS/HFP candidates} as a new MPS candidate. The separation between three of its components is $3.4^{\second}$ \citep{1984Sci...223...46L}. Based on our results, the radio variability properties of the lensed quasars show no peculiarities compared to other quasars.

\section[\appendixname~\thesection]{Light curves and radio spectra}

 \begin{figure}
 \centerline{\includegraphics[width=190mm]{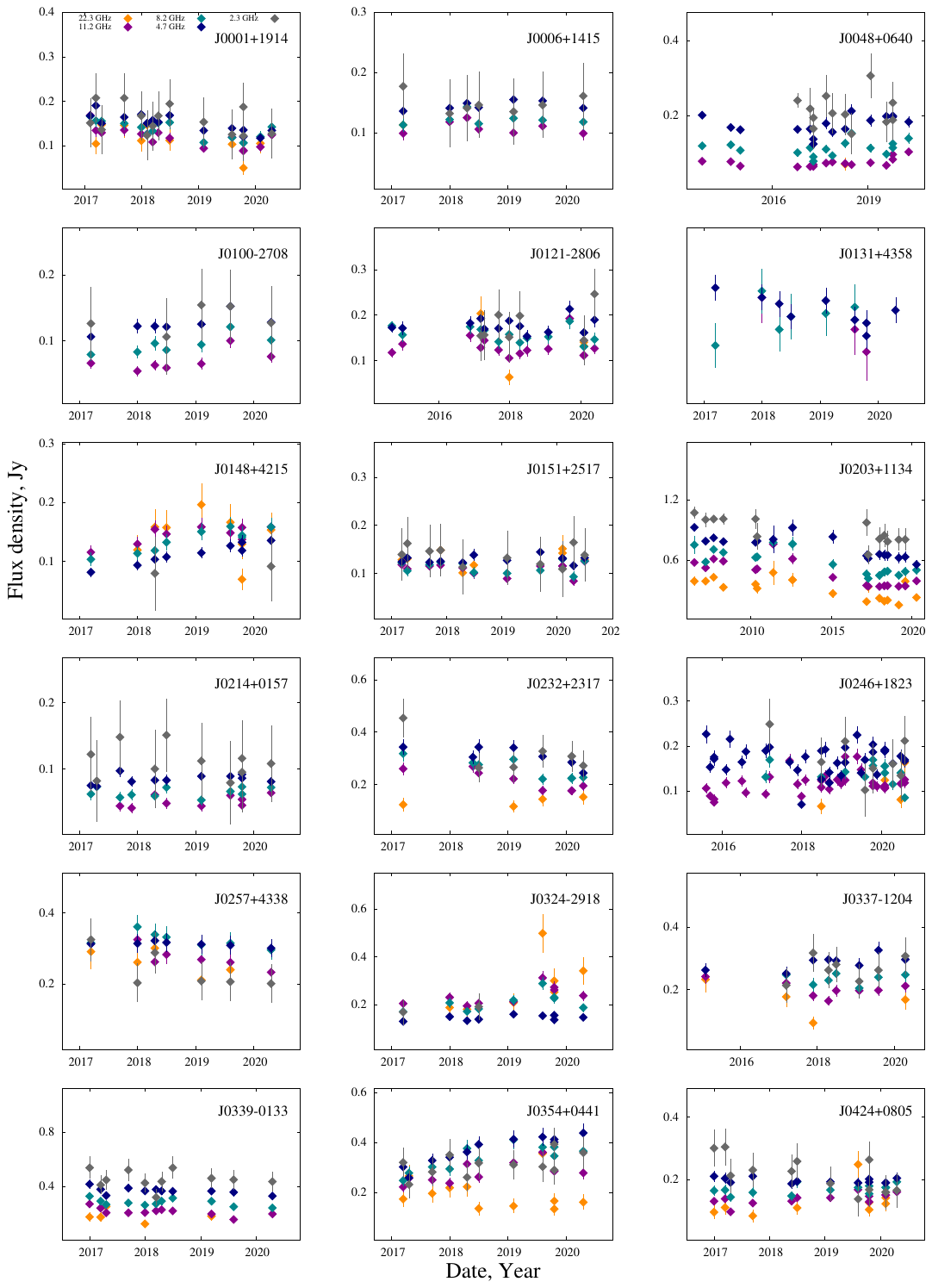}}
 \caption{The RATAN-600 light curves, the measurements at 22.3 GHz are colored by orange, at 11.2 GHz by magenta, at 8.2 GHz by green, at 4.7 GHz by blue.}
 \label{fig:C1}
 \end{figure}

 \clearpage
 \newpage

\begin{figure}
\centerline{\includegraphics[width=190mm]{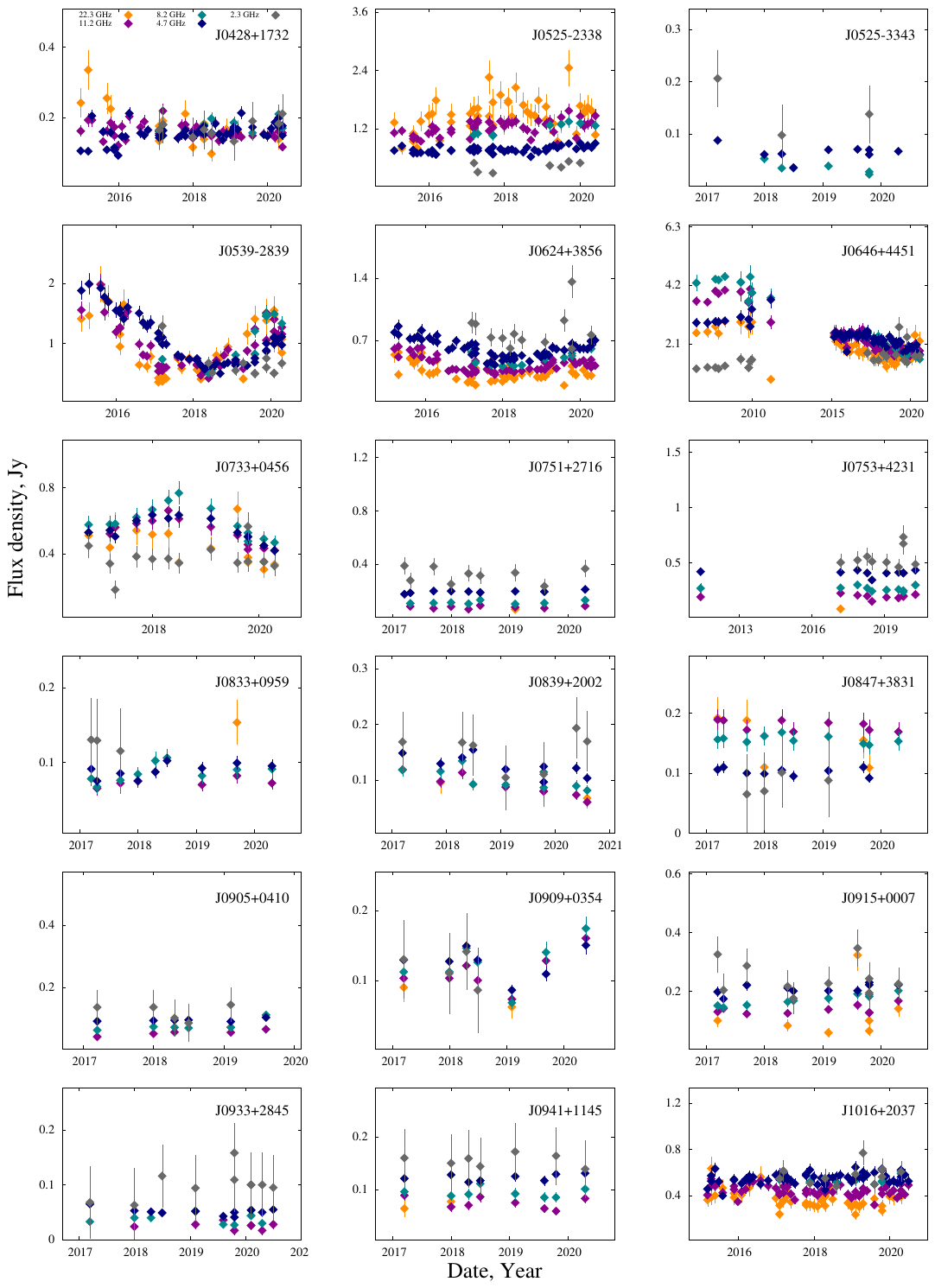}}
\caption{The RATAN-600 light curves, the measurements at 22.3 GHz are colored by orange, at 11.2 GHz by magenta, at 8.2 GHz by green, at 4.7 GHz by blue.}
\label{fig:C2}
\end{figure}

\clearpage
\newpage

\begin{figure}
\centerline{\includegraphics[width=190mm]{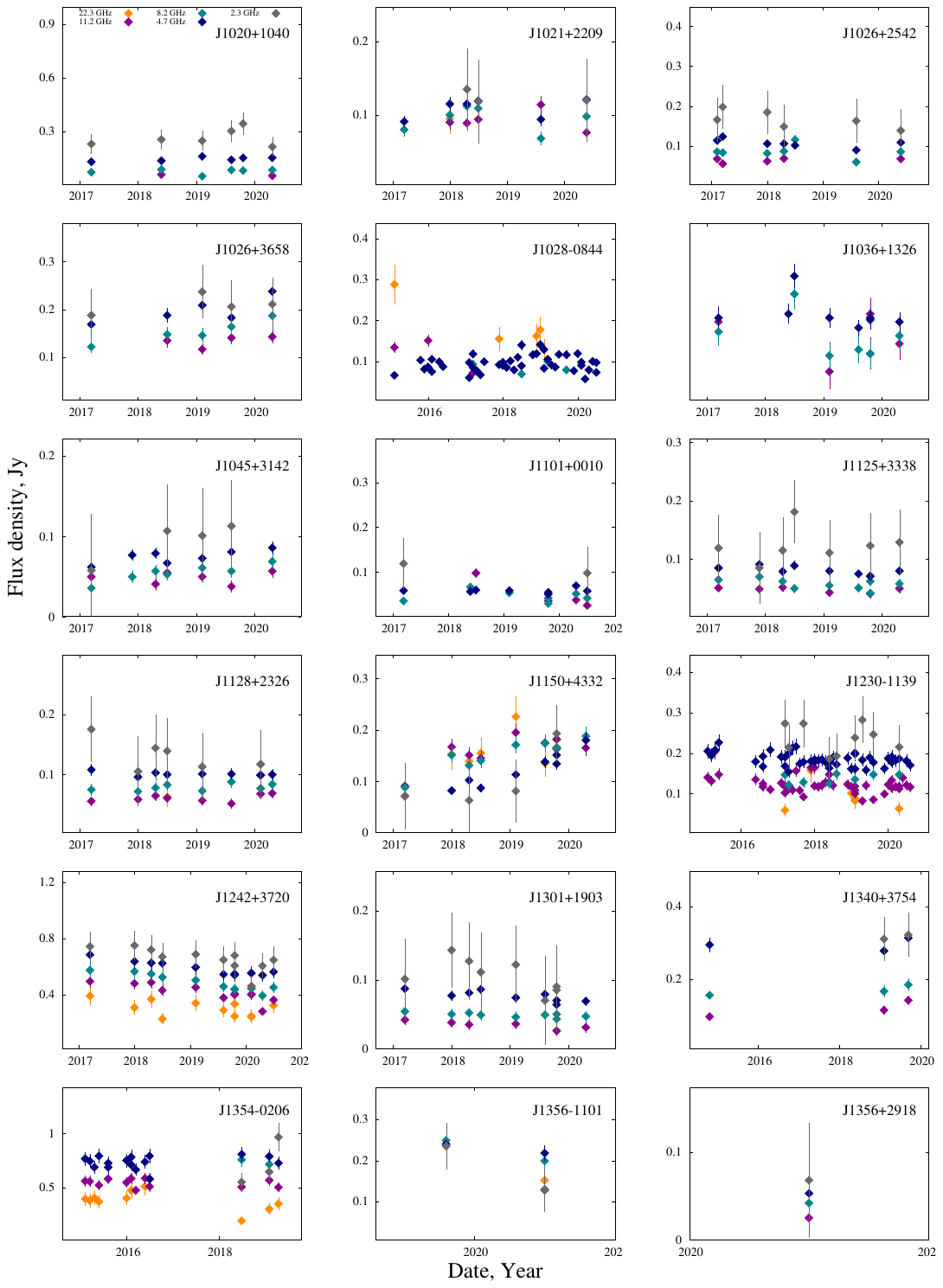}}
\caption{The RATAN-600 light curves, the measurements at 22.3 GHz are colored by orange, at 11.2 GHz by magenta, at 8.2 GHz by green, at 4.7 GHz by blue.}
\label{fig:C3}
\end{figure}

\clearpage
\newpage

\begin{figure}
\centerline{\includegraphics[width=190mm]{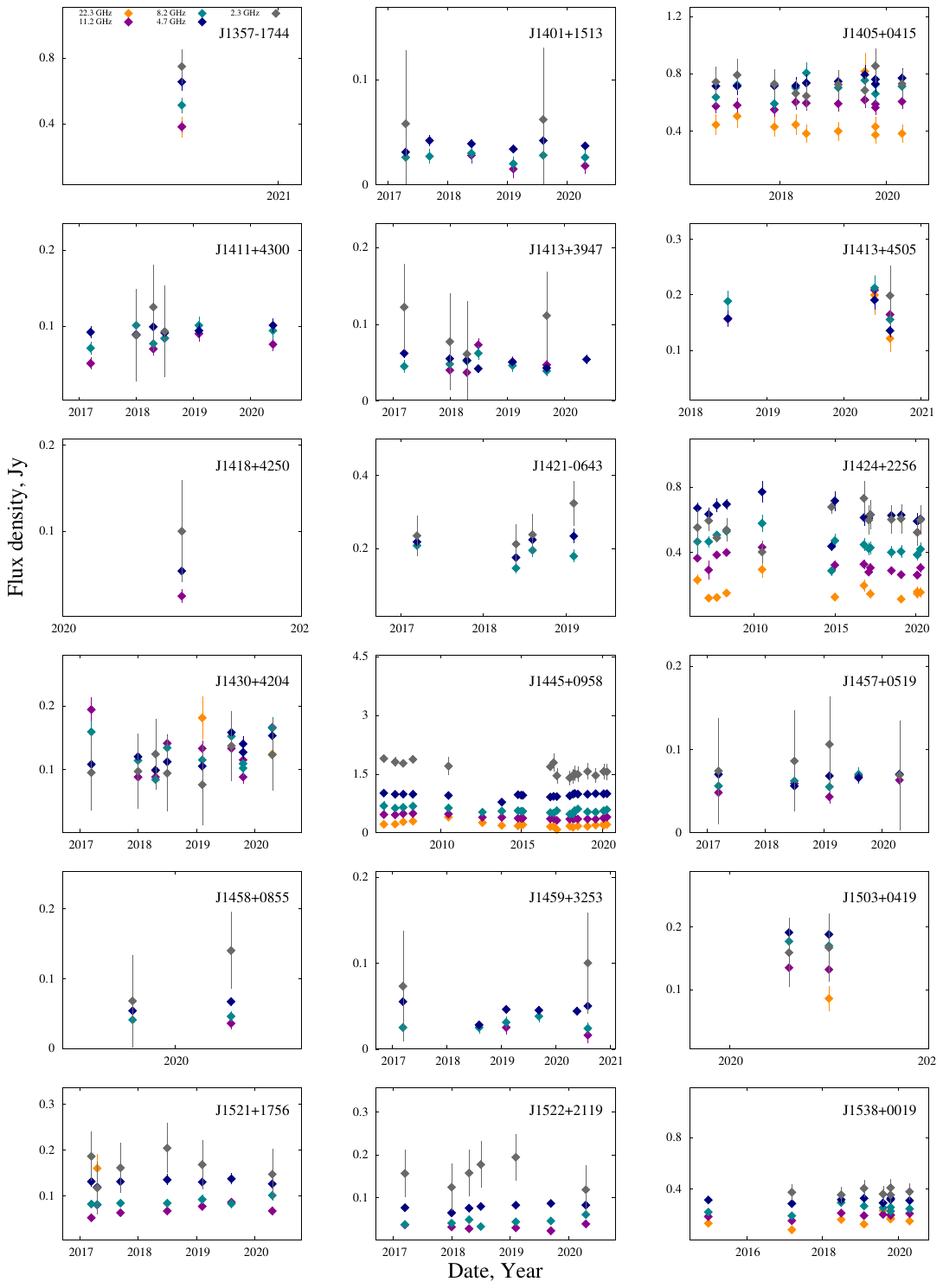}}
\caption{The RATAN-600 light curves, the measurements at 22.3 GHz are colored by orange, at 11.2 GHz by magenta, at 8.2 GHz by green, at 4.7 GHz by blue.}
\label{fig:C4}
\end{figure}

\clearpage
\newpage

\begin{figure}
\centerline{\includegraphics[width=190mm]{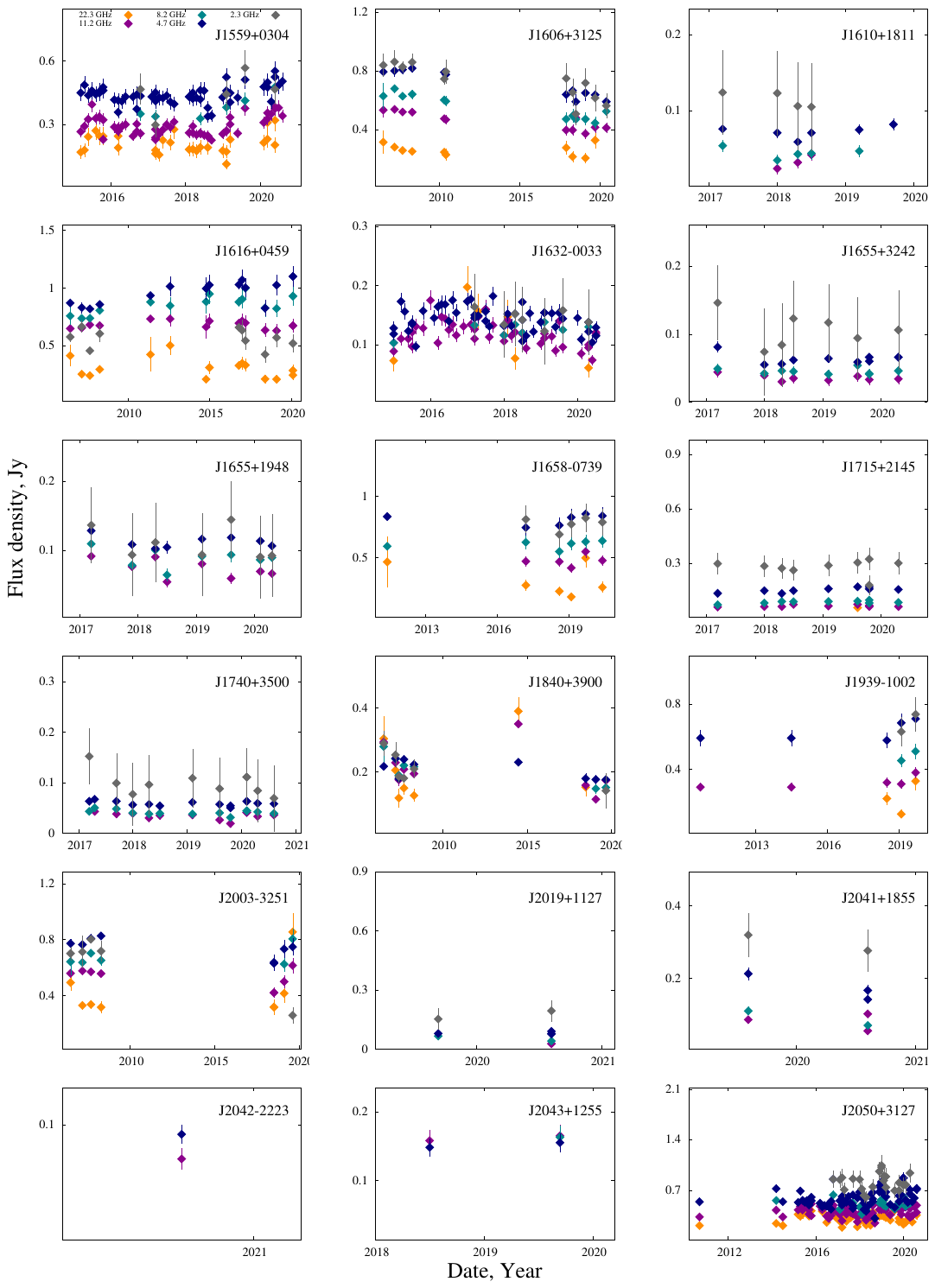}}
\caption{The RATAN-600 light curves, the measurements at 22.3 GHz are colored by orange, at 11.2 GHz by magenta, at 8.2 GHz by green, at 4.7 GHz by blue.}
\label{fig:C5}
\end{figure}

\clearpage
\newpage

\begin{figure}
\centerline{\includegraphics[width=190mm]{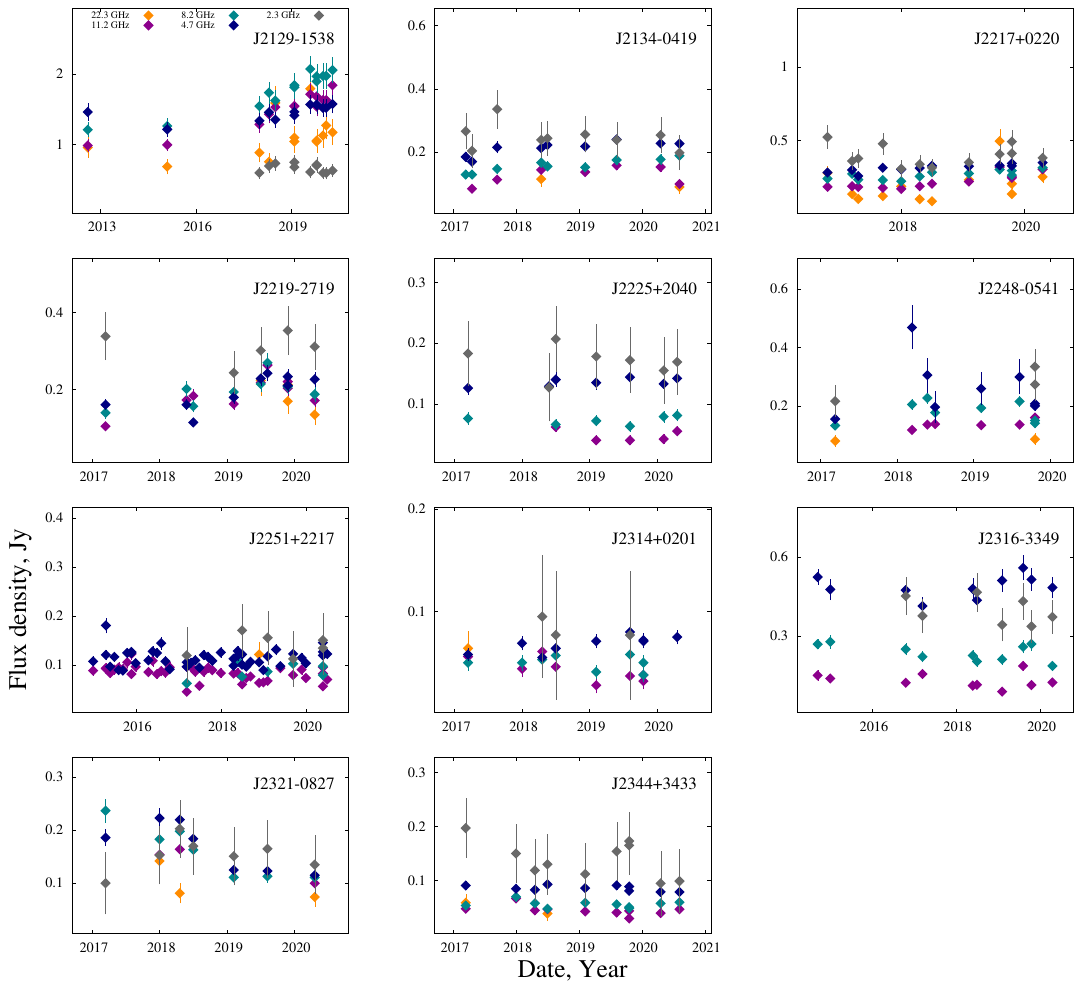}}
\caption{The RATAN-600 light curves, the measurements at 22.3 GHz are colored by orange, at 11.2 GHz by magenta, at 8.2 GHz by green, at 4.7 GHz by blue.}
\label{fig:C6}
\end{figure}

\newpage

\begin{figure}
\centerline{\includegraphics[width=190mm]{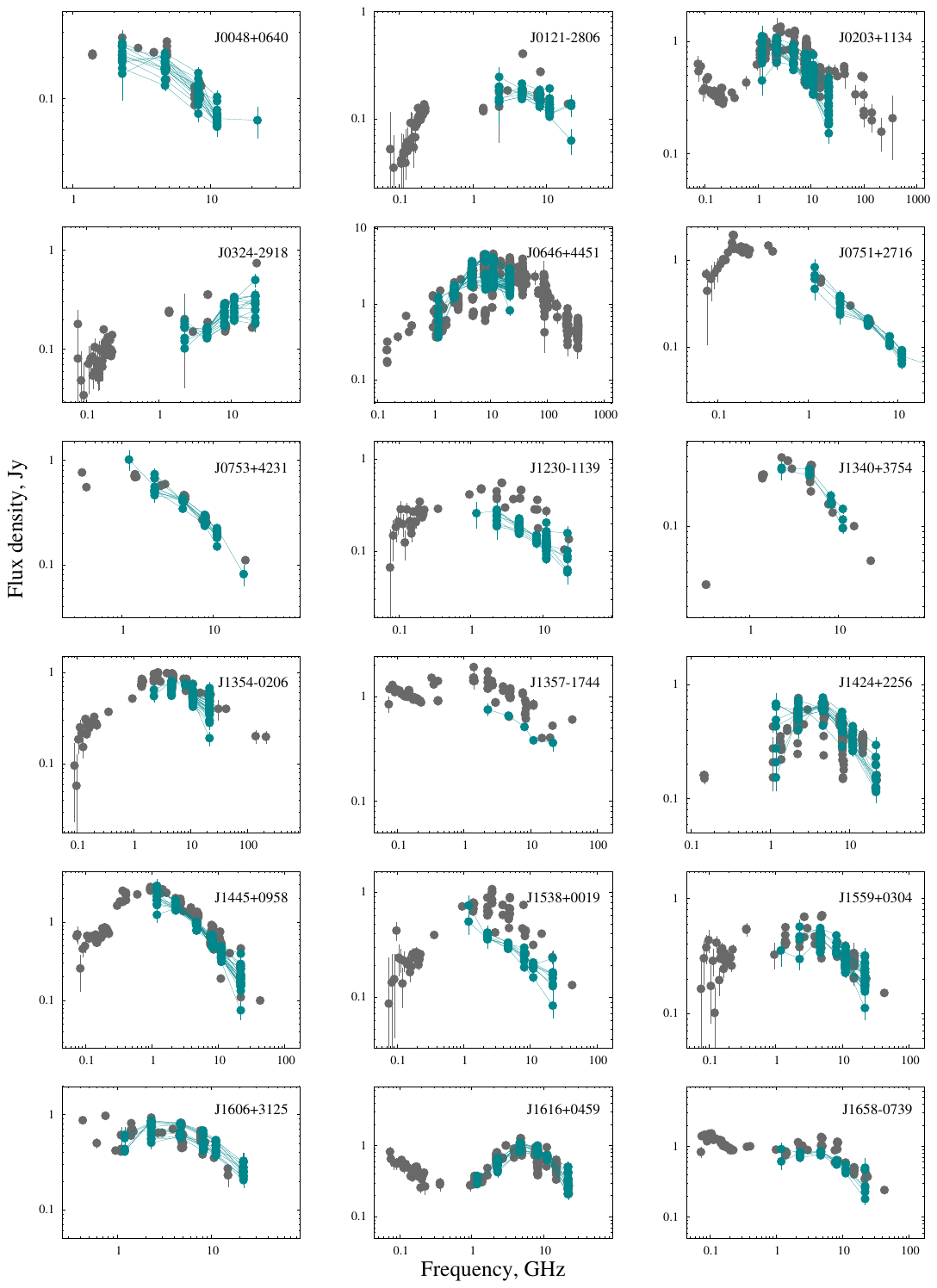}}
\caption{The radio spectra of  PS quasars constructed using the RATAN-600 (green) and literature data from CATS (grey).}
\label{fig:B1}
\end{figure}

\begin{figure}
\centerline{\includegraphics[width=190mm]{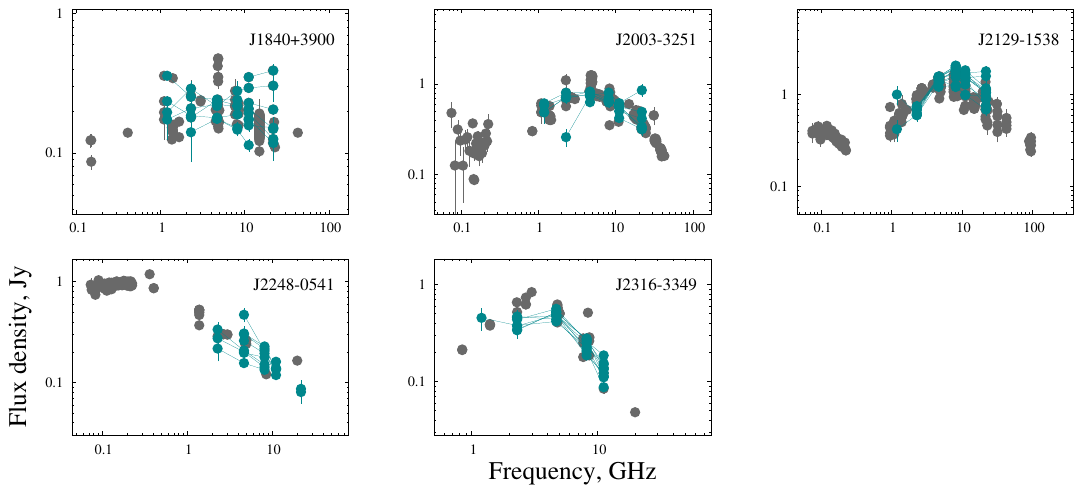}}
\caption{The radio spectra of PS quasars constructed using the RATAN-600 (green) and literature data from CATS (grey).}
\label{fig:B2}
\end{figure}

\begin{adjustwidth}{-\extralength}{0cm}

\reftitle{References}

\bibliography{mingaliev}

\begin{thebibliography}{999}

\bibitem[{Schmidt}(1963)]{1963Natur.197.1040S}
{Schmidt}, M.
\newblock {3C 273 : A Star-Like Object with Large Red-Shift}.
\newblock {\em \nat} {\bf 1963}, {\em 197},~1040.
\newblock {\url{https://doi.org/10.1038/1971040a0}}.

\bibitem[{Matthews} and {Sandage}(1963)]{1963ApJ...138...30M}
{Matthews}, T.A.; {Sandage}, A.R.
\newblock {Optical Identification of 3C 48, 3C 196, and 3C 286 with Stellar Objects.}
\newblock {\em \apj} {\bf 1963}, {\em 138},~30.
\newblock {\url{https://doi.org/10.1086/147615}}.

\bibitem[{Dent}(1965)]{1965Sci...148.1458D}
{Dent}, W.A.
\newblock {Quasi-Stellar Sources: Variation in the Radio Emission of 3C 273}.
\newblock {\em Science} {\bf 1965}, {\em 148},~1458--1460.
\newblock {\url{https://doi.org/10.1126/science.148.3676.1458}}.

\bibitem[{Kellermann} and {Pauliny-Toth}(1968)]{1968ARA&A...6..417K}
{Kellermann}, K.I.; {Pauliny-Toth}, I.I.K.
\newblock {Variable Radio Sources}.
\newblock {\em \araa} {\bf 1968}, {\em 6},~417.
\newblock {\url{https://doi.org/10.1146/annurev.aa.06.090168.002221}}.

\bibitem[{Rees}(1970)]{1970Natur.227.1303R}
{Rees}, M.J.
\newblock {Extragalactic Variable Radio Sources}.
\newblock {\em \nat} {\bf 1970}, {\em 227},~1303--1306.
\newblock {\url{https://doi.org/10.1038/2271303a0}}.

\bibitem[{Marscher} and {Gear}(1985)]{1985ApJ...298..114M}
{Marscher}, A.P.; {Gear}, W.K.
\newblock {Models for high-frequency radio outbursts in extragalactic sources, with application to the early 1983 millimeter-to-infrared flare of 3C 273.}
\newblock {\em \apj} {\bf 1985}, {\em 298},~114--127.
\newblock {\url{https://doi.org/10.1086/163592}}.

\bibitem[{Nyland} et~al.(2021){Nyland}, {Dong}, {Patil}, {Lacy}, {Kimball}, {Hallinan}, {Sarbadhicary}, {Polisensky}, {Kassim}, {Peters}, {Clarke}, {Mukherjee}, {van Velzen}, and {Baldassare}]{2021IAUS..359...27N}
{Nyland}, K.; {Dong}, D.; {Patil}, P.; {Lacy}, M.; {Kimball}, A.; {Hallinan}, G.; {Sarbadhicary}, S.; {Polisensky}, E.; {Kassim}, N.; {Peters}, W.;  et~al.
\newblock {Variable radio AGN at high redshift identified in the VLA Sky Survey}.
\newblock In Proceedings of the Galaxy Evolution and Feedback across Different Environments; {Storchi Bergmann}, T.; {Forman}, W.; {Overzier}, R.; {Riffel}, R., Eds.,  2021, Vol. 359, {\em IAU Symposium}, pp. 27--32,  \href{http://xxx.lanl.gov/abs/2005.04734}{{\normalfont [arXiv:astro-ph.GA/2005.04734]}}.
\newblock {\url{https://doi.org/10.1017/S1743921320001921}}.

\bibitem[{Gergely} and {Biermann}(2009)]{2009ApJ...697.1621G}
{Gergely}, L.{\'A}.; {Biermann}, P.L.
\newblock {The Spin-Flip Phenomenon in Supermassive Black hole binary mergers}.
\newblock {\em \apj} {\bf 2009}, {\em 697},~1621--1633,  \href{http://xxx.lanl.gov/abs/0704.1968}{{\normalfont [arXiv:astro-ph/0704.1968]}}.
\newblock {\url{https://doi.org/10.1088/0004-637X/697/2/1621}}.

\bibitem[{Horton} et~al.(2020){Horton}, {Krause}, and {Hardcastle}]{2020MNRAS.499.5765H}
{Horton}, M.A.; {Krause}, M.G.H.; {Hardcastle}, M.J.
\newblock {3D hydrodynamic simulations of large-scale precessing jets: radio morphology}.
\newblock {\em \mnras} {\bf 2020}, {\em 499},~5765--5781,  \href{http://xxx.lanl.gov/abs/2010.00480}{{\normalfont [arXiv:astro-ph.GA/2010.00480]}}.
\newblock {\url{https://doi.org/10.1093/mnras/staa3020}}.

\bibitem[{Villata} and {Raiteri}(1999)]{1999A&A...347...30V}
{Villata}, M.; {Raiteri}, C.M.
\newblock {Helical jets in blazars. I. The case of MKN 501}.
\newblock {\em \aap} {\bf 1999}, {\em 347},~30--36.

\bibitem[{Jarvis} et~al.(2019){Jarvis}, {Harrison}, {Thomson}, {Circosta}, {Mainieri}, {Alexander}, {Edge}, {Lansbury}, {Molyneux}, and {Mullaney}]{2019MNRAS.485.2710J}
{Jarvis}, M.E.; {Harrison}, C.M.; {Thomson}, A.P.; {Circosta}, C.; {Mainieri}, V.; {Alexander}, D.M.; {Edge}, A.C.; {Lansbury}, G.B.; {Molyneux}, S.J.; {Mullaney}, J.R.
\newblock {Prevalence of radio jets associated with galactic outflows and feedback from quasars}.
\newblock {\em \mnras} {\bf 2019}, {\em 485},~2710--2730,  \href{http://xxx.lanl.gov/abs/1902.07727}{{\normalfont [arXiv:astro-ph.GA/1902.07727]}}.
\newblock {\url{https://doi.org/10.1093/mnras/stz556}}.

\bibitem[{An} et~al.(2013){An}, {Baan}, {Wang}, {Wang}, and {Hong}]{2013MNRAS.434.3487A}
{An}, T.; {Baan}, W.A.; {Wang}, J.Y.; {Wang}, Y.; {Hong}, X.Y.
\newblock {Periodic radio variabilities in NRAO 530: a jet-disc connection?}
\newblock {\em \mnras} {\bf 2013}, {\em 434},~3487--3496.
\newblock {\url{https://doi.org/10.1093/mnras/stt1265}}.

\bibitem[{Wang} et~al.(2014){Wang}, {An}, {Baan}, and {Lu}]{2014MNRAS.443...58W}
{Wang}, J.Y.; {An}, T.; {Baan}, W.A.; {Lu}, X.L.
\newblock {Periodic radio variabilities of the blazar 1156+295: harmonic oscillations}.
\newblock {\em \mnras} {\bf 2014}, {\em 443},~58--66.
\newblock {\url{https://doi.org/10.1093/mnras/stu1135}}.

\bibitem[{Walker}(1998)]{1998MNRAS.294..307W}
{Walker}, M.A.
\newblock {Interstellar scintillation of compact extragalactic radio sources}.
\newblock {\em \mnras} {\bf 1998}, {\em 294},~307--311.
\newblock {\url{https://doi.org/10.1046/j.1365-8711.1998.01238.x}}.

\bibitem[{Koay} et~al.(2019){Koay}, {Jauncey}, {Hovatta}, {Kiehlmann}, {Bignall}, {Max-Moerbeck}, {Pearson}, {Readhead}, {Reeves}, {Reynolds}, and {Vedantham}]{2019MNRAS.489.5365K}
{Koay}, J.Y.; {Jauncey}, D.L.; {Hovatta}, T.; {Kiehlmann}, S.; {Bignall}, H.E.; {Max-Moerbeck}, W.; {Pearson}, T.J.; {Readhead}, A.C.S.; {Reeves}, R.; {Reynolds}, C.;  et~al.
\newblock {The presence of interstellar scintillation in the 15 GHz interday variability of 1158 OVRO-monitored blazars}.
\newblock {\em \mnras} {\bf 2019}, {\em 489},~5365--5380,  \href{http://xxx.lanl.gov/abs/1909.01566}{{\normalfont [arXiv:astro-ph.GA/1909.01566]}}.
\newblock {\url{https://doi.org/10.1093/mnras/stz2488}}.

\bibitem[{Fiedler} et~al.(1987){Fiedler}, {Dennison}, {Johnston}, and {Hewish}]{1987Natur.326..675F}
{Fiedler}, R.L.; {Dennison}, B.; {Johnston}, K.J.; {Hewish}, A.
\newblock {Extreme scattering events caused by compact structures in the interstellar medium}.
\newblock {\em \nat} {\bf 1987}, {\em 326},~675--678.
\newblock {\url{https://doi.org/10.1038/326675a0}}.

\bibitem[{Pushkarev} et~al.(2013){Pushkarev}, {Kovalev}, {Lister}, {Hovatta}, {Savolainen}, {Aller}, {Aller}, {Ros}, {Zensus}, {Richards}, {Max-Moerbeck}, and {Readhead}]{2013A&A...555A..80P}
{Pushkarev}, A.B.; {Kovalev}, Y.Y.; {Lister}, M.L.; {Hovatta}, T.; {Savolainen}, T.; {Aller}, M.F.; {Aller}, H.D.; {Ros}, E.; {Zensus}, J.A.; {Richards}, J.L.;  et~al.
\newblock {VLBA observations of a rare multiple quasar imaging event caused by refraction in the interstellar medium}.
\newblock {\em \aap} {\bf 2013}, {\em 555},~A80,  \href{http://xxx.lanl.gov/abs/1305.6005}{{\normalfont [arXiv:astro-ph.CO/1305.6005]}}.
\newblock {\url{https://doi.org/10.1051/0004-6361/201321484}}.

\bibitem[{De Breuck} et~al.(2000){De Breuck}, {van Breugel}, {R{\"o}ttgering}, and {Miley}]{2000A&AS..143..303D}
{De Breuck}, C.; {van Breugel}, W.; {R{\"o}ttgering}, H.J.A.; {Miley}, G.
\newblock {A sample of 669 ultra steep spectrum radio sources to find high redshift radio galaxies}.
\newblock {\em \aaps} {\bf 2000}, {\em 143},~303--333,  \href{http://xxx.lanl.gov/abs/astro-ph/0002297}{{\normalfont [arXiv:astro-ph/astro-ph/0002297]}}.
\newblock {\url{https://doi.org/10.1051/aas:2000181}}.

\bibitem[{De Breuck} et~al.(2006){De Breuck}, {Klamer}, {Johnston}, {Hunstead}, {Bryant}, {Rocca-Volmerange}, and {Sadler}]{2006MNRAS.366...58D}
{De Breuck}, C.; {Klamer}, I.; {Johnston}, H.; {Hunstead}, R.W.; {Bryant}, J.; {Rocca-Volmerange}, B.; {Sadler}, E.M.
\newblock {A search for distant radio galaxies from SUMSS and NVSS - II. Optical spectroscopy$^{1*}$}.
\newblock {\em \mnras} {\bf 2006}, {\em 366},~58--72,  \href{http://xxx.lanl.gov/abs/astro-ph/0511169}{{\normalfont [arXiv:astro-ph/astro-ph/0511169]}}.
\newblock {\url{https://doi.org/10.1111/j.1365-2966.2005.09799.x}}.

\bibitem[{Coppejans} et~al.(2016){Coppejans}, {Frey}, {Cseh}, {M{\"u}ller}, {Paragi}, {Falcke}, {Gab{\'a}nyi}, {Gurvits}, {An}, and {Titov}]{2016MNRAS.463.3260C}
{Coppejans}, R.; {Frey}, S.; {Cseh}, D.; {M{\"u}ller}, C.; {Paragi}, Z.; {Falcke}, H.; {Gab{\'a}nyi}, K.{\'E}.; {Gurvits}, L.I.; {An}, T.; {Titov}, O.
\newblock {On the nature of bright compact radio sources at z $\gt$ 4.5}.
\newblock {\em \mnras} {\bf 2016}, {\em 463},~3260--3275,  \href{http://xxx.lanl.gov/abs/1609.00575}{{\normalfont [1609.00575]}}.
\newblock {\url{https://doi.org/10.1093/mnras/stw2236}}.

\bibitem[{O'Dea} et~al.(1991){O'Dea}, {Baum}, and {Stanghellini}]{1991ApJ...380...66O}
{O'Dea}, C.P.; {Baum}, S.A.; {Stanghellini}, C.
\newblock {What are the gigahertz peaked-spectrum radio sources?}
\newblock {\em \apj} {\bf 1991}, {\em 380},~66--77.
\newblock {\url{https://doi.org/10.1086/170562}}.

\bibitem[{Coppejans} et~al.(2016){Coppejans}, {Cseh}, {van Velzen}, {Falcke}, {Intema}, {Paragi}, {M{\"u}ller}, {Williams}, {Frey}, {Gurvits}, and {K{\"o}rding}]{2016MNRAS.459.2455C}
{Coppejans}, R.; {Cseh}, D.; {van Velzen}, S.; {Falcke}, H.; {Intema}, H.T.; {Paragi}, Z.; {M{\"u}ller}, C.; {Williams}, W.L.; {Frey}, S.; {Gurvits}, L.I.;  et~al.
\newblock {What are the megahertz peaked-spectrum sources?}
\newblock {\em \mnras} {\bf 2016}, {\em 459},~2455--2471,  \href{http://xxx.lanl.gov/abs/1604.00171}{{\normalfont [1604.00171]}}.
\newblock {\url{https://doi.org/10.1093/mnras/stw799}}.

\bibitem[{O'Dea} and {Saikia}(2021)]{2021A&ARv..29....3O}
{O'Dea}, C.P.; {Saikia}, D.J.
\newblock {Compact steep-spectrum and peaked-spectrum radio sources}.
\newblock {\em \aapr} {\bf 2021}, {\em 29},~3,  \href{http://xxx.lanl.gov/abs/2009.02750}{{\normalfont [arXiv:astro-ph.GA/2009.02750]}}.
\newblock {\url{https://doi.org/10.1007/s00159-021-00131-w}}.

\bibitem[{Fanti} et~al.(1990){Fanti}, {Fanti}, {Schilizzi}, {Spencer}, {Nan Rendong}, {Parma}, {van Breugel}, and {Venturi}]{1990A&A...231..333F}
{Fanti}, R.; {Fanti}, C.; {Schilizzi}, R.T.; {Spencer}, R.E.; {Nan Rendong}.; {Parma}, P.; {van Breugel}, W.J.M.; {Venturi}, T.
\newblock {On the nature of compact steep spectrum radio sources.}
\newblock {\em \aap} {\bf 1990}, {\em 231},~333--346.

\bibitem[{Snellen} et~al.(2000){Snellen}, {Schilizzi}, {Miley}, {de Bruyn}, {Bremer}, and {R{\"o}ttgering}]{2000MNRAS.319..445S}
{Snellen}, I.A.G.; {Schilizzi}, R.T.; {Miley}, G.K.; {de Bruyn}, A.G.; {Bremer}, M.N.; {R{\"o}ttgering}, H.J.A.
\newblock {On the evolution of young radio-loud AGN}.
\newblock {\em \mnras} {\bf 2000}, {\em 319},~445--456,  \href{http://xxx.lanl.gov/abs/astro-ph/0002130}{{\normalfont [arXiv:astro-ph/astro-ph/0002130]}}.
\newblock {\url{https://doi.org/10.1046/j.1365-8711.2000.03935.x}}.

\bibitem[{An} and {Baan}(2012)]{2012ApJ...760...77A}
{An}, T.; {Baan}, W.A.
\newblock {The Dynamic Evolution of Young Extragalactic Radio Sources}.
\newblock {\em \apj} {\bf 2012}, {\em 760},~77,  \href{http://xxx.lanl.gov/abs/1211.1760}{{\normalfont [arXiv:astro-ph.CO/1211.1760]}}.
\newblock {\url{https://doi.org/10.1088/0004-637X/760/1/77}}.

\bibitem[{Readhead} et~al.(1996){Readhead}, {Taylor}, {Xu}, {Pearson}, {Wilkinson}, and {Polatidis}]{1996ApJ...460..612R}
{Readhead}, A.C.S.; {Taylor}, G.B.; {Xu}, W.; {Pearson}, T.J.; {Wilkinson}, P.N.; {Polatidis}, A.G.
\newblock {The Statistics and Ages of Compact Symmetric Objects}.
\newblock {\em \apj} {\bf 1996}, {\em 460},~612--633.
\newblock {\url{https://doi.org/10.1086/176996}}.

\bibitem[{Coppejans} et~al.(2015){Coppejans}, {Cseh}, {Williams}, {van Velzen}, and {Falcke}]{2015MNRAS.450.1477C}
{Coppejans}, R.; {Cseh}, D.; {Williams}, W.L.; {van Velzen}, S.; {Falcke}, H.
\newblock {Megahertz peaked-spectrum sources in the Bo{\"o}tes field I - a route towards finding high-redshift AGN}.
\newblock {\em \mnras} {\bf 2015}, {\em 450},~1477--1485.
\newblock {\url{https://doi.org/10.1093/mnras/stv681}}.

\bibitem[{Coppejans} et~al.(2017){Coppejans}, {van Velzen}, {Intema}, {M{\"u}ller}, {Frey}, {Coppejans}, {Cseh}, {Williams}, {Falcke}, {K{\"o}rding}, {Orr{\'u}}, {Paragi}, and {Gab{\'a}nyi}]{2017MNRAS.467.2039C}
{Coppejans}, R.; {van Velzen}, S.; {Intema}, H.T.; {M{\"u}ller}, C.; {Frey}, S.; {Coppejans}, D.L.; {Cseh}, D.; {Williams}, W.L.; {Falcke}, H.; {K{\"o}rding}, E.G.;  et~al.
\newblock {Radio spectra of bright compact sources at z \&gt; 4.5}.
\newblock {\em \mnras} {\bf 2017}, {\em 467},~2039--2060,  \href{http://xxx.lanl.gov/abs/1701.06622}{{\normalfont [arXiv:astro-ph.GA/1701.06622]}}.
\newblock {\url{https://doi.org/10.1093/mnras/stx215}}.

\bibitem[{Sotnikova} et~al.(2021){Sotnikova}, {Mikhailov}, {Mufakharov}, {Mingaliev}, {Bursov}, {Semenova}, {Stolyarov}, {Udovitskiy}, {Kudryashova}, and {Erkenov}]{2021MNRAS.508.2798S}
{Sotnikova}, Y.; {Mikhailov}, A.; {Mufakharov}, T.; {Mingaliev}, M.; {Bursov}, N.; {Semenova}, T.; {Stolyarov}, V.; {Udovitskiy}, R.; {Kudryashova}, A.; {Erkenov}, A.
\newblock {High-redshift quasars at z {\ensuremath{\geq}} 3 - I. Radio spectra}.
\newblock {\em \mnras} {\bf 2021}, {\em 508},~2798--2814,  \href{http://xxx.lanl.gov/abs/2109.14029}{{\normalfont [arXiv:astro-ph.GA/2109.14029]}}.
\newblock {\url{https://doi.org/10.1093/mnras/stab2114}}.

\bibitem[{Stanghellini} et~al.(1998){Stanghellini}, {O'Dea}, {Dallacasa}, {Baum}, {Fanti}, and {Fanti}]{1998A&AS..131..303S}
{Stanghellini}, C.; {O'Dea}, C.P.; {Dallacasa}, D.; {Baum}, S.A.; {Fanti}, R.; {Fanti}, C.
\newblock {A complete sample of GHz-peaked-spectrum radio sources and its radio properties}.
\newblock {\em \aaps} {\bf 1998}, {\em 131},~303--315,  \href{http://xxx.lanl.gov/abs/arXiv:astro-ph/9803222}{{\normalfont [arXiv:astro-ph/9803222]}}.

\bibitem[{O'Dea}(1998)]{1998PASP..110..493O}
{O'Dea}, C.P.
\newblock {The Compact Steep-Spectrum and Gigahertz Peaked-Spectrum Radio Sources}.
\newblock {\em \pasp} {\bf 1998}, {\em 110},~493--532.
\newblock {\url{https://doi.org/10.1086/316162}}.

\bibitem[{Stanghellini}(1999)]{1999MmSAI..70..117S}
{Stanghellini}, C.
\newblock {Radio variability and polarization of GHz-peaked-spectrum sources.}
\newblock {\em \memsai} {\bf 1999}, {\em 70},~117--120.

\bibitem[{Tornikoski} et~al.(2009){Tornikoski}, {Torniainen}, {L{\"a}hteenm{\"a}ki}, {Hovatta}, {Nieppola}, {Turunen}, {Lainela}, {Valtaoja}, {Aller}, {Aller}, {Mingaliev}, and {Trushkin}]{2009AN....330..128T}
{Tornikoski}, M.; {Torniainen}, I.; {L{\"a}hteenm{\"a}ki}, A.; {Hovatta}, T.; {Nieppola}, E.; {Turunen}, M.; {Lainela}, M.; {Valtaoja}, E.; {Aller}, M.F.; {Aller}, H.D.;  et~al.
\newblock {Long-term radio behaviour of GPS sources and candidates}.
\newblock {\em Astronomische Nachrichten} {\bf 2009}, {\em 330},~128--132.
\newblock {\url{https://doi.org/10.1002/asna.200811139}}.

\bibitem[{Mingaliev} et~al.(2012){Mingaliev}, {Sotnikova}, {Torniainen}, {Tornikoski}, and {Udovitskiy}]{2012A&A...544A..25M}
{Mingaliev}, M.G.; {Sotnikova}, Y.V.; {Torniainen}, I.; {Tornikoski}, M.; {Udovitskiy}, R.Y.
\newblock {Multifrequency study of GHz-peaked spectrum sources and candidates with the RATAN-600 radio telescope}.
\newblock {\em \aap} {\bf 2012}, {\em 544},~A25.
\newblock {\url{https://doi.org/10.1051/0004-6361/201118506}}.

\bibitem[{Mingaliev} et~al.(2013){Mingaliev}, {Sotnikova}, {Mufakharov}, {Erkenov}, and {Udovitskiy}]{2013AstBu..68..262M}
{Mingaliev}, M.G.; {Sotnikova}, Y.V.; {Mufakharov}, T.V.; {Erkenov}, A.K.; {Udovitskiy}, R.Y.
\newblock {Gigahertz-peaked spectrum (GPS) galaxies and quasars}.
\newblock {\em Astrophysical Bulletin} {\bf 2013}, {\em 68},~262--272.
\newblock {\url{https://doi.org/10.1134/S1990341313030036}}.

\bibitem[{Sotnikova} et~al.(2019){Sotnikova}, {Mufakharov}, {Majorova}, {Mingaliev}, {Udovitskii}, {Bursov}, and {Semenova}]{2019AstBu..74..348S}
{Sotnikova}, Y.V.; {Mufakharov}, T.V.; {Majorova}, E.K.; {Mingaliev}, M.G.; {Udovitskii}, R.Y.; {Bursov}, N.N.; {Semenova}, T.A.
\newblock {Multifrequency Study of GHz-peaked Spectrum Sources}.
\newblock {\em Astrophysical Bulletin} {\bf 2019}, {\em 74},~348--364,  \href{http://xxx.lanl.gov/abs/1911.12769}{{\normalfont [arXiv:astro-ph.GA/1911.12769]}}.
\newblock {\url{https://doi.org/10.1134/S1990341319040023}}.

\bibitem[{Sotnikova} et~al.(2021){Sotnikova}, {Mufakharov}, {Mingaliev}, and {Mikhailov}]{2021AN....342.1195S}
{Sotnikova}, Y.; {Mufakharov}, T.; {Mingaliev}, M.; {Mikhailov}, A.
\newblock {Multifrequency study of gigahertz‑peaked spectrum sources with RATAN ‑600}.
\newblock {\em Astronomische Nachrichten} {\bf 2021}, {\em 342},~1195--1199,  \href{http://xxx.lanl.gov/abs/2111.04828}{{\normalfont [arXiv:astro-ph.GA/2111.04828]}}.
\newblock {\url{https://doi.org/10.1002/asna.20210052}}.

\bibitem[{Dallacasa} and {Orienti}(2016)]{2016AN....337..120D}
{Dallacasa}, D.; {Orienti}, M.
\newblock {Radio spectra of High Frequency Peakers}.
\newblock {\em Astronomische Nachrichten} {\bf 2016}, {\em 337},~120.
\newblock {\url{https://doi.org/10.1002/asna.201512756}}.

\bibitem[{Kovalev}(2005)]{2005BaltA..14..413K}
{Kovalev}, Y.Y.
\newblock {``TEMPORARY Gps/hfp'' Radio Sources}.
\newblock {\em Baltic Astronomy} {\bf 2005}, {\em 14},~413--416.

\bibitem[{Orienti} et~al.(2010){Orienti}, {Dallacasa}, and {Stanghellini}]{2010MNRAS.408.1075O}
{Orienti}, M.; {Dallacasa}, D.; {Stanghellini}, C.
\newblock {Spectral variability in faint high-frequency peakers}.
\newblock {\em \mnras} {\bf 2010}, {\em 408},~1075--1088,  \href{http://xxx.lanl.gov/abs/1006.2049}{{\normalfont [arXiv:astro-ph.CO/1006.2049]}}.
\newblock {\url{https://doi.org/10.1111/j.1365-2966.2010.17179.x}}.

\bibitem[{Condon} et~al.(1998){Condon}, {Cotton}, {Greisen}, {Yin}, {Perley}, {Taylor}, and {Broderick}]{1998AJ....115.1693C}
{Condon}, J.J.; {Cotton}, W.D.; {Greisen}, E.W.; {Yin}, Q.F.; {Perley}, R.A.; {Taylor}, G.B.; {Broderick}, J.J.
\newblock {The NRAO VLA Sky Survey}.
\newblock {\em \aj} {\bf 1998}, {\em 115},~1693--1716.
\newblock {\url{https://doi.org/10.1086/300337}}.

\bibitem[{Intema} et~al.(2017){Intema}, {Jagannathan}, {Mooley}, and {Frail}]{2017A&A...598A..78I}
{Intema}, H.T.; {Jagannathan}, P.; {Mooley}, K.P.; {Frail}, D.A.
\newblock {The GMRT 150 MHz all-sky radio survey. First alternative data release TGSS ADR1}.
\newblock {\em \aap} {\bf 2017}, {\em 598},~A78,  \href{http://xxx.lanl.gov/abs/1603.04368}{{\normalfont [arXiv:astro-ph.CO/1603.04368]}}.
\newblock {\url{https://doi.org/10.1051/0004-6361/201628536}}.

\bibitem[{Hurley-Walker} et~al.(2017){Hurley-Walker}, {Callingham}, {Hancock}, {Franzen}, {Hindson}, {Kapi{\'n}ska}, {Morgan}, {Offringa}, {Wayth}, {Wu}, {Zheng}, {Murphy}, {Bell}, {Dwarakanath}, {For}, {Gaensler}, {Johnston-Hollitt}, {Lenc}, {Procopio}, {Staveley-Smith}, {Ekers}, {Bowman}, {Briggs}, {Cappallo}, {Deshpande}, {Greenhill}, {Hazelton}, {Kaplan}, {Lonsdale}, {McWhirter}, {Mitchell}, {Morales}, {Morgan}, {Oberoi}, {Ord}, {Prabu}, {Shankar}, {Srivani}, {Subrahmanyan}, {Tingay}, {Webster}, {Williams}, and {Williams}]{2017MNRAS.464.1146H}
{Hurley-Walker}, N.; {Callingham}, J.R.; {Hancock}, P.J.; {Franzen}, T.M.O.; {Hindson}, L.; {Kapi{\'n}ska}, A.D.; {Morgan}, J.; {Offringa}, A.R.; {Wayth}, R.B.; {Wu}, C.;  et~al.
\newblock {GaLactic and Extragalactic All-sky Murchison Widefield Array (GLEAM) survey - I. A low-frequency extragalactic catalogue}.
\newblock {\em \mnras} {\bf 2017}, {\em 464},~1146--1167,  \href{http://xxx.lanl.gov/abs/1610.08318}{{\normalfont [arXiv:astro-ph.GA/1610.08318]}}.
\newblock {\url{https://doi.org/10.1093/mnras/stw2337}}.

\bibitem[{Rengelink} et~al.(1997){Rengelink}, {Tang}, {de Bruyn}, {Miley}, {Bremer}, {Roettgering}, and {Bremer}]{1997A&AS..124..259R}
{Rengelink}, R.B.; {Tang}, Y.; {de Bruyn}, A.G.; {Miley}, G.K.; {Bremer}, M.N.; {Roettgering}, H.J.A.; {Bremer}, M.A.R.
\newblock {The Westerbork Northern Sky Survey (WENSS), I. A 570 square degree Mini-Survey around the North Ecliptic Pole}.
\newblock {\em \aaps} {\bf 1997}, {\em 124},~259--280.
\newblock {\url{https://doi.org/10.1051/aas:1997358}}.

\bibitem[{Giommi} et~al.(2019){Giommi}, {Brandt}, {Barres de Almeida}, {Pollock}, {Arneodo}, {Chang}, {Civitarese}, {De Angelis}, {D'Elia}, {Del Rio Vera}, {Di Pippo}, {Middei}, {Penacchioni}, {Perri}, {Ruffini}, {Sahakyan}, and {Turriziani}]{2019A&A...631A.116G}
{Giommi}, P.; {Brandt}, C.H.; {Barres de Almeida}, U.; {Pollock}, A.M.T.; {Arneodo}, F.; {Chang}, Y.L.; {Civitarese}, O.; {De Angelis}, M.; {D'Elia}, V.; {Del Rio Vera}, J.;  et~al.
\newblock {Open Universe for Blazars: a new generation of astronomical products based on 14 years of Swift-XRT data}.
\newblock {\em \aap} {\bf 2019}, {\em 631},~A116,  \href{http://xxx.lanl.gov/abs/1904.06043}{{\normalfont [arXiv:astro-ph.HE/1904.06043]}}.
\newblock {\url{https://doi.org/10.1051/0004-6361/201935646}}.

\bibitem[{Massaro} et~al.(2015){Massaro}, {Maselli}, {Leto}, {Marchegiani}, {Perri}, {Giommi}, and {Piranomonte}]{2015Ap&SS.357...75M}
{Massaro}, E.; {Maselli}, A.; {Leto}, C.; {Marchegiani}, P.; {Perri}, M.; {Giommi}, P.; {Piranomonte}, S.
\newblock {The 5th edition of the Roma-BZCAT. A short presentation}.
\newblock {\em \apss} {\bf 2015}, {\em 357},~75,  \href{http://xxx.lanl.gov/abs/1502.07755}{{\normalfont [arXiv:astro-ph.HE/1502.07755]}}.
\newblock {\url{https://doi.org/10.1007/s10509-015-2254-2}}.

\bibitem[{Albareti} et~al.(2015){Albareti}, {Comparat}, {Guti{\'e}rrez}, {Prada}, {P{\^a}ris}, {Schlegel}, {L{\'o}pez-Corredoira}, {Schneider}, {Manchado}, {Garc{\'\i}a-Hern{\'a}ndez}, {Petitjean}, and {Ge}]{2015MNRAS.452.4153A}
{Albareti}, F.D.; {Comparat}, J.; {Guti{\'e}rrez}, C.M.; {Prada}, F.; {P{\^a}ris}, I.; {Schlegel}, D.; {L{\'o}pez-Corredoira}, M.; {Schneider}, D.P.; {Manchado}, A.; {Garc{\'\i}a-Hern{\'a}ndez}, D.A.;  et~al.
\newblock {Constraint on the time variation of the fine-structure constant with the SDSS-III/BOSS DR12 quasar sample}.
\newblock {\em \mnras} {\bf 2015}, {\em 452},~4153--4168,  \href{http://xxx.lanl.gov/abs/1501.00560}{{\normalfont [arXiv:astro-ph.CO/1501.00560]}}.
\newblock {\url{https://doi.org/10.1093/mnras/stv1406}}.

\bibitem[{P{\^a}ris} et~al.(2018){P{\^a}ris}, {Petitjean}, {Aubourg}, {Myers}, {Streblyanska}, {Lyke}, {Anderson}, {Armengaud}, {Bautista}, {Blanton}, {Blomqvist}, {Brinkmann}, {Brownstein}, {Brandt}, {Burtin}, {Dawson}, {de la Torre}, {Georgakakis}, {Gil-Mar{\'\i}n}, {Green}, {Hall}, {Kneib}, {LaMassa}, {Le Goff}, {MacLeod}, {Mariappan}, {McGreer}, {Merloni}, {Noterdaeme}, {Palanque-Delabrouille}, {Percival}, {Ross}, {Rossi}, {Schneider}, {Seo}, {Tojeiro}, {Weaver}, {Weijmans}, {Y{\`e}che}, {Zarrouk}, and {Zhao}]{2018A&A...613A..51P}
{P{\^a}ris}, I.; {Petitjean}, P.; {Aubourg}, {\'E}.; {Myers}, A.D.; {Streblyanska}, A.; {Lyke}, B.W.; {Anderson}, S.F.; {Armengaud}, {\'E}.; {Bautista}, J.; {Blanton}, M.R.;  et~al.
\newblock {The Sloan Digital Sky Survey Quasar Catalog: Fourteenth data release}.
\newblock {\em \aap} {\bf 2018}, {\em 613},~A51,  \href{http://xxx.lanl.gov/abs/1712.05029}{{\normalfont [arXiv:astro-ph.GA/1712.05029]}}.
\newblock {\url{https://doi.org/10.1051/0004-6361/201732445}}.

\bibitem[{Afanas'ev} et~al.(2003){Afanas'ev}, {Dodonov}, {Moiseev}, {Gorshkov}, {Konnikova}, and {Mingaliev}]{2003ARep...47..458A}
{Afanas'ev}, V.L.; {Dodonov}, S.N.; {Moiseev}, A.V.; {Gorshkov}, A.G.; {Konnikova}, V.K.; {Mingaliev}, M.G.
\newblock {Radio and Optical Spectral Studies of Radio Sources}.
\newblock {\em Astronomy Reports} {\bf 2003}, {\em 47},~458--466.
\newblock {\url{https://doi.org/10.1134/1.1583772}}.

\bibitem[{Verkhodanov} et~al.(2005){Verkhodanov}, {Trushkin}, {Andernach}, and {Chernenkov}]{2005BSAO...58..118V}
{Verkhodanov}, O.V.; {Trushkin}, S.A.; {Andernach}, H.; {Chernenkov}, V.N.
\newblock {Current status of the CATS database.}
\newblock {\em Bulletin of the Special Astrophysics Observatory} {\bf 2005}, {\em 58},~118--129,  \href{http://xxx.lanl.gov/abs/0705.2959}{{\normalfont [arXiv:astro-ph/0705.2959]}}.

\bibitem[{Verkhodanov} et~al.(2009){Verkhodanov}, {Trushkin}, {Andernach}, and {Chernenkov}]{2009DatSJ...8...34V}
{Verkhodanov}, O.V.; {Trushkin}, S.A.; {Andernach}, H.; {Chernenkov}, V.N.
\newblock {The CATS Service: An Astrophysical Research Tool}.
\newblock {\em Data Science Journal} {\bf 2009}, {\em 8},~34--40,  \href{http://xxx.lanl.gov/abs/0901.3118}{{\normalfont [arXiv:astro-ph.IM/0901.3118]}}.
\newblock {\url{https://doi.org/10.2481/dsj.8.34}}.

\bibitem[{Aller} et~al.(1992){Aller}, {Aller}, and {Hughes}]{1992ApJ...399...16A}
{Aller}, M.F.; {Aller}, H.D.; {Hughes}, P.A.
\newblock {Pearson-Readhead survey sources - Properties of the centimeter-wavelength flux and polarization of a complete radio sample}.
\newblock {\em \apj} {\bf 1992}, {\em 399},~16--28.
\newblock {\url{https://doi.org/10.1086/171898}}.

\bibitem[{Kraus} et~al.(2003){Kraus}, {Krichbaum}, {Wegner}, {Witzel}, {Cim{\`o}}, {Quirrenbach}, {Britzen}, {Fuhrmann}, {Lobanov}, {Naundorf}, {Otterbein}, {Peng}, {Risse}, {Ros}, and {Zensus}]{2003A&A...401..161K}
{Kraus}, A.; {Krichbaum}, T.P.; {Wegner}, R.; {Witzel}, A.; {Cim{\`o}}, G.; {Quirrenbach}, A.; {Britzen}, S.; {Fuhrmann}, L.; {Lobanov}, A.P.; {Naundorf}, C.E.;  et~al.
\newblock {Intraday variability in compact extragalactic radio sources. II. Observations with the Effelsberg 100 m radio telescope}.
\newblock {\em \aap} {\bf 2003}, {\em 401},~161--172.
\newblock {\url{https://doi.org/10.1051/0004-6361:20030118}}.

\bibitem[{Vaughan} et~al.(2003){Vaughan}, {Edelson}, {Warwick}, and {Uttley}]{2003MNRAS.345.1271V}
{Vaughan}, S.; {Edelson}, R.; {Warwick}, R.S.; {Uttley}, P.
\newblock {On characterizing the variability properties of X-ray light curves from active galaxies}.
\newblock {\em \mnras} {\bf 2003}, {\em 345},~1271--1284,  \href{http://xxx.lanl.gov/abs/astro-ph/0307420}{{\normalfont [arXiv:astro-ph/astro-ph/0307420]}}.
\newblock {\url{https://doi.org/10.1046/j.1365-2966.2003.07042.x}}.

\bibitem[{Tornikoski} et~al.(2000){Tornikoski}, {Lainela}, and {Valtaoja}]{2000AJ....120.2278T}
{Tornikoski}, M.; {Lainela}, M.; {Valtaoja}, E.
\newblock {The High Radio Frequency Spectra and Variability of Southern Flat-Spectrum Radio Sources}.
\newblock {\em \aj} {\bf 2000}, {\em 120},~2278--2283.
\newblock {\url{https://doi.org/10.1086/316809}}.

\bibitem[{Nieppola} et~al.(2007){Nieppola}, {Tornikoski}, {L{\"a}hteenm{\"a}ki}, {Valtaoja}, {Hakala}, {Hovatta}, {Kotiranta}, {Nummila}, {Ojala}, {Parviainen}, {Ranta}, {Saloranta}, {Torniainen}, and {Tr{\"o}ller}]{2007AJ....133.1947N}
{Nieppola}, E.; {Tornikoski}, M.; {L{\"a}hteenm{\"a}ki}, A.; {Valtaoja}, E.; {Hakala}, T.; {Hovatta}, T.; {Kotiranta}, M.; {Nummila}, S.; {Ojala}, T.; {Parviainen}, M.;  et~al.
\newblock {37 GHz Observations of a Large Sample of BL Lacertae Objects}.
\newblock {\em \aj} {\bf 2007}, {\em 133},~1947--1953,  \href{http://xxx.lanl.gov/abs/0705.0887}{{\normalfont [0705.0887]}}.
\newblock {\url{https://doi.org/10.1086/512609}}.

\bibitem[{Khabibullina} et~al.(2024){Khabibullina}, {Mikhailov}, {Sotnikova}, {Mufakharov}, {Mingaliev}, {Kudryashova}, {Bursov}, {Stolyarov}, and {Udovitskij}]{2024arXiv240202283K}
{Khabibullina}, M.; {Mikhailov}, A.; {Sotnikova}, Y.; {Mufakharov}, T.; {Mingaliev}, M.; {Kudryashova}, A.; {Bursov}, N.; {Stolyarov}, V.; {Udovitskij}, R.
\newblock {Radio properties of high-redshift galaxies at $z \geq 1$}.
\newblock {\em arXiv e-prints} {\bf 2024}, p. arXiv:2402.02283,  \href{http://xxx.lanl.gov/abs/2402.02283}{{\normalfont [arXiv:astro-ph.GA/2402.02283]}}.

\bibitem[{Morgan} et~al.(2018){Morgan}, {Macquart}, {Ekers}, {Chhetri}, {Tokumaru}, {Manoharan}, {Tremblay}, {Bisi}, and {Jackson}]{2018MNRAS.473.2965M}
{Morgan}, J.S.; {Macquart}, J.P.; {Ekers}, R.; {Chhetri}, R.; {Tokumaru}, M.; {Manoharan}, P.K.; {Tremblay}, S.; {Bisi}, M.M.; {Jackson}, B.V.
\newblock {Interplanetary Scintillation with the Murchison Widefield Array I: a sub-arcsecond survey over 900 deg$^{2}$ at 79 and 158 MHz}.
\newblock {\em \mnras} {\bf 2018}, {\em 473},~2965--2983,  \href{http://xxx.lanl.gov/abs/1709.00312}{{\normalfont [arXiv:astro-ph.IM/1709.00312]}}.
\newblock {\url{https://doi.org/10.1093/mnras/stx2284}}.

\bibitem[{Narayan}(1992)]{1992RSPTA.341..151N}
{Narayan}, R.
\newblock {The Physics of Pulsar Scintillation}.
\newblock {\em Philosophical Transactions of the Royal Society of London Series A} {\bf 1992}, {\em 341},~151--165.
\newblock {\url{https://doi.org/10.1098/rsta.1992.0090}}.

\bibitem[{Kellermann} and {Owen}(1988)]{1988gera.book..563K}
{Kellermann}, K.I.; {Owen}, F.N.
\newblock {Radio galaxies and quasars.} In {\em Galactic and Extragalactic Radio Astronomy}; {Kellermann}, K.I.; {Verschuur}, G.L., Eds.;  1988; pp. 563--602.

\bibitem[{Krezinger} et~al.(2022){Krezinger}, {Perger}, {Gab{\'a}nyi}, {Frey}, {Gurvits}, {Paragi}, {An}, {Zhang}, {Cao}, and {Sbarrato}]{2022ApJS..260...49K}
{Krezinger}, M.; {Perger}, K.; {Gab{\'a}nyi}, K.{\'E}.; {Frey}, S.; {Gurvits}, L.I.; {Paragi}, Z.; {An}, T.; {Zhang}, Y.; {Cao}, H.; {Sbarrato}, T.
\newblock {Radio-loud Quasars above Redshift 4: Very Long Baseline Interferometry (VLBI) Imaging of an Extended Sample}.
\newblock {\em \apjs} {\bf 2022}, {\em 260},~49,  \href{http://xxx.lanl.gov/abs/2204.02114}{{\normalfont [arXiv:astro-ph.GA/2204.02114]}}.
\newblock {\url{https://doi.org/10.3847/1538-4365/ac63b8}}.

\bibitem[{Frey} et~al.(2015){Frey}, {Paragi}, {Fogasy}, and {Gurvits}]{2015MNRAS.446.2921F}
{Frey}, S.; {Paragi}, Z.; {Fogasy}, J.O.; {Gurvits}, L.I.
\newblock {The first estimate of radio jet proper motion at z > 5}.
\newblock {\em \mnras} {\bf 2015}, {\em 446},~2921--2928,  \href{http://xxx.lanl.gov/abs/1410.8101}{{\normalfont [arXiv:astro-ph.GA/1410.8101]}}.
\newblock {\url{https://doi.org/10.1093/mnras/stu2294}}.

\bibitem[{Sbarrato} et~al.(2012){Sbarrato}, {Ghisellini}, {Nardini}, {Tagliaferri}, {Foschini}, {Ghirlanda}, {Tavecchio}, {Greiner}, {Rau}, and {Gehrels}]{2012MNRAS.426L..91S}
{Sbarrato}, T.; {Ghisellini}, G.; {Nardini}, M.; {Tagliaferri}, G.; {Foschini}, L.; {Ghirlanda}, G.; {Tavecchio}, F.; {Greiner}, J.; {Rau}, A.; {Gehrels}, N.
\newblock {SDSS J102623.61+254259.5: the second most distant blazar at z = 5.3}.
\newblock {\em \mnras} {\bf 2012}, {\em 426},~L91--L95,  \href{http://xxx.lanl.gov/abs/1208.3467}{{\normalfont [arXiv:astro-ph.CO/1208.3467]}}.
\newblock {\url{https://doi.org/10.1111/j.1745-3933.2012.01332.x}}.

\bibitem[{Zhang} et~al.(2020){Zhang}, {An}, and {Frey}]{2020SciBu..65..525Z}
{Zhang}, Y.; {An}, T.; {Frey}, S.
\newblock {Fast jet proper motion discovered in a blazar at z=4.72}.
\newblock {\em Science Bulletin} {\bf 2020}, {\em 65},~525--530,  \href{http://xxx.lanl.gov/abs/1912.12597}{{\normalfont [arXiv:astro-ph.HE/1912.12597]}}.
\newblock {\url{https://doi.org/10.1016/j.scib.2020.01.008}}.

\bibitem[{Hook} et~al.(2002){Hook}, {McMahon}, {Shaver}, and {Snellen}]{2002A&A...391..509H}
{Hook}, I.M.; {McMahon}, R.G.; {Shaver}, P.A.; {Snellen}, I.A.G.
\newblock {Discovery of radio-loud quasars with redshifts above 4 from the PMN sample}.
\newblock {\em \aap} {\bf 2002}, {\em 391},~509--517,  \href{http://xxx.lanl.gov/abs/astro-ph/0207101}{{\normalfont [arXiv:astro-ph/astro-ph/0207101]}}.
\newblock {\url{https://doi.org/10.1051/0004-6361:20020869}}.

\bibitem[{Hook} and {McMahon}(1998)]{1998MNRAS.294L...7H}
{Hook}, I.M.; {McMahon}, R.G.
\newblock {Discovery of radio-loud quasars with z=4.72 and z=4.010}.
\newblock {\em \mnras} {\bf 1998}, {\em 294},~L7--L12,  \href{http://xxx.lanl.gov/abs/astro-ph/9801026}{{\normalfont [arXiv:astro-ph/astro-ph/9801026]}}.
\newblock {\url{https://doi.org/10.1046/j.1365-8711.1998.01368.x10.1111/j.1365-8711.1998.01368.x}}.

\bibitem[{Celotti} and {Ghisellini}(2008)]{2008MNRAS.385..283C}
{Celotti}, A.; {Ghisellini}, G.
\newblock {The power of blazar jets}.
\newblock {\em \mnras} {\bf 2008}, {\em 385},~283--300,  \href{http://xxx.lanl.gov/abs/0711.4112}{{\normalfont [arXiv:astro-ph/0711.4112]}}.
\newblock {\url{https://doi.org/10.1111/j.1365-2966.2007.12758.x}}.

\bibitem[{Zickgraf} et~al.(1997){Zickgraf}, {Voges}, {Krautter}, {Thiering}, {Appenzeller}, {Mujica}, and {Serrano}]{1997A&A...323L..21Z}
{Zickgraf}, F.J.; {Voges}, W.; {Krautter}, J.; {Thiering}, I.; {Appenzeller}, I.; {Mujica}, R.; {Serrano}, A.
\newblock {Identification of a complete sample of northern ROSAT All-Sky Survey X-ray sources. V. Discovery of a z=4.28 QSO near the RASS source RX J1028.6-0844.}
\newblock {\em \aap} {\bf 1997}, {\em 323},~L21--L24.

\bibitem[{Fiedler} et~al.(1987){Fiedler}, {Waltman}, {Spencer}, {Johnston}, {Angerhofer}, {Florkowski}, {Josties}, {Klepczynski}, {McCarthy}, and {Matsakis}]{1987ApJS...65..319F}
{Fiedler}, R.L.; {Waltman}, E.B.; {Spencer}, J.H.; {Johnston}, K.J.; {Angerhofer}, P.E.; {Florkowski}, D.R.; {Josties}, F.J.; {Klepczynski}, W.J.; {McCarthy}, D.D.; {Matsakis}, D.N.
\newblock {Daily Observations of Compact Extragalactic Radio Sources at 2695 and 8085 MHz, 1979--1985}.
\newblock {\em \apjs} {\bf 1987}, {\em 65},~319.
\newblock {\url{https://doi.org/10.1086/191228}}.

\bibitem[{Lazio} et~al.(2001){Lazio}, {Waltman}, {Ghigo}, {Fiedler}, {Foster}, and {Johnston}]{2001ApJS..136..265L}
{Lazio}, T.J.W.; {Waltman}, E.B.; {Ghigo}, F.D.; {Fiedler}, R.L.; {Foster}, R.S.; {Johnston}, K.J.
\newblock {A Dual-Frequency, Multiyear Monitoring Program of Compact Radio Sources}.
\newblock {\em \apjs} {\bf 2001}, {\em 136},~265--392,  \href{http://xxx.lanl.gov/abs/astro-ph/0105433}{{\normalfont [arXiv:astro-ph/astro-ph/0105433]}}.
\newblock {\url{https://doi.org/10.1086/322531}}.

\bibitem[{Vagnetti} et~al.(2011){Vagnetti}, {Turriziani}, and {Trevese}]{2011A&A...536A..84V}
{Vagnetti}, F.; {Turriziani}, S.; {Trevese}, D.
\newblock {Ensemble X-ray variability of active galactic nuclei from serendipitous source catalogues}.
\newblock {\em \aap} {\bf 2011}, {\em 536},~A84,  \href{http://xxx.lanl.gov/abs/1110.4768}{{\normalfont [arXiv:astro-ph.CO/1110.4768]}}.
\newblock {\url{https://doi.org/10.1051/0004-6361/201118072}}.

\bibitem[{Vagnetti} et~al.(2016){Vagnetti}, {Middei}, {Antonucci}, {Paolillo}, and {Serafinelli}]{2016A&A...593A..55V}
{Vagnetti}, F.; {Middei}, R.; {Antonucci}, M.; {Paolillo}, M.; {Serafinelli}, R.
\newblock {Ensemble X-ray variability of active galactic nuclei. II. Excess variance and updated structure function}.
\newblock {\em \aap} {\bf 2016}, {\em 593},~A55,  \href{http://xxx.lanl.gov/abs/1607.02629}{{\normalfont [arXiv:astro-ph.GA/1607.02629]}}.
\newblock {\url{https://doi.org/10.1051/0004-6361/201629057}}.

\bibitem[Arthur and Vassilvitskii(2007)]{1283494}
Arthur, D.; Vassilvitskii, S.
\newblock k-means++: the advantages of careful seeding.
\newblock In Proceedings of the SODA '07: Proceedings of the eighteenth annual ACM-SIAM symposium on Discrete algorithms, Philadelphia, PA, USA,  2007; pp. 1027--1035.

\bibitem[{Kohonen}(2001)]{2001som..book.....K}
{Kohonen}, T.
\newblock {\em {Self-Organizing Maps}};  2001.

\bibitem[{Torniainen} et~al.(2005){Torniainen}, {Tornikoski}, {Ter{\"a}sranta}, {Aller}, and {Aller}]{2005A&A...435..839T}
{Torniainen}, I.; {Tornikoski}, M.; {Ter{\"a}sranta}, H.; {Aller}, M.F.; {Aller}, H.D.
\newblock {Long term variability of gigahertz-peaked spectrum sources and candidates}.
\newblock {\em \aap} {\bf 2005}, {\em 435},~839--856.
\newblock {\url{https://doi.org/10.1051/0004-6361:20041886}}.

\bibitem[{Torniainen} et~al.(2007){Torniainen}, {Tornikoski}, {L{\"a}hteenm{\"a}ki}, {Aller}, {Aller}, and {Mingaliev}]{2007A&A...469..451T}
{Torniainen}, I.; {Tornikoski}, M.; {L{\"a}hteenm{\"a}ki}, A.; {Aller}, M.F.; {Aller}, H.D.; {Mingaliev}, M.G.
\newblock {Radio continuum spectra of gigahertz-peaked spectrum galaxies}.
\newblock {\em \aap} {\bf 2007}, {\em 469},~451--457.
\newblock {\url{https://doi.org/10.1051/0004-6361:20066892}}.

\bibitem[{Snellen} et~al.(1998){Snellen}, {Schilizzi}, {de Bruyn}, {Miley}, {Rengelink}, {Roettgering}, and {Bremer}]{1998A&AS..131..435S}
{Snellen}, I.A.G.; {Schilizzi}, R.T.; {de Bruyn}, A.G.; {Miley}, G.K.; {Rengelink}, R.B.; {Roettgering}, H.J.; {Bremer}, M.N.
\newblock {A new sample of faint Gigahertz Peaked Spectrum radio sources}.
\newblock {\em \aaps} {\bf 1998}, {\em 131},~435--449,  \href{http://xxx.lanl.gov/abs/arXiv:astro-ph/9803140}{{\normalfont [arXiv:astro-ph/9803140]}}.
\newblock {\url{https://doi.org/10.1051/aas:1998281}}.

\bibitem[{Bicknell} et~al.(1997){Bicknell}, {Dopita}, and {O'Dea}]{1997ApJ...485..112B}
{Bicknell}, G.V.; {Dopita}, M.A.; {O'Dea}, C.P.
\newblock {Unification of the Radio and Optical Properties of Gigahertz Peak Spectrum and Compact Steep-Spectrum Radio Sources}.
\newblock {\em \aj} {\bf 1997}, {\em 485},~112--124.
\newblock {\url{https://doi.org/10.1086/304400}}.

\bibitem[{Fanti} et~al.(1995){Fanti}, {Fanti}, {Dallacasa}, {Schilizzi}, {Spencer}, and {Stanghellini}]{1995A&A...302..317F}
{Fanti}, C.; {Fanti}, R.; {Dallacasa}, D.; {Schilizzi}, R.T.; {Spencer}, R.E.; {Stanghellini}, C.
\newblock {Are compact steep-spectrum sources young?}
\newblock {\em \aap} {\bf 1995}, {\em 302},~317.

\bibitem[{van Breugel} et~al.(1984){van Breugel}, {Miley}, and {Heckman}]{1984AJ.....89....5V}
{van Breugel}, W.; {Miley}, G.; {Heckman}, T.
\newblock {Studies of kiloparsec-scale, steep-spectrum radio cores. I. VLA maps.}
\newblock {\em \aj} {\bf 1984}, {\em 89},~5--22.
\newblock {\url{https://doi.org/10.1086/113480}}.

\bibitem[{Bai} and {Lee}(2005)]{2005JKAS...38..125B}
{Bai}, J.M.; {Lee}, M.G.
\newblock {GPS Quasars as Special Blazars}.
\newblock {\em Journal of Korean Astronomical Society} {\bf 2005}, {\em 38},~125--+.

\bibitem[{Labiano} et~al.(2006){Labiano}, {Vermeulen}, {Barthel}, {O'Dea}, {Gallimore}, {Baum}, and {de Vries}]{2006A&A...447..481L}
{Labiano}, A.; {Vermeulen}, R.C.; {Barthel}, P.D.; {O'Dea}, C.P.; {Gallimore}, J.F.; {Baum}, S.; {de Vries}, W.
\newblock {H I absorption in <ASTROBJ>3C 49</ASTROBJ> and <ASTROBJ>3C 268.3</ASTROBJ>. Probing the environment of compact steep spectrum and GHz peaked spectrum sources}.
\newblock {\em \aap} {\bf 2006}, {\em 447},~481--487,  \href{http://xxx.lanl.gov/abs/astro-ph/0510563}{{\normalfont [arXiv:astro-ph/astro-ph/0510563]}}.
\newblock {\url{https://doi.org/10.1051/0004-6361:20053856}}.

\bibitem[{Orienti} et~al.(2007){Orienti}, {Dallacasa}, and {Stanghellini}]{2007A&A...461..923O}
{Orienti}, M.; {Dallacasa}, D.; {Stanghellini}, C.
\newblock {Constraining the spectral age of very asymmetric CSOs. Evidence of the influence of the ambient medium}.
\newblock {\em \aap} {\bf 2007}, {\em 461},~923--929,  \href{http://xxx.lanl.gov/abs/astro-ph/0610359}{{\normalfont [arXiv:astro-ph/astro-ph/0610359]}}.
\newblock {\url{https://doi.org/10.1051/0004-6361:20066122}}.

\bibitem[{Dicken} et~al.(2012){Dicken}, {Tadhunter}, {Axon}, {Morganti}, {Robinson}, {Kouwenhoven}, {Spoon}, {Kharb}, {Inskip}, {Holt}, {Ramos Almeida}, and {Nesvadba}]{2012ApJ...745..172D}
{Dicken}, D.; {Tadhunter}, C.; {Axon}, D.; {Morganti}, R.; {Robinson}, A.; {Kouwenhoven}, M.B.N.; {Spoon}, H.; {Kharb}, P.; {Inskip}, K.J.; {Holt}, J.;  et~al.
\newblock {Spitzer Mid-IR Spectroscopy of Powerful 2 Jy and 3CRR Radio Galaxies. I. Evidence against a Strong Starburst-AGN Connection in Radio-loud AGN}.
\newblock {\em \apj} {\bf 2012}, {\em 745},~172,  \href{http://xxx.lanl.gov/abs/1111.4476}{{\normalfont [arXiv:astro-ph.CO/1111.4476]}}.
\newblock {\url{https://doi.org/10.1088/0004-637X/745/2/172}}.

\bibitem[{Baum} et~al.(1990){Baum}, {O'Dea}, {Murphy}, and {de~Bruyn}]{1990A&A...232...19B}
{Baum}, S.A.; {O'Dea}, C.P.; {Murphy}, D.W.; {de~Bruyn}, A.G.
\newblock {0108+388 - A compact double source with surprising properties}.
\newblock {\em \aap} {\bf 1990}, {\em 232},~19--26.

\bibitem[{Reynolds} and {Begelman}(1997)]{1997ApJ...487L.135R}
{Reynolds}, C.S.; {Begelman}, M.C.
\newblock {Intermittant Radio Galaxies and Source Statistics}.
\newblock {\em \apjl} {\bf 1997}, {\em 487},~L135--L138,  \href{http://xxx.lanl.gov/abs/astro-ph/9707221}{{\normalfont [arXiv:astro-ph/astro-ph/9707221]}}.
\newblock {\url{https://doi.org/10.1086/310894}}.

\bibitem[{Stanghellini} et~al.(2005){Stanghellini}, {O'Dea}, {Dallacasa}, {Cassaro}, {Baum}, {Fanti}, and {Fanti}]{2005A&A...443..891S}
{Stanghellini}, C.; {O'Dea}, C.P.; {Dallacasa}, D.; {Cassaro}, P.; {Baum}, S.A.; {Fanti}, R.; {Fanti}, C.
\newblock {Extended emission around GPS radio sources}.
\newblock {\em \aap} {\bf 2005}, {\em 443},~891--902,  \href{http://xxx.lanl.gov/abs/astro-ph/0507499}{{\normalfont [arXiv:astro-ph/astro-ph/0507499]}}.
\newblock {\url{https://doi.org/10.1051/0004-6361:20042226}}.

\bibitem[{Readhead} et~al.(2023){Readhead}, {Ravi}, {Blandford}, {Sullivan}, {Somalwar}, {Liodakis}, {Lister}, {Taylor}, {Wilkinson}, {Globus}, {Kiehlmann}, {Lawrence}, {O'Neill}, {Pavlidou}, {Pearson}, {Sheldahl}, {Siemiginowska}, and {Tassis}]{2023arXiv230311361R}
{Readhead}, A.C.S.; {Ravi}, V.; {Blandford}, R.D.; {Sullivan}, A.G.; {Somalwar}, J.; {Liodakis}, I.; {Lister}, M.L.; {Taylor}, G.B.; {Wilkinson}, P.N.; {Globus}, N.;  et~al.
\newblock {The Evolution of Compact Symmetric Objects -- A Possible Connection with Tidal Disruption Events}.
\newblock {\em arXiv e-prints} {\bf 2023}, p. arXiv:2303.11361,  \href{http://xxx.lanl.gov/abs/2303.11361}{{\normalfont [arXiv:astro-ph.HE/2303.11361]}}.
\newblock {\url{https://doi.org/10.48550/arXiv.2303.11361}}.

\bibitem[{Torniainen}(2008)]{2008PhDT.......182T}
{Torniainen}, I.
\newblock {Multifrequency studies of gigahertz-peaked spectrum sources and candidates}.
\newblock PhD thesis, Helsinki University of Technology, Finland,  2008.

\bibitem[{O'Dea} et~al.(1990){O'Dea}, {Baum}, and {Morris}]{1990A&AS...82..261O}
{O'Dea}, C.P.; {Baum}, S.A.; {Morris}, G.B.
\newblock {CCD observations of gigahertz-peaked-spectrum radio sources}.
\newblock {\em \aaps} {\bf 1990}, {\em 82},~261--272.

\bibitem[{de Vries} et~al.(1997){de Vries}, {Barthel}, and {O'Dea}]{1997A&A...321..105D}
{de Vries}, W.H.; {Barthel}, P.D.; {O'Dea}, C.P.
\newblock {Radio spectra of Gigahertz Peaked Spectrum radio sources}.
\newblock {\em \aap} {\bf 1997}, {\em 321},~105--110.

\bibitem[{Edwards} and {Tingay}(2004)]{2004A&A...424...91E}
{Edwards}, P.G.; {Tingay}, S.J.
\newblock {New candidate GHz peaked spectrum and compact steep spectrum sources}.
\newblock {\em \aap} {\bf 2004}, {\em 424},~91--106,  \href{http://xxx.lanl.gov/abs/arXiv:astro-ph/0407010}{{\normalfont [arXiv:astro-ph/0407010]}}.
\newblock {\url{https://doi.org/10.1051/0004-6361:20035749}}.

\bibitem[{Dallacasa} et~al.(2002){Dallacasa}, {Stanghellini}, {Centonza}, and {Furnari}]{2002NewAR..46..299D}
{Dallacasa}, D.; {Stanghellini}, C.; {Centonza}, M.; {Furnari}, G.
\newblock {High frequency peakers}.
\newblock {\em \nar} {\bf 2002}, {\em 46},~299--302.
\newblock {\url{https://doi.org/10.1016/S1387-6473(01)00198-1}}.

\bibitem[{Orienti}(2009)]{2009AN....330..167O}
{Orienti}, M.
\newblock {High frequency peakers}.
\newblock {\em Astronomische Nachrichten} {\bf 2009}, {\em 330},~167.
\newblock {\url{https://doi.org/10.1002/asna.200811147}}.

\bibitem[{Orienti} et~al.(2007){Orienti}, {Dallacasa}, and {Stanghellini}]{2007A&A...475..813O}
{Orienti}, M.; {Dallacasa}, D.; {Stanghellini}, C.
\newblock {Constraining the nature of high frequency peakers. The spectral variability}.
\newblock {\em \aap} {\bf 2007}, {\em 475},~813--820,  \href{http://xxx.lanl.gov/abs/0708.3979}{{\normalfont [arXiv:astro-ph/0708.3979]}}.
\newblock {\url{https://doi.org/10.1051/0004-6361:20078105}}.

\bibitem[{Orienti} and {Dallacasa}(2008)]{2008A&A...479..409O}
{Orienti}, M.; {Dallacasa}, D.
\newblock {Constraining the nature of high frequency peakers. II. Polarization properties}.
\newblock {\em \aap} {\bf 2008}, {\em 479},~409--415,  \href{http://xxx.lanl.gov/abs/0712.3207}{{\normalfont [arXiv:astro-ph/0712.3207]}}.
\newblock {\url{https://doi.org/10.1051/0004-6361:20078572}}.

\bibitem[{Paragi} et~al.(1999){Paragi}, {Frey}, {Gurvits}, {Kellermann}, {Schilizzi}, {McMahon}, {Hook}, and {Pauliny-Toth}]{1999A&A...344...51P}
{Paragi}, Z.; {Frey}, S.; {Gurvits}, L.I.; {Kellermann}, K.I.; {Schilizzi}, R.T.; {McMahon}, R.G.; {Hook}, I.M.; {Pauliny-Toth}, I.I.K.
\newblock {VLBI imaging of extremely high redshift quasars at 5 GHz}.
\newblock {\em \aap} {\bf 1999}, {\em 344},~51--60,  \href{http://xxx.lanl.gov/abs/astro-ph/9901396}{{\normalfont [arXiv:astro-ph/astro-ph/9901396]}}.

\bibitem[{Zhang} et~al.(2022){Zhang}, {An}, {Frey}, {Gab{\'a}nyi}, and {Sotnikova}]{2022ApJ...937...19Z}
{Zhang}, Y.; {An}, T.; {Frey}, S.; {Gab{\'a}nyi}, K.{\'E}.; {Sotnikova}, Y.
\newblock {Radio Jet Proper-motion Analysis of Nine Distant Quasars above Redshift 3.5}.
\newblock {\em \apj} {\bf 2022}, {\em 937},~19,  \href{http://xxx.lanl.gov/abs/2209.10760}{{\normalfont [arXiv:astro-ph.HE/2209.10760]}}.
\newblock {\url{https://doi.org/10.3847/1538-4357/ac87f8}}.

\bibitem[{Plavin} et~al.(2022){Plavin}, {Kovalev}, and {Pushkarev}]{2022ApJS..260....4P}
{Plavin}, A.V.; {Kovalev}, Y.Y.; {Pushkarev}, A.B.
\newblock {Direction of Parsec-scale Jets for 9220 Active Galactic Nuclei}.
\newblock {\em \apjs} {\bf 2022}, {\em 260},~4,  \href{http://xxx.lanl.gov/abs/2203.13750}{{\normalfont [arXiv:astro-ph.HE/2203.13750]}}.
\newblock {\url{https://doi.org/10.3847/1538-4365/ac6352}}.

\bibitem[{Gobeille} et~al.(2014){Gobeille}, {Wardle}, and {Cheung}]{2014arXiv1406.4797G}
{Gobeille}, D.B.; {Wardle}, J.F.C.; {Cheung}, C.C.
\newblock {VLA Observations of a Complete Sample of Radio Loud Quasars between redshifts 2.5 and 5.28: I. high-redshift sample summary and the radio images}.
\newblock {\em arXiv e-prints} {\bf 2014}, p. arXiv:1406.4797,  \href{http://xxx.lanl.gov/abs/1406.4797}{{\normalfont [arXiv:astro-ph.GA/1406.4797]}}.

\bibitem[{An} et~al.(2022){An}, {Wang}, {Zhang}, {Aditya}, {Hong}, and {Cui}]{2022MNRAS.511.4572A}
{An}, T.; {Wang}, A.; {Zhang}, Y.; {Aditya}, J.N.H.S.; {Hong}, X.; {Cui}, L.
\newblock {A compact symmetric radio source born at one-tenth the current age of the Universe}.
\newblock {\em \mnras} {\bf 2022}, {\em 511},~4572--4581,  \href{http://xxx.lanl.gov/abs/2109.10599}{{\normalfont [arXiv:astro-ph.GA/2109.10599]}}.
\newblock {\url{https://doi.org/10.1093/mnras/stac205}}.

\bibitem[{Garrett} et~al.(1994){Garrett}, {Muxlow}, {Patnaik}, and {Walsh}]{1994MNRAS.269..902G}
{Garrett}, M.A.; {Muxlow}, T.W.B.; {Patnaik}, A.R.; {Walsh}, D.
\newblock {MERLIN observations of the gravitational lens system 2016+112.}
\newblock {\em \mnras} {\bf 1994}, {\em 269},~902--906.
\newblock {\url{https://doi.org/10.1093/mnras/269.4.902}}.

\bibitem[{Cheng} et~al.(2023){Cheng}, {An}, {Wang}, and {Jaiswal}]{2023Galax..11...42C}
{Cheng}, X.; {An}, T.; {Wang}, A.; {Jaiswal}, S.
\newblock {High-Frequency and High-Resolution VLBI Observations of GHz Peaked Spectrum Objects}.
\newblock {\em Galaxies} {\bf 2023}, {\em 11},~42,  \href{http://xxx.lanl.gov/abs/2303.03066}{{\normalfont [arXiv:astro-ph.HE/2303.03066]}}.
\newblock {\url{https://doi.org/10.3390/galaxies11020042}}.

\bibitem[{Orienti} and {Dallacasa}(2014)]{2014MNRAS.438..463O}
{Orienti}, M.; {Dallacasa}, D.
\newblock {Physical properties of young radio sources: VLBA observations of high-frequency peaking radio sources}.
\newblock {\em \mnras} {\bf 2014}, {\em 438},~463--475,  \href{http://xxx.lanl.gov/abs/1311.2999}{{\normalfont [arXiv:astro-ph.CO/1311.2999]}}.
\newblock {\url{https://doi.org/10.1093/mnras/stt2217}}.

\bibitem[{Callingham} et~al.(2017){Callingham}, {Ekers}, {Gaensler}, {Line}, {Hurley-Walker}, {Sadler}, {Tingay}, {Hancock}, {Bell}, {Dwarakanath}, {For}, {Franzen}, {Hindson}, {Johnston-Hollitt}, {Kapi{\'n}ska}, {Lenc}, {McKinley}, {Morgan}, {Offringa}, {Procopio}, {Staveley-Smith}, {Wayth}, {Wu}, and {Zheng}]{2017ApJ...836..174C}
{Callingham}, J.R.; {Ekers}, R.D.; {Gaensler}, B.M.; {Line}, J.L.B.; {Hurley-Walker}, N.; {Sadler}, E.M.; {Tingay}, S.J.; {Hancock}, P.J.; {Bell}, M.E.; {Dwarakanath}, K.S.;  et~al.
\newblock {Extragalactic Peaked-spectrum Radio Sources at Low Frequencies}.
\newblock {\em \apj} {\bf 2017}, {\em 836},~174,  \href{http://xxx.lanl.gov/abs/1701.02771}{{\normalfont [arXiv:astro-ph.GA/1701.02771]}}.
\newblock {\url{https://doi.org/10.3847/1538-4357/836/2/174}}.

\bibitem[{O'Dea} et~al.(1990){O'Dea}, {Baum}, {Stanghellini}, {Morris}, {Patnaik}, and {Gopal-Krishna}]{1990A&AS...84..549O}
{O'Dea}, C.P.; {Baum}, S.A.; {Stanghellini}, C.; {Morris}, G.B.; {Patnaik}, A.R.; {Gopal-Krishna}.
\newblock {Multifrequency VLA observations of GHz-peaked-spectrum radio cores}.
\newblock {\em \aaps} {\bf 1990}, {\em 84},~549--562.

\bibitem[{O'Dea} et~al.(1986){O'Dea}, {Dent}, {Kinzel}, and {Balonek}]{1986AJ.....92.1262O}
{O'Dea}, C.P.; {Dent}, W.A.; {Kinzel}, W.M.; {Balonek}, T.J.
\newblock {Multifrequency radio observations of the variable quasars 0133+476, 0235+164, 1749+096, and 2131-021.}
\newblock {\em \aj} {\bf 1986}, {\em 92},~1262--1271.
\newblock {\url{https://doi.org/10.1086/114260}}.

\bibitem[{Planck Collaboration} et~al.(2011){Planck Collaboration}, {Ade}, {Aghanim}, {Angelakis}, {Arnaud}, {Ashdown}, {Aumont}, {Baccigalupi}, {Balbi}, {Banday}, {Barreiro}, {Bartlett}, {Battaner}, {Benabed}, {Beno{\^\i}t}, {Bernard}, {Bersanelli}, {Bhatia}, {Bonaldi}, {Bonavera}, {Bond}, {Borrill}, {Bouchet}, {Bucher}, {Burigana}, {Cabella}, {Cappellini}, {Cardoso}, {Catalano}, {Cay{\'o}n}, {Challinor}, {Chamballu}, {Chary}, {Chen}, {Chiang}, {Christensen}, {Clements}, {Colombi}, {Couchot}, {Coulais}, {Crill}, {Cuttaia}, {Danese}, {Davies}, {Davis}, {de Bernardis}, {de Gasperis}, {de Rosa}, {de Zotti}, {Delabrouille}, {Delouis}, {D{\'e}sert}, {Dickinson}, {Donzelli}, {Dor{\'e}}, {D{\"o}rl}, {Douspis}, {Dupac}, {Efstathiou}, {En{\ss}lin}, {Finelli}, {Forni}, {Frailis}, {Franceschi}, {Fuhrmann}, {Galeotta}, {Ganga}, {Giard}, {Giardino}, {Giraud-H{\'e}raud}, {Gonz{\'a}lez-Nuevo}, {G{\'o}rski}, {Gratton}, {Gregorio}, {Gruppuso}, {Harrison}, {Henrot-Versill{\'e}}, {Herranz}, {Hildebrandt}, {Hivon}, {Hobson},
  {Holmes}, {Hovest}, {Hoyland}, {Huffenberger}, {Huynh}, {Jaffe}, {Juvela}, {Keih{\"a}nen}, {Keskitalo}, {Kisner}, {Kneissl}, {Knox}, {Krichbaum}, {Kurki-Suonio}, {Lagache}, {L{\"a}hteenm{\"a}ki}, {Lamarre}, {Lasenby}, {Laureijs}, {Lavonen}, {Lawrence}, {Leach}, {Leahy}, {Leonardi}, {Le{\'o}n-Tavares}, {Linden-V{\o}rnle}, {L{\'o}pez-Caniego}, {Lubin}, {Mac{\'\i}as-P{\'e}rez}, {Maffei}, {Maino}, {Mandolesi}, {Mann}, {Maris}, {Marleau}, {Mart{\'\i}nez-Gonz{\'a}lez}, {Masi}, {Massardi}, {Matarrese}, {Matthai}, {Mazzotta}, {Meinhold}, {Melchiorri}, {Mendes}, {Mennella}, {Mingaliev}, {Miville-Desch{\^e}nes}, {Moneti}, {Montier}, {Morgante}, {Mortlock}, {Munshi}, {Murphy}, {Naselsky}, {Natoli}, {Nestoras}, {Netterfield}, {Nieppola}, {N{\o}rgaard-Nielsen}, {Noviello}, {Novikov}, {Novikov}, {Osborne}, {Pajot}, {Paladini}, {Partridge}, {Pasian}, {Patanchon}, {Pearson}, {Perdereau}, {Perotto}, {Perrotta}, {Piacentini}, {Piat}, {Pierpaoli}, {Plaszczynski}, {Platania}, {Pointecouteau}, {Polenta}, {Ponthieu}, {Poutanen},
  {Pr{\'e}zeau}, {Procopio}, {Prunet}, {Puget}, {Rachen}, {Reach}, {Rebolo}, {Reinecke}, {Renault}, {Ricciardi}, {Riller}, {Riquelme}, {Ristorcelli}, {Rocha}, {Rosset}, {Rowan-Robinson}, {Rubi{\~n}o-Mart{\'\i}n}, {Rusholme}, {Sajina}, {Sandri}, {Savolainen}, {Scott}, {Seiffert}, {Sievers}, {Smoot}, {Sotnikova}, {Starck}, {Stivoli}, {Stolyarov}, {Sudiwala}, {Sygnet}, {Tammi}, {Tauber}, {Terenzi}, {Toffolatti}, {Tomasi}, {Tornikoski}, {Torre}, {Tristram}, {Tuovinen}, {T{\"u}rler}, {Turunen}, {Umana}, {Ungerechts}, {Valenziano}, {Varis}, {Vielva}, {Villa}, {Vittorio}, {Wade}, {Wandelt}, {Wilkinson}, {Yvon}, {Zacchei}, {Zensus}, and {Zonca}]{2011A&A...536A..14P}
{Planck Collaboration}.; {Ade}, P.A.R.; {Aghanim}, N.; {Angelakis}, E.; {Arnaud}, M.; {Ashdown}, M.; {Aumont}, J.; {Baccigalupi}, C.; {Balbi}, A.; {Banday}, A.J.;  et~al.
\newblock {Planck early results. XIV. ERCSC validation and extreme radio sources}.
\newblock {\em \aap} {\bf 2011}, {\em 536},~A14,  \href{http://xxx.lanl.gov/abs/1101.1721}{{\normalfont [arXiv:astro-ph.CO/1101.1721]}}.
\newblock {\url{https://doi.org/10.1051/0004-6361/201116475}}.

\bibitem[{O'Dea}(1990)]{1990MNRAS.245P..20O}
{O'Dea}, C.P.
\newblock {Do quasars with radio spectra peaked at gigahertz frequencies have extremely high redshifts?}
\newblock {\em \mnras} {\bf 1990}, {\em 245},~20P.

\bibitem[{Marecki} et~al.(1999){Marecki}, {Falcke}, {Niezgoda}, {Garrington}, and {Patnaik}]{1999A&AS..135..273M}
{Marecki}, A.; {Falcke}, H.; {Niezgoda}, J.; {Garrington}, S.T.; {Patnaik}, A.R.
\newblock {Gigahertz Peaked Spectrum sources from the Jodrell Bank-VLA Astrometric Survey. I. Sources in the region 35degr {\lt}= delta {\lt}= 75degr}.
\newblock {\em \aaps} {\bf 1999}, {\em 135},~273--289.

\bibitem[{Dallacasa} et~al.(2000){Dallacasa}, {Stanghellini}, {Centonza}, and {Fanti}]{2000A&A...363..887D}
{Dallacasa}, D.; {Stanghellini}, C.; {Centonza}, M.; {Fanti}, R.
\newblock {High frequency peakers. I. The bright sample}.
\newblock {\em \aap} {\bf 2000}, {\em 363},~887--900,  \href{http://xxx.lanl.gov/abs/astro-ph/0012428}{{\normalfont [arXiv:astro-ph/astro-ph/0012428]}}.

\bibitem[{Tinti} et~al.(2005){Tinti}, {Dallacasa}, {de Zotti}, {Celotti}, and {Stanghellini}]{2005A&A...432...31T}
{Tinti}, S.; {Dallacasa}, D.; {de Zotti}, G.; {Celotti}, A.; {Stanghellini}, C.
\newblock {High Frequency Peakers: Young radio sources or flaring blazars?}
\newblock {\em \aap} {\bf 2005}, {\em 432},~31--43,  \href{http://xxx.lanl.gov/abs/astro-ph/0410663}{{\normalfont [arXiv:astro-ph/astro-ph/0410663]}}.
\newblock {\url{https://doi.org/10.1051/0004-6361:20041620}}.

\bibitem[{Mufakharov} et~al.(2021){Mufakharov}, {Mikhailov}, {Sotnikova}, {Mingaliev}, {Stolyarov}, {Erkenov}, {Nizhelskij}, and {Tsybulev}]{2021MNRAS.503.4662M}
{Mufakharov}, T.; {Mikhailov}, A.; {Sotnikova}, Y.; {Mingaliev}, M.; {Stolyarov}, V.; {Erkenov}, A.; {Nizhelskij}, N.; {Tsybulev}, P.
\newblock {Flux-density measurements of the high-redshift blazar PSO J047.4478+27.2992 at 4.7 and 8.2 GHz with RATAN-600}.
\newblock {\em \mnras} {\bf 2021}, {\em 503},~4662--4666,  \href{http://xxx.lanl.gov/abs/2011.12072}{{\normalfont [arXiv:astro-ph.HE/2011.12072]}}.
\newblock {\url{https://doi.org/10.1093/mnras/staa3688}}.

\bibitem[{Healey} et~al.(2007){Healey}, {Romani}, {Taylor}, {Sadler}, {Ricci}, {Murphy}, {Ulvestad}, and {Winn}]{2007ApJS..171...61H}
{Healey}, S.E.; {Romani}, R.W.; {Taylor}, G.B.; {Sadler}, E.M.; {Ricci}, R.; {Murphy}, T.; {Ulvestad}, J.S.; {Winn}, J.N.
\newblock {CRATES: An All-Sky Survey of Flat-Spectrum Radio Sources}.
\newblock {\em \apjs} {\bf 2007}, {\em 171},~61--71,  \href{http://xxx.lanl.gov/abs/astro-ph/0702346}{{\normalfont [arXiv:astro-ph/astro-ph/0702346]}}.
\newblock {\url{https://doi.org/10.1086/513742}}.

\bibitem[{Kimball} et~al.(2011){Kimball}, {Ivezi{\'c}}, {Wiita}, and {Schneider}]{2011AJ....141..182K}
{Kimball}, A.E.; {Ivezi{\'c}}, {\v{Z}}.; {Wiita}, P.J.; {Schneider}, D.P.
\newblock {Correlations of Quasar Optical Spectra with Radio Morphology}.
\newblock {\em \aj} {\bf 2011}, {\em 141},~182,  \href{http://xxx.lanl.gov/abs/1103.4791}{{\normalfont [arXiv:astro-ph.CO/1103.4791]}}.
\newblock {\url{https://doi.org/10.1088/0004-6256/141/6/182}}.

\bibitem[{Gordon} et~al.(2016){Gordon}, {Jacobs}, {Beasley}, {Peck}, {Gaume}, {Charlot}, {Fey}, {Ma}, {Titov}, and {Boboltz}]{2016AJ....151..154G}
{Gordon}, D.; {Jacobs}, C.; {Beasley}, A.; {Peck}, A.; {Gaume}, R.; {Charlot}, P.; {Fey}, A.; {Ma}, C.; {Titov}, O.; {Boboltz}, D.
\newblock {Second Epoch VLBA Calibrator Survey Observations: VCS-II}.
\newblock {\em \aj} {\bf 2016}, {\em 151},~154.
\newblock {\url{https://doi.org/10.3847/0004-6256/151/6/154}}.

\bibitem[{D'Abrusco} et~al.(2014){D'Abrusco}, {Massaro}, {Paggi}, {Smith}, {Masetti}, {Landoni}, and {Tosti}]{2014ApJS..215...14D}
{D'Abrusco}, R.; {Massaro}, F.; {Paggi}, A.; {Smith}, H.A.; {Masetti}, N.; {Landoni}, M.; {Tosti}, G.
\newblock {The WISE Blazar-like Radio-loud Sources: An All-sky Catalog of Candidate {\ensuremath{\gamma}}-ray Blazars}.
\newblock {\em \apjs} {\bf 2014}, {\em 215},~14,  \href{http://xxx.lanl.gov/abs/1410.0029}{{\normalfont [arXiv:astro-ph.HE/1410.0029]}}.
\newblock {\url{https://doi.org/10.1088/0067-0049/215/1/14}}.

\bibitem[{Petrov} et~al.(2006){Petrov}, {Kovalev}, {Fomalont}, and {Gordon}]{VCS4}
{Petrov}, L.; {Kovalev}, Y.Y.; {Fomalont}, E.B.; {Gordon}, D.
\newblock {The Fourth VLBA Calibrator Survey: VCS4}.
\newblock {\em \aj} {\bf 2006}, {\em 131},~1872--1879.
\newblock {\url{https://doi.org/10.1086/499947}}.

\bibitem[{Amirkhanyan} and {Mikhailov}(2006)]{2006Ap.....49..184A}
{Amirkhanyan}, V.R.; {Mikhailov}, V.P.
\newblock {The radio source Z0254+43: z = 4.067}.
\newblock {\em Astrophysics} {\bf 2006}, {\em 49},~184--193.
\newblock {\url{https://doi.org/10.1007/s10511-006-0019-x}}.

\bibitem[{Patnaik} et~al.(1992){Patnaik}, {Browne}, {Wilkinson}, and {Wrobel}]{1992MNRAS.254..655P}
{Patnaik}, A.R.; {Browne}, I.W.A.; {Wilkinson}, P.N.; {Wrobel}, J.M.
\newblock {Interferometer phase calibration sources - I. The region 35deg <= Dec. <= 75deg.}
\newblock {\em \mnras} {\bf 1992}, {\em 254},~655.
\newblock {\url{https://doi.org/10.1093/mnras/254.4.655}}.

\bibitem[{Lovell} et~al.(2008){Lovell}, {Rickett}, {Macquart}, {Jauncey}, {Bignall}, {Kedziora-Chudczer}, {Ojha}, {Pursimo}, {Dutka}, {Senkbeil}, and {Shabala}]{2008ApJ...689..108L}
{Lovell}, J.E.J.; {Rickett}, B.J.; {Macquart}, J.P.; {Jauncey}, D.L.; {Bignall}, H.E.; {Kedziora-Chudczer}, L.; {Ojha}, R.; {Pursimo}, T.; {Dutka}, M.; {Senkbeil}, C.;  et~al.
\newblock {The Micro-Arcsecond Scintillation-Induced Variability (MASIV) Survey. II. The First Four Epochs}.
\newblock {\em \apj} {\bf 2008}, {\em 689},~108--126,  \href{http://xxx.lanl.gov/abs/0808.1140}{{\normalfont [arXiv:astro-ph/0808.1140]}}.
\newblock {\url{https://doi.org/10.1086/592485}}.

\bibitem[{Chhetri} et~al.(2018){Chhetri}, {Morgan}, {Ekers}, {Macquart}, {Sadler}, {Giroletti}, {Callingham}, and {Tingay}]{2018MNRAS.474.4937C}
{Chhetri}, R.; {Morgan}, J.; {Ekers}, R.D.; {Macquart}, J.P.; {Sadler}, E.M.; {Giroletti}, M.; {Callingham}, J.R.; {Tingay}, S.J.
\newblock {Interplanetary scintillation studies with the Murchison Widefield Array - II. Properties of sub-arcsecond compact sources at low radio frequencies}.
\newblock {\em \mnras} {\bf 2018}, {\em 474},~4937--4955,  \href{http://xxx.lanl.gov/abs/1711.00393}{{\normalfont [arXiv:astro-ph.HE/1711.00393]}}.
\newblock {\url{https://doi.org/10.1093/mnras/stx2864}}.

\bibitem[{Ross} et~al.(2021){Ross}, {Callingham}, {Hurley-Walker}, {Seymour}, {Hancock}, {Franzen}, {Morgan}, {White}, {Bell}, and {Patil}]{2021MNRAS.501.6139R}
{Ross}, K.; {Callingham}, J.R.; {Hurley-Walker}, N.; {Seymour}, N.; {Hancock}, P.; {Franzen}, T.M.O.; {Morgan}, J.; {White}, S.V.; {Bell}, M.E.; {Patil}, P.
\newblock {Spectral variability of radio sources at low frequencies}.
\newblock {\em \mnras} {\bf 2021}, {\em 501},~6139--6155,  \href{http://xxx.lanl.gov/abs/2012.01842}{{\normalfont [arXiv:astro-ph.GA/2012.01842]}}.
\newblock {\url{https://doi.org/10.1093/mnras/staa3795}}.

\bibitem[{Nyland} et~al.(2020){Nyland}, {Dong}, {Patil}, {Lacy}, {van Velzen}, {Kimball}, {Sarbadhicary}, {Hallinan}, {Baldassare}, {Clarke}, {Goulding}, {Greene}, {Hughes}, {Kassim}, {Kunert-Bajraszewska}, {Maccarone}, {Mooley}, {Mukherjee}, {Peters}, {Petrov}, {Polisensky}, {Rujopakarn}, {Whittle}, and {Vaccari}]{2020ApJ...905...74N}
{Nyland}, K.; {Dong}, D.Z.; {Patil}, P.; {Lacy}, M.; {van Velzen}, S.; {Kimball}, A.E.; {Sarbadhicary}, S.K.; {Hallinan}, G.; {Baldassare}, V.; {Clarke}, T.E.;  et~al.
\newblock {Quasars That Have Transitioned from Radio-quiet to Radio-loud on Decadal Timescales Revealed by VLASS and FIRST}.
\newblock {\em \apj} {\bf 2020}, {\em 905},~74,  \href{http://xxx.lanl.gov/abs/2011.08872}{{\normalfont [arXiv:astro-ph.GA/2011.08872]}}.
\newblock {\url{https://doi.org/10.3847/1538-4357/abc341}}.

\bibitem[{Chih Hsu} et~al.(2023){Chih Hsu}, {Koay}, {Matsushita}, {Hwang}, {Hovatta}, {Kiehlmann}, {Readhead}, {Max-Moerbeck}, and {Reeves}]{2023arXiv230809626C}
{Chih Hsu}, P.; {Koay}, J.Y.; {Matsushita}, S.; {Hwang}, C.Y.; {Hovatta}, T.; {Kiehlmann}, S.; {Readhead}, A.; {Max-Moerbeck}, W.; {Reeves}, R.
\newblock {Milliarcsecond Core Size Dependence of the Radio Variability of Blazars}.
\newblock {\em arXiv e-prints} {\bf 2023}, p. arXiv:2308.09626,  \href{http://xxx.lanl.gov/abs/2308.09626}{{\normalfont [arXiv:astro-ph.GA/2308.09626]}}.
\newblock {\url{https://doi.org/10.48550/arXiv.2308.09626}}.

\bibitem[{Kiehlmann} et~al.(2023{\natexlab{a}}){Kiehlmann}, {Readhead}, {O'Neill}, {Wilkinson}, {Lister}, {Liodakis}, {Bruzewski}, {Pearson}, {Sheldahl}, {Siemiginowska}, {Tassis}, and {Taylor}]{2023arXiv230311359K}
{Kiehlmann}, S.; {Readhead}, A.C.S.; {O'Neill}, S.; {Wilkinson}, P.N.; {Lister}, M.L.; {Liodakis}, I.; {Bruzewski}, S.; {Pearson}, T.J.; {Sheldahl}, E.; {Siemiginowska}, A.;  et~al.
\newblock {Compact Symmetric Objects: A Distinct Population of Jetted Active Galaxies}.
\newblock {\em arXiv e-prints} {\bf 2023}, p. arXiv:2303.11359,  \href{http://xxx.lanl.gov/abs/2303.11359}{{\normalfont [arXiv:astro-ph.HE/2303.11359]}}.
\newblock {\url{https://doi.org/10.48550/arXiv.2303.11359}}.

\bibitem[{Kiehlmann} et~al.(2023{\natexlab{b}}){Kiehlmann}, {Lister}, {Readhead}, {Liodakis}, {O'Neill}, {Pearson}, {Sheldahl}, {Siemiginowska}, {Tassis}, {Taylor}, and {Wilkinson}]{2023arXiv230311357K}
{Kiehlmann}, S.; {Lister}, M.L.; {Readhead}, A.C.S.; {Liodakis}, I.; {O'Neill}, S.; {Pearson}, T.J.; {Sheldahl}, E.; {Siemiginowska}, A.; {Tassis}, K.; {Taylor}, G.B.;  et~al.
\newblock {Towards a Comprehensive Catalog of Bona Fide Compact Symmetric Objects}.
\newblock {\em arXiv e-prints} {\bf 2023}, p. arXiv:2303.11357,  \href{http://xxx.lanl.gov/abs/2303.11357}{{\normalfont [arXiv:astro-ph.HE/2303.11357]}}.
\newblock {\url{https://doi.org/10.48550/arXiv.2303.11357}}.

\bibitem[{Perger} et~al.(2021){Perger}, {Frey}, {Schwartz}, {Gab{\'a}nyi}, {Gurvits}, and {Paragi}]{2021ApJ...915...98P}
{Perger}, K.; {Frey}, S.; {Schwartz}, D.A.; {Gab{\'a}nyi}, K.{\'E}.; {Gurvits}, L.I.; {Paragi}, Z.
\newblock {Multi-scale Radio and X-Ray Structure of the High-redshift Quasar PMN J0909+0354}.
\newblock {\em \apj} {\bf 2021}, {\em 915},~98,  \href{http://xxx.lanl.gov/abs/2105.06307}{{\normalfont [arXiv:astro-ph.GA/2105.06307]}}.
\newblock {\url{https://doi.org/10.3847/1538-4357/ac0144}}.

\bibitem[Pedregosa et~al.(2011)Pedregosa, Varoquaux, Gramfort, Michel, Thirion, Grisel, Blondel, Prettenhofer, Weiss, Dubourg, Vanderplas, Passos, Cournapeau, Brucher, Perrot, and {{\'E}}douard Duchesnay]{JMLR:v12:pedregosa11a}
Pedregosa, F.; Varoquaux, G.; Gramfort, A.; Michel, V.; Thirion, B.; Grisel, O.; Blondel, M.; Prettenhofer, P.; Weiss, R.; Dubourg, V.;  et~al.
\newblock Scikit-learn: Machine Learning in Python.
\newblock {\em Journal of Machine Learning Research} {\bf 2011}, {\em 12},~2825--2830.

\bibitem[Bengfort and Bilbro(2019)]{Bengfort2019}
Bengfort, B.; Bilbro, R.
\newblock Yellowbrick: Visualizing the Scikit-Learn Model Selection Process.
\newblock {\em Journal of Open Source Software} {\bf 2019}, {\em 4},~1075.
\newblock {\url{https://doi.org/10.21105/joss.01075}}.

\bibitem[van~der Maaten and Hinton(2008)]{vanDerMaaten2008}
van~der Maaten, L.; Hinton, G.
\newblock Visualizing Data using {t-SNE}.
\newblock {\em Journal of Machine Learning Research} {\bf 2008}, {\em 9},~2579--2605.

\bibitem[Wittek et~al.(2017)Wittek, Gao, Lim, and Zhao]{JSSv078i09}
Wittek, P.; Gao, S.C.; Lim, I.S.; Zhao, L.
\newblock somoclu: An Efficient Parallel Library for Self-Organizing Maps.
\newblock {\em Journal of Statistical Software} {\bf 2017}, {\em 78},~1–21.
\newblock {\url{https://doi.org/10.18637/jss.v078.i09}}.

\bibitem[{Gorshkov} et~al.(2012){Gorshkov}, {Konnikova}, and {Mingaliev}]{2012ARep...56..345G}
{Gorshkov}, A.G.; {Konnikova}, V.K.; {Mingaliev}, M.G.
\newblock {Long-term variability of a complete sample of flat-spectrum radio sources at declinations 10{\textdegree}-12{\textdegree}30' (J2000)}.
\newblock {\em Astronomy Reports} {\bf 2012}, {\em 56},~345--362.
\newblock {\url{https://doi.org/10.1134/S106377291204004X}}.

\bibitem[{Drinkwater} et~al.(1997){Drinkwater}, {Webster}, {Francis}, {Condon}, {Ellison}, {Jauncey}, {Lovell}, {Peterson}, and {Savage}]{1997MNRAS.284...85D}
{Drinkwater}, M.J.; {Webster}, R.L.; {Francis}, P.J.; {Condon}, J.J.; {Ellison}, S.L.; {Jauncey}, D.L.; {Lovell}, J.; {Peterson}, B.A.; {Savage}, A.
\newblock {The Parkes Half-Jansky Flat-Spectrum Sample}.
\newblock {\em \mnras} {\bf 1997}, {\em 284},~85--125,  \href{http://xxx.lanl.gov/abs/astro-ph/9609019}{{\normalfont [arXiv:astro-ph/astro-ph/9609019]}}.
\newblock {\url{https://doi.org/10.1093/mnras/284.1.85}}.

\bibitem[{Bennett} et~al.(1986){Bennett}, {Lawrence}, {Burke}, {Hewitt}, and {Mahoney}]{1986ApJS...61....1B}
{Bennett}, C.L.; {Lawrence}, C.R.; {Burke}, B.F.; {Hewitt}, J.N.; {Mahoney}, J.
\newblock {The MIT--Green Bank (MG) 5 GHz Survey}.
\newblock {\em \apjs} {\bf 1986}, {\em 61},~1.
\newblock {\url{https://doi.org/10.1086/191108}}.

\bibitem[{Torniainen} et~al.(2008){Torniainen}, {Tornikoski}, {Turunen}, {Lainela}, {L{\"a}hteenm{\"a}ki}, {Hovatta}, {Mingaliev}, {Aller}, and {Aller}]{2008A&A...482..483T}
{Torniainen}, I.; {Tornikoski}, M.; {Turunen}, M.; {Lainela}, M.; {L{\"a}hteenm{\"a}ki}, A.; {Hovatta}, T.; {Mingaliev}, M.G.; {Aller}, M.F.; {Aller}, H.D.
\newblock {Cluster analyses of gigahertz-peaked spectrum sources with self-organising maps}.
\newblock {\em \aap} {\bf 2008}, {\em 482},~483--498.
\newblock {\url{https://doi.org/10.1051/0004-6361:20079222}}.

\bibitem[{Jeong} et~al.(2016){Jeong}, {Sohn}, {Chung}, {Park}, and {Park}]{2016AN....337..130J}
{Jeong}, Y.; {Sohn}, B.W.; {Chung}, A.; {Park}, S.; {Park}, P.
\newblock {Identifying High Frequency Peakers using the Korean VLBI Network}.
\newblock {\em Astronomische Nachrichten} {\bf 2016}, {\em 337},~130.
\newblock {\url{https://doi.org/10.1002/asna.201512278}}.

\bibitem[{Wright} et~al.(1996){Wright}, {Griffith}, {Hunt}, {Troup}, {Burke}, and {Ekers}]{1996ApJS..103..145W}
{Wright}, A.E.; {Griffith}, M.R.; {Hunt}, A.J.; {Troup}, E.; {Burke}, B.F.; {Ekers}, R.D.
\newblock {The Parkes-MIT-NRAO (PMN) Surveys. VIII. Source Catalog for the Zenith Survey (-37.0 degrees < delta < -29.0 degrees )}.
\newblock {\em \apjs} {\bf 1996}, {\em 103},~145.
\newblock {\url{https://doi.org/10.1086/192272}}.

\bibitem[{Brinkmann} et~al.(1995){Brinkmann}, {Siebert}, {Reich}, {Fuerst}, {Reich}, {Voges}, {Truemper}, and {Wielebinski}]{1995A&AS..109..147B}
{Brinkmann}, W.; {Siebert}, J.; {Reich}, W.; {Fuerst}, E.; {Reich}, P.; {Voges}, W.; {Truemper}, J.; {Wielebinski}, R.
\newblock {The ROSAT AGN content of the 87GB 5 GHz survey: bulk properties of previously optically identified sources.}
\newblock {\em \aaps} {\bf 1995}, {\em 109},~147--170.

\bibitem[{Mu{\~n}oz} et~al.(2003){Mu{\~n}oz}, {Falco}, {Kochanek}, {Leh{\'a}r}, and {Mediavilla}]{2003ApJ...594..684M}
{Mu{\~n}oz}, J.A.; {Falco}, E.E.; {Kochanek}, C.S.; {Leh{\'a}r}, J.; {Mediavilla}, E.
\newblock {The Redshift Distribution of Flat-Spectrum Radio Sources}.
\newblock {\em \apj} {\bf 2003}, {\em 594},~684--694,  \href{http://xxx.lanl.gov/abs/astro-ph/0305382}{{\normalfont [arXiv:astro-ph/astro-ph/0305382]}}.
\newblock {\url{https://doi.org/10.1086/377077}}.

\bibitem[{Snios} et~al.(2021){Snios}, {Schwartz}, {Siemiginowska}, {Sobolewska}, {Birkinshaw}, {Cheung}, {Gobeille}, {Marshall}, {Migliori}, {Wardle}, and {Worrall}]{2021ApJ...914..130S}
{Snios}, B.; {Schwartz}, D.A.; {Siemiginowska}, A.; {Sobolewska}, M.; {Birkinshaw}, M.; {Cheung}, C.C.; {Gobeille}, D.B.; {Marshall}, H.L.; {Migliori}, G.; {Wardle}, J.F.C.;  et~al.
\newblock {Discovery of Candidate X-Ray Jets in High-redshift Quasars}.
\newblock {\em \apj} {\bf 2021}, {\em 914},~130,  \href{http://xxx.lanl.gov/abs/2102.12609}{{\normalfont [arXiv:astro-ph.HE/2102.12609]}}.
\newblock {\url{https://doi.org/10.3847/1538-4357/abfe64}}.

\bibitem[{Riley} et~al.(1999){Riley}, {Waldram}, and {Riley}]{1999MNRAS.306...31R}
{Riley}, J.M.W.; {Waldram}, E.M.; {Riley}, J.M.
\newblock {The 7C survey of radio sources at 151 MHz: 33 regions in the range 7h < RA < 17h, 30 deg <Dec. <58 deg}.
\newblock {\em \mnras} {\bf 1999}, {\em 306},~31--34.
\newblock {\url{https://doi.org/10.1046/j.1365-8711.1999.02416.x}}.

\bibitem[{Liu} and {Zhang}(2002)]{2002A&A...381..757L}
{Liu}, F.K.; {Zhang}, Y.H.
\newblock {A new list of extra-galactic radio jets}.
\newblock {\em \aap} {\bf 2002}, {\em 381},~757--760,  \href{http://xxx.lanl.gov/abs/astro-ph/0212477}{{\normalfont [arXiv:astro-ph/astro-ph/0212477]}}.
\newblock {\url{https://doi.org/10.1051/0004-6361:20011572}}.

\bibitem[{Yang} et~al.(2006){Yang}, {Gurvits}, {Lobanov}, {Frey}, and {Hong}]{2006evn..confE..86Y}
{Yang}, J.; {Gurvits}, L.; {Lobanov}, A.; {Frey}, S.; {Hong}, X.
\newblock {Dual frequency VSOP imaging of a high redshift quasar PKS 1402+044}.
\newblock In Proceedings of the Proceedings of the 8th European VLBI Network Symposium; {Baan}, W.; {Bachiller}, R.; {Booth}, R.; {Charlot}, P.; {Diamond}, P.; {Garrett}, M.; {Hong}, X.; {Jonas}, J.; {Kus}, A.; {Mantovani}, F.;  et~al., Eds.,  2006, p.~86,  \href{http://xxx.lanl.gov/abs/astro-ph/0611425}{{\normalfont [arXiv:astro-ph/astro-ph/0611425]}}.

\bibitem[{Dodson} et~al.(2008){Dodson}, {Fomalont}, {Wiik}, {Horiuchi}, {Hirabayashi}, {Edwards}, {Murata}, {Asaki}, {Moellenbrock}, {Scott}, {Taylor}, {Gurvits}, {Paragi}, {Frey}, {Shen}, {Lovell}, {Tingay}, {Rioja}, {Fodor}, {Lister}, {Mosoni}, {Coldwell}, {Piner}, and {Yang}]{2008ApJS..175..314D}
{Dodson}, R.; {Fomalont}, E.B.; {Wiik}, K.; {Horiuchi}, S.; {Hirabayashi}, H.; {Edwards}, P.G.; {Murata}, Y.; {Asaki}, Y.; {Moellenbrock}, G.A.; {Scott}, W.K.;  et~al.
\newblock {The VSOP 5 GHz Active Galactic Nucleus Survey. V. Imaging Results for the Remaining 140 Sources}.
\newblock {\em \apjs} {\bf 2008}, {\em 175},~314--355,  \href{http://xxx.lanl.gov/abs/0710.5707}{{\normalfont [arXiv:astro-ph/0710.5707]}}.
\newblock {\url{https://doi.org/10.1086/525025}}.

\bibitem[{Lawrence} et~al.(1984){Lawrence}, {Schneider}, {Schmidt}, {Bennett}, {Hewitt}, {Burke}, {Turner}, and {Gunn}]{1984Sci...223...46L}
{Lawrence}, C.R.; {Schneider}, D.P.; {Schmidt}, M.; {Bennett}, C.L.; {Hewitt}, J.N.; {Burke}, B.F.; {Turner}, E.L.; {Gunn}, J.E.
\newblock {Discovery of a New Gravitational Lens System}.
\newblock {\em Science} {\bf 1984}, {\em 223},~46--49.
\newblock {\url{https://doi.org/10.1126/science.223.4631.46}}.

\bibitem[{More} et~al.(2009){More}, {McKean}, {More}, {Porcas}, {Koopmans}, and {Garrett}]{2009MNRAS.394..174M}
{More}, A.; {McKean}, J.P.; {More}, S.; {Porcas}, R.W.; {Koopmans}, L.V.E.; {Garrett}, M.A.
\newblock {The role of luminous substructure in the gravitational lens system MG 2016+112}.
\newblock {\em \mnras} {\bf 2009}, {\em 394},~174--190,  \href{http://xxx.lanl.gov/abs/0810.5341}{{\normalfont [arXiv:astro-ph/0810.5341]}}.
\newblock {\url{https://doi.org/10.1111/j.1365-2966.2008.14342.x}}.

\bibitem[{Wu} et~al.(2013){Wu}, {Brandt}, {Miller}, {Garmire}, {Schneider}, and {Vignali}]{2013ApJ...763..109W}
{Wu}, J.; {Brandt}, W.N.; {Miller}, B.P.; {Garmire}, G.P.; {Schneider}, D.P.; {Vignali}, C.
\newblock {An X-Ray and Multiwavelength Survey of Highly Radio-loud Quasars at z > 4: Jet-linked Emission in the Brightest Radio Beacons of the Early Universe}.
\newblock {\em \apj} {\bf 2013}, {\em 763},~109,  \href{http://xxx.lanl.gov/abs/1301.0012}{{\normalfont [arXiv:astro-ph.CO/1301.0012]}}.
\newblock {\url{https://doi.org/10.1088/0004-637X/763/2/109}}.

\bibitem[{Zhu} et~al.(2019){Zhu}, {Brandt}, {Wu}, {Garmire}, and {Miller}]{2019MNRAS.482.2016Z}
{Zhu}, S.F.; {Brandt}, W.N.; {Wu}, J.; {Garmire}, G.P.; {Miller}, B.P.
\newblock {Investigating the X-ray enhancements of highly radio-loud quasars at z > 4}.
\newblock {\em \mnras} {\bf 2019}, {\em 482},~2016--2038,  \href{http://xxx.lanl.gov/abs/1810.06572}{{\normalfont [arXiv:astro-ph.HE/1810.06572]}}.
\newblock {\url{https://doi.org/10.1093/mnras/sty2832}}.

\bibitem[{Lagattuta} et~al.(2010){Lagattuta}, {Auger}, and {Fassnacht}]{2010ApJ...716L.185L}
{Lagattuta}, D.J.; {Auger}, M.W.; {Fassnacht}, C.D.
\newblock {Adaptive Optics Observations of B0128+437: A Low-mass, High-redshift Gravitational Lens}.
\newblock {\em \apjl} {\bf 2010}, {\em 716},~L185--L189,  \href{http://xxx.lanl.gov/abs/0912.2344}{{\normalfont [arXiv:astro-ph.CO/0912.2344]}}.
\newblock {\url{https://doi.org/10.1088/2041-8205/716/2/L185}}.

\bibitem[{Phillips} et~al.(2000){Phillips}, {Norbury}, {Koopmans}, {Browne}, {Jackson}, {Wilkinson}, {Biggs}, {Blandford}, {de Bruyn}, {Fassnacht}, {Helbig}, {Mao}, {Marlow}, {Myers}, {Pearson}, {Readhead}, {Rusin}, and {Xanthopoulos}]{2000MNRAS.319L...7P}
{Phillips}, P.M.; {Norbury}, M.A.; {Koopmans}, L.V.E.; {Browne}, I.W.A.; {Jackson}, N.J.; {Wilkinson}, P.N.; {Biggs}, A.D.; {Blandford}, R.D.; {de Bruyn}, A.G.; {Fassnacht}, C.D.;  et~al.
\newblock {A new quadruple gravitational lens system: CLASS B0128+437}.
\newblock {\em \mnras} {\bf 2000}, {\em 319},~L7--L11,  \href{http://xxx.lanl.gov/abs/astro-ph/0009334}{{\normalfont [arXiv:astro-ph/astro-ph/0009334]}}.
\newblock {\url{https://doi.org/10.1046/j.1365-8711.2000.04033.x}}.

\bibitem[{Tonry} and {Kochanek}(1999)]{1999AJ....117.2034T}
{Tonry}, J.L.; {Kochanek}, C.S.
\newblock {Redshifts of the Gravitational Lenses MG 0414+0534 and MG 0751+2716}.
\newblock {\em \aj} {\bf 1999}, {\em 117},~2034--2038,  \href{http://xxx.lanl.gov/abs/astro-ph/9809063}{{\normalfont [arXiv:astro-ph/astro-ph/9809063]}}.
\newblock {\url{https://doi.org/10.1086/300834}}.

\bibitem[{Lehar} et~al.(1997){Lehar}, {Burke}, {Conner}, {Falco}, {Fletcher}, {Irwin}, {McMahon}, {Muslow}, and {Schechter}]{1997AJ....114...48L}
{Lehar}, J.; {Burke}, B.F.; {Conner}, S.R.; {Falco}, E.E.; {Fletcher}, A.B.; {Irwin}, M.; {McMahon}, R.G.; {Muslow}, T.W.B.; {Schechter}, P.L.
\newblock {The Gravitationally Lensed Radio Source MG 0751+2716}.
\newblock {\em \aj} {\bf 1997}, {\em 114},~48--53,  \href{http://xxx.lanl.gov/abs/astro-ph/9702191}{{\normalfont [arXiv:astro-ph/astro-ph/9702191]}}.
\newblock {\url{https://doi.org/10.1086/118451}}.

\bibitem[{Rusin} et~al.(2001){Rusin}, {Kochanek}, {Norbury}, {Falco}, {Impey}, {Leh{\'a}r}, {McLeod}, {Rix}, {Keeton}, {Mu{\~n}oz}, and {Peng}]{2001ApJ...557..594R}
{Rusin}, D.; {Kochanek}, C.S.; {Norbury}, M.; {Falco}, E.E.; {Impey}, C.D.; {Leh{\'a}r}, J.; {McLeod}, B.A.; {Rix}, H.W.; {Keeton}, C.R.; {Mu{\~n}oz}, J.A.;  et~al.
\newblock {B1359+154: A Six-Image Lens Produced by a z \raisebox{-0.5ex}\textasciitilde= 1 Compact Group of Galaxies}.
\newblock {\em \apj} {\bf 2001}, {\em 557},~594--604,  \href{http://xxx.lanl.gov/abs/astro-ph/0011505}{{\normalfont [arXiv:astro-ph/astro-ph/0011505]}}.
\newblock {\url{https://doi.org/10.1086/322251}}.

\bibitem[{Tonry}(1998)]{1998AJ....115....1T}
{Tonry}, J.L.
\newblock {Redshifts of the Gravitational Lenses B1422+231 and PG 1115+080}.
\newblock {\em \aj} {\bf 1998}, {\em 115},~1--5,  \href{http://xxx.lanl.gov/abs/astro-ph/9706199}{{\normalfont [arXiv:astro-ph/astro-ph/9706199]}}.
\newblock {\url{https://doi.org/10.1086/300170}}.

\bibitem[{Patnaik} et~al.(1992){Patnaik}, {Browne}, {Walsh}, {Chaffee}, and {Foltz}]{1992MNRAS.259P...1P}
{Patnaik}, A.R.; {Browne}, I.W.A.; {Walsh}, D.; {Chaffee}, F.H.; {Foltz}, C.B.
\newblock {B 1422+231 : a new gravitationally lensed system at Z = 3.62.}
\newblock {\em \mnras} {\bf 1992}, {\em 259},~1P--4.
\newblock {\url{https://doi.org/10.1093/mnras/259.1.1P}}.

\bibitem[{Mao} and {Schneider}(1998)]{1998MNRAS.295..587M}
{Mao}, S.; {Schneider}, P.
\newblock {Evidence for substructure in lens galaxies?}
\newblock {\em \mnras} {\bf 1998}, {\em 295},~587--594,  \href{http://xxx.lanl.gov/abs/astro-ph/9707187}{{\normalfont [arXiv:astro-ph/astro-ph/9707187]}}.
\newblock {\url{https://doi.org/10.1046/j.1365-8711.1998.01319.x}}.

\bibitem[{Ofek} et~al.(2006){Ofek}, {Maoz}, {Rix}, {Kochanek}, and {Falco}]{2006ApJ...641...70O}
{Ofek}, E.O.; {Maoz}, D.; {Rix}, H.W.; {Kochanek}, C.S.; {Falco}, E.E.
\newblock {Spectroscopic Redshifts for Seven Lens Galaxies}.
\newblock {\em \apj} {\bf 2006}, {\em 641},~70--77,  \href{http://xxx.lanl.gov/abs/astro-ph/0510465}{{\normalfont [arXiv:astro-ph/astro-ph/0510465]}}.
\newblock {\url{https://doi.org/10.1086/500403}}.

\bibitem[{Winn} et~al.(2002){Winn}, {Morgan}, {Hewitt}, {Kochanek}, {Lovell}, {Patnaik}, {Pindor}, {Schechter}, and {Schommer}]{2002AJ....123...10W}
{Winn}, J.N.; {Morgan}, N.D.; {Hewitt}, J.N.; {Kochanek}, C.S.; {Lovell}, J.E.J.; {Patnaik}, A.R.; {Pindor}, B.; {Schechter}, P.L.; {Schommer}, R.A.
\newblock {PMN J1632-0033: A New Gravitationally Lensed Quasar}.
\newblock {\em \aj} {\bf 2002}, {\em 123},~10--19,  \href{http://xxx.lanl.gov/abs/astro-ph/0104092}{{\normalfont [arXiv:astro-ph/astro-ph/0104092]}}.
\newblock {\url{https://doi.org/10.1086/338094}}.

\bibitem[{Winn} et~al.(2004){Winn}, {Rusin}, and {Kochanek}]{2004Natur.427..613W}
{Winn}, J.N.; {Rusin}, D.; {Kochanek}, C.S.
\newblock {The central image of a gravitationally lensed quasar}.
\newblock {\em \nat} {\bf 2004}, {\em 427},~613--615,  \href{http://xxx.lanl.gov/abs/astro-ph/0312136}{{\normalfont [arXiv:astro-ph/astro-ph/0312136]}}.
\newblock {\url{https://doi.org/10.1038/nature02279}}.

\end{thebibliography}

\PublishersNote{}
\end{adjustwidth}
\end{document}